\documentclass{article}\usepackage[paper=a4paper,top=20mm, bottom=30mm, inner=20mm, outer=20mm]{geometry}

\usepackage{comment}

\usepackage{natbib}

\usepackage{mathtools,amssymb,mathrsfs,bm,centernot,siunitx}
\usepackage[scr=boondox]{mathalpha}	
\usepackage{graphicx}
\usepackage{caption,subcaption,float}
\usepackage{tabularx,multirow}
\usepackage[inline]{enumitem}

\usepackage{xcolor}

\usepackage{calc,etoolbox}

\usepackage[inline]{enumitem}

\makeatletter

\@ifclassloaded{article}{
	\numberwithin{equation}{section}
	\numberwithin{figure}{section}
	\numberwithin{table}{section}
	
	\def\fps@figure{!tp}
}{}

\@ifclassloaded{agujournal2019}{}{
	\@ifpackageloaded{hyperref}{}{\usepackage[hidelinks]{hyperref}}
}
\@ifpackageloaded{cleveref}{}{
	\usepackage[compress]{cleveref}
	
	\crefformat{section}{\S#2#1#3}
	\crefformat{subsection}{\S#2#1#3}
	\crefformat{subsubsection}{\S#2#1#3}
	\crefrangeformat{section}{\S\S#3#1#4 to~#5#2#6}
	\crefmultiformat{section}{\S#2#1#3}{ and~\S#2#1#3}{, \S#2#1#3}{ and~\S#2#1#3}
	
	\@ifclassloaded{jfm}{
		\crefname{figure}{figure}{figures}
		\Crefname{figure}{Figure}{Figures}
	}{
		\crefname{figure}{fig.}{figs.}
		\Crefname{figure}{Fig.}{Figs.}
	}
	
	\crefname{equation}{}{}
	\crefformat{equation}{(#2#1#3)}
	\crefmultiformat{equation}{(#2#1#3)}{ and (#2#1#3)}{, (#2#1#3)}{ and (#2#1#3)}
}

\@ifclassloaded{WileyNJD-v2}{}{
	\usepackage[title]{appendix}
	\newenvironment{myappendices}
	{
			\begin{appendices}
				\crefalias{section}{appsec}
			}{
		\end{appendices}
	}
	
	\crefname{appsec}{Appendix}{Appendices}
	\Crefname{appsec}{Appendix}{Appendices}
}

\usepackage{xspace}

\DeclareRobustCommand\onedot{\futurelet\@let@token\@onedot}
\def\@onedot{\ifx\@let@token.\else.\null\fi\xspace}

\def\eg{e.g\onedot} 
\def\ie{i.e\onedot}

\def\sf{s.f\onedot}

\@ifclassloaded{svjour3}{}{
	\newcommand{\note}[2][self]{\textnormal{\normalsize{\textcolor{red}{\textsc{Note to #1: #2}}}}}
}

\newcommand{\floattag}[1]{%
	\@namedef{the\@captype}{#1}%
	\@namedef{theH\@captype}{#1}%
	\addtocounter{\@captype}{-1}}

\makeatother
\input{Templates/StandardMathsMacros}
\graphicspath{{Figures/}}

\usepackage{caption}

\newcommand{\ElTu}[1]{ \mbox{$\mbox{$#1$}_{\text{ET}}$} }

\newlength{\multlinedwidth}
\begin{document}

\title{Gravity current energetics and particle suspension}

\author{Edward W.G. Skevington and Robert M. Dorrell}

\maketitle

\begin{abstract}
	Gravity currents are a ubiquitous density driven flow occurring in both the natural environment and in industry. They include: seafloor turbidity currents, primary vectors of sediment, nutrient and pollutant transport; cold fronts; and hazardous gas spills. However, while the energetics are critical for their evolution and particle suspension, 
	they are included in system scale models only crudely,
	so we cannot yet predict and explain the dynamics and run-out of such real-world flows. Herein, a novel depth-averaged framework is developed to capture the evolution of volume, concentration, momentum, and turbulent kinetic energy 
	from direct integrals of the full governing equations. 
	For the first time, we show 
	the connection between the vertical profiles, the evolution of the depth-averaged flow, and the energetics.
	The viscous dissipation of mean-flow energy near the bed makes a leading order contribution, and an energetic approach to entrainment captures detrainment of fluid through particle settling.
	These observations allow a reconsideration of particle suspension, advancing over 50-years of research.
	We find that the new formulation can describe the full evolution of a shallow dilute current, with the accuracy depending primarily on closures for the profiles and source terms. Critically, this enables accurate and computationally efficient hazard risk analysis and earth surface modelling.
\end{abstract}

\section{Introduction} \label{sec:intro}

Gravity currents are fluid flows driven across a horizontal or shallowly sloped boundary by a density difference with the surrounding fluid. These include industrial accidents such as the spreading of toxic gas \citep{ar_Rottman_1985}, the failure of carbon dioxide pipelines \citep{ar_Liu_2019}, and oil spills \citep{ar_Hoult_1972}, along with environmental flows such as cold fronts, katabatic winds, salinity currents \citep{ar_Simpson_1982,bk_Simpson_GCEL} and currents within frozen lakes \citep{ar_Jansen_2021}. In submarine environments, suspended particle loads cause an excess of density over the surrounding ambient water, generating a gravity current. These particle-driven turbidity currents play a dominant role in oceanic transport processes, distributing particulates, nutrients, and pollutants from the continental margin to the deep ocean, preserving a record of paleo-environments, and posing a hazard to submarine infrastructure such as cables and pipes \citep{ar_Carter_2015,ar_Hsu_2008}. The run-out of these currents is impressive, some currents traversing thousands of kilometres \citep{ar_Lewis_1994,ar_Savoye_2009}, and the cumulative deposits can be enormous, up to $10^7 \, \si{km^3}$ \citep{ar_Curray_2002}. Consequently, the dynamics of these currents is of practical interest \citep{ar_Reece_2024}.

\subsection{Background}

To maintain their excess density, turbidity currents must suspend their particle load \citep{ar_Wells_2021}. Early investigation into the energetics of particle suspension was performed by \citet{ar_Knapp_1938} and \citet{ar_Bagnold_1962}. They investigated the auto-suspension  for steady currents on sloped beds, where an arbitrarily large amount of particles of a particular size can be transported by a current. These authors observed that, for flow downhill, the driving force of the current was enhanced by having an additional particle in suspension due to the mass of the particle. When the energy required to lift a particle against settling was below the energy provided by the increased driving force then having the particle in suspension was a net gain to the current's energy, and auto-suspension was possible. However, this analysis gives no thought to the means by which the work done by the downslope force turns into work uplifting particles, which will involve energy losses so that the Knapp-Bagnold condition is necessary but not sufficient. As discussed by \citet{ar_Bagnold_1966} the uplift of particles is turbulent in nature. Consequently, it is understood that the downslope force maintains the speed of the current against Reynolds stresses, the energy lost to Reynolds stresses transferred to entrained fluid and the turbulent kinetic energy (TKE), and this turbulence uplifts the particles as a buoyancy-flux. In turbidity currents, the bulk energetics of this process has been captured in the model of \citet{ar_Parker_1986} by depth-averaging the governing equations. The model captures the self-acceleration that can happen when the work by downslope gravity exceeds the energy required to suspend particles, resulting in an igniting current that progressively increases in both volume and sediment mass. 

After four decades the model presented by \citet{ar_Parker_1986}, and similar depth-average shallow-water models \citep{ar_Bonnecaze_1993}, remain a foundation for the theoretical understanding of turbidity currents \citep{ar_Wells_2021,ar_Wahab_2022,ar_Talling_2023}. The depth-average model itself captures the body of the current, which is well approximated as hydrostatic; the non-hydrostatic front of the current must be captured separately as a boundary condition \citep{ar_Benjamin_1968,ar_Ungarish_2018}. Depth-average models in general are used across the breadth of gravity current research \citep{bk_Stoker_WW,ar_Ellison_1959,ar_Huppert_2006,ar_Meiburg_2015,bk_Ungarish_GCIv2}. These models are mathematically simple, and can therefore be used to analyse idealised gravity current dynamics such as collision with obstacles \citep{ar_Skevington_F001_Draining,ar_Skevington_F003_CritWall_Overtopping,ar_Skevington_F004_Backflow,ar_Skevington_F002_SemiInf_Wall} or flow over an edge \citep{ar_Momen_2017,ar_Ungarish_2019,ar_Skevington_F005_RadSuper}. In addition to the conceptual insights they provide, there is a substantial reduction in complexity from a 3-dimensional model which would be simulated directly, or through a large-eddy or Reynolds-averaged approach, to a depth-average model. This simplicity results in a substantial increase in the spatial and temporal scales which can be simulated. For example, \cite{ar_Wahab_2022} were able to simulate the morphodynamic evolution of the submarine fans generated by turbidity currents over geophysical scales. Similarly, vast numbers of large-scale events must be simulated for the hazard forecasting and risk management of other classes of gravity currents, such as hazardous gas spills, avalanches, and pyroclastic density currents. Thus, it is important to continue the development of the depth-average modelling framework to ensure accurate prediction.

For accurate prediction of particle-driven currents, depth-average models must accurately capture the energetics. The energy expended to hold the particles in suspension consumes the TKE of the current. It is entirely possible for the sinks of TKE from uplift and viscous dissipation to exceed production, at which point the turbulent energy begins to decrease. In equilibrium simulations, it has been shown that there is a sharp threshold, with the magnitude and distribution of TKE varying weakly with settling velocity until a total collapse of the turbulence at a critical value \citep{ar_Shringarpure_2012,ar_Cantero_2012} which has been confirmed empirically \citep{ar_Eggenhuisen_2017}. Out of equilibrium, being overloaded with particles does not necessarily result in a collapse of turbulence, instead causing deposition to a reduced sediment load \citep{ar_Dorrell_2013}. There is evidence that after transition onto a shallow slope the self-acceleration feedback sometimes runs in reverse, a reduction in driving force causing less turbulence causing particle deposition causing less driving force, resulting in a sudden deposition of the full transported load \citep{ar_Talling_2007}.

\subsection{Motivation}

We ask the question: are the energetics predicted by \citet{ar_Parker_1986} reliable? There is a substantial simplification in the derivation of the model to a top-hat profile, wherein the velocity, concentration, and TKE are uniform up to some depth, above which they discontinuously vanish. A more general approach incorporates the key effects of the profiles through shape-factors. Top-hat models are not realistic, \citet{ar_Parker_1987} show that the shape-factors change by $38\%$ in real experimental flows, increasing to $45\%$ in the work of \citet{ar_Islam_2010}. Moreover, it was shown in \cite{ar_Dorrell_2014} that including the shape in a depth-average model results in considerably different predictions. While the simplification to top-hat models is not always present \citep[\eg][]{ar_Sher_2015,ar_Negretti_2017}, the models being used to predict the bulk energetics of gravity currents do not account for realistic profiles of velocity and density, and the considerable differences caused by shape factors indicates that the energetics are not reliably captured.  For the discussion going forward, it will be important that the derivation of \cite{ar_Parker_1986} eliminates any explicit inclusion of turbulence beyond a quantification of the TKE. They show that, under the top-hat assumption, there is a consistency relationship between the turbulent production (which passes energy from the mean-flow kinetic energy to the TKE) and other properties of the flow (basal drag and entrainment of ambient). A similar consistency relationship exists for the buoyancy-flux (energy from TKE to gravitational potential energy (GPE)). Thus, it is not possible to specify either of these effects, and instead they are implied by the model. In \citet{ar_Parker_1987} the assumption of a top-hat flow structure is relaxed, but no equivalent consistency relationships for the turbulence are provided. This makes it appear as though no such relationships exist and the turbulence requires some additional closure. However, this cannot be the case: there is nothing special about the top-hat model except for its analytical simplicity. The consistency relationships do exist, they are just complicated and unstated. To upgrade the modelling framework to one which includes the vertical profiles, a new model is required similar to \citet{ar_Parker_1987} but which eliminates the turbulent production and uplift like in \citet{ar_Parker_1986}. We also require explicit expressions for the consistency relationships, to ensure model closures such as vertical profiles and entrainment produce the correct energetics.

Work by \citet{ar_Toniolo_2006} highlighted another connected deficiency in the standard formulation: how entrainment is modelled. The particle settling velocity should reduce the extent to which fluid is entrained, and in the non-turbulent case fluid should be detrained \citep{ar_Toniolo_2006,ar_Dorrell_2010}. This deficiency arises because we are parametrising the effect of the turbulent energetics in mixing the fluid, and not the turbulent energetics themselves: previous authors have stressed that the physical origin of entrainment is the turbulent buoyancy-flux \citep{ar_Strang_2001,ar_Odier_2014}. \Citet{ar_Arneborg_2007} and \citet{ar_Wells_2010} have formulated a top-hat model of compositional currents where the buoyancy-flux replaces entrainment. Would an interpretation of entrainment in terms of buoyancy-flux give rise to the particulate effects discussed by \citet{ar_Toniolo_2006}, \citet{ar_Pittaluga_2018}, and \citet{ar_Ma_2024}?

The present paper answers the questions above. First, we build a new model which allows for the specification of vertical structure, but otherwise requires an identical set of closures to \citet{ar_Parker_1986}, and we present the consistency relationships for the implied turbulent processes. Thus, this model is a direct generalisation. By representing entrainment in the same way, this first model inherits the same problems when it comes to closing entrainment, and questions arise around the effect of particle settling. We present an alternative model where we shift from requiring a closure for entrainment to a closure for buoyancy-flux, which is enabled by our consistency relationships. This alternative model naturally includes the type of particulate effects demonstrated by \citet{ar_Toniolo_2006}.

Exploration of the energetics is further motivated by the results of \citet{ar_Skevington_F006_TCFlowPower} who demonstrated that the top-hat model \citep{ar_Parker_1986} has a missing source of energy, and postulated that these arise from the flow profiles. Here we show that this is in part true, the profiles of velocity and density give rise to a large number of additional terms which cannot be neglected at leading order. However, a pseudo-equilibrium balance can be established wherein the effects of profiles, while important, does not resolve the energetic imbalance. Instead, this is resolved by two changes: the change in how entrainment is incorporated into the model, and the inclusion of the viscous dissipation of mean-flow energy.

\subsection{Structure}

This paper is structured as follows. We first derive the depth-average model in \cref{sec:model}, starting from a three dimensional system of equations (\cref{sec:model_3D}), averaging over depth (\cref{sec:model_avg}), and discussing how to interpret the equations as a predictive model (\cref{sec:model_interp}). We then derive the rates of transfer between the different energies in the current (\cref{sec:production_dervation}), which are the consistency relationships for model closures (\cref{sec:production_discussion}). The full model is compared to simulations, showing the importance of flow shape (\cref{sec:quasiequ_simulations}) and mean-flow dissipation (\cref{sec:quasiequ_meandiss}). 
We then discuss the difficulties of defining the depth and entrainment in gravity currents (\cref{sec:depth_defn}), an alternative approach in terms of buoyancy-flux (\cref{sec:depth_entrinment}), and how this give rise to alternative model which incorporates particulate effects (\cref{sec:depth_volfree}).
Particle auto-suspension in turbidity currents is discussed in \cref{sec:suspension_efficiency}.
We highlight future research directions in \cref{sec:future} 
and conclude in \cref{sec:conclusion}. Appendices are  provided discussing the dimensional scales within gravity currents (\cref{app:deriv_channel_scales}), comparing to the variables used by \cite{ar_Ellison_1959} (\cref{app:ETshape}), the depth-average (\cref{app:depth_average}), and the depth-rescaling symmetry group (\cref{app:depth_group}). The supplementary material includes details of the ensemble averaging in \cref{sec:model_3D} and algebraic manipulations in \cref{sec:production_dervation}.
\section{The generalised depth-average model} \label{sec:model}

\begin{figure}
	\centering
	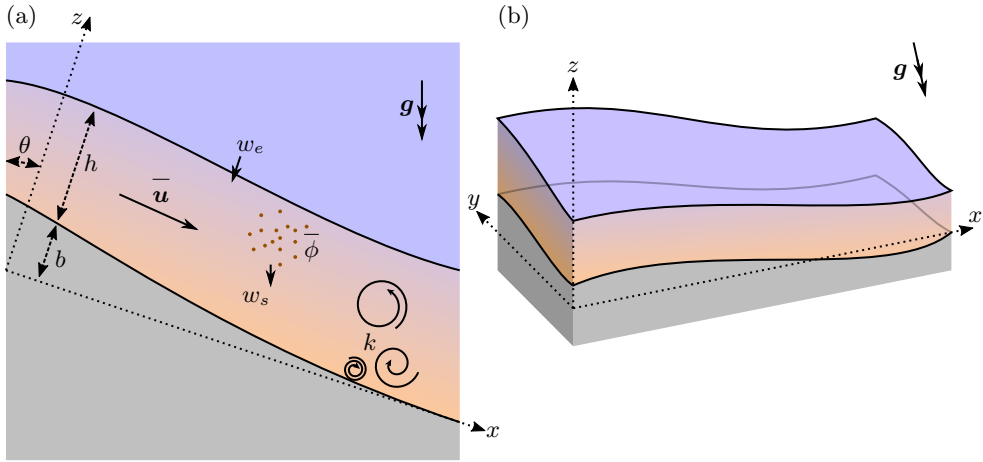
	\caption{The configuration of the turbidity current, with the bed in grey, ambient in blue, and current in brown fading toward blue in the less concentrated upper regions. (a) A two dimensional slice oriented with the vertical direction up for the case when $x$ is the downslope direction. (b) A three dimensional view oriented with respect to the coordinate system.}
	\label{fig:configuration}
\end{figure}

Here we carefully derive the generalised depth-averaged model for a gravity current as depicted in \cref{fig:configuration}. The model is constructed following similar arguments to \cite{ar_Parker_1986}.  The derivation is presented to enable us to detail the reduced set of assumptions about the shape of the velocity, density, and turbulence profiles. We begin in \cref{sec:model_3D} with the three-dimensional system of equations \cref{eqn:3Dsys_scales}, which have been simplified by the scaling analysis in \cref{app:deriv_channel_scales}. Then, in \cref{sec:model_avg}, we average the simplified three-dimensional system over the depth using the results in \cref{app:depth_average} to obtain the generalised system
\cref{eqn:2Dsys_vol} to \cref{eqn:2Dsys_egy} 
which is interpreted in \cref{sec:model_interp}. 

The system of equations is derived for a particle-driven turbidity current. However, the essential feature that we rely upon is that the excess density of the current is linearly dependent on some scalar, $\phi$, which is advected by the current up to some velocity offset. A thermal or salinity current could be captured by the model by taking $\phi$ to be the temperature or salinity anomaly and neglecting settling and erosion.

\subsection{The three dimensional system} \label{sec:model_3D}

We define a coordinate system $(\vecb{x},t)=(x,y,z,t)$, where $t$ is time. The $z$ direction is approximately bed-normal with $b(x,y)$ the bed elevation (slowly varying) and the current exists in the region $z>b$. The coordinate system is not necessarily so that $z$ is vertical, so that the first two components of gravity $\vecb{g}$ need not be zero. We denote the angle of $z$ to the vertical by $\theta$ so that $g_3 = - g \cos\theta$. In later sections (\cref{sec:quasiequ,sec:suspension_efficiency}) we restrict ourselves to a two-dimensional bottom current where $0<\theta<\pi/2$, $g_1 = g \sin\theta$, and $g_2 = 0$ as depicted in \cref{fig:configuration}; we do not make these assumptions in the derivation. Velocity is denoted by $\vecb{u} = (u,v,w)$ and volumetric concentration of particles by $\phi$.

Our goal is to represent each property of the flow field as a depth-averaged quantity multiplied by a relatively steady shape-function (\ie slowly varying in $x,y,t$). Due to the turbulent nature of the flow each variable, say $f$, has high wavenumber and high frequency fluctuations. For this reason we introduce Reynolds averaging, specifically ensemble averaging, with the average of a function $f$ denoted $\reavg{f}$ and fluctuation $\reflc{f} = f - \reavg{f}$ \citep[\eg][]{bk_Drew_TMF,bk_Pope_TF}. The averaged variables, $\reavg{f}$, are sufficiently smooth to have their profiles represented by a slowly varying shape.

The (constant) density of water is $\rho_f$ and the density of the particles is $\rho_s$ ($\rho_s>\rho_f$ for bottom currents); it is assumed that the only density variations come from the concentration so that the current density is $\reavg{\rho} = \rho_f + \reavg{\phi} (\rho_s-\rho_f)$. It is assumed that the 
particles are non-cohesive, dilute ($\reavg{\phi} \ll 1$) and Boussinesq ($R \reavg{\phi} \ll 1$ where $R = (\rho_s-\rho_f)/\rho_f$)  
so that the bed elevation can be treated as constant in time and there are no high-concentration effects on the flow such as hindered settling or concentration dependent rheology.

To ensure that the the model only includes leading order effects we introduce dimensional scales. We employ a time-scale $\mathscr{T}$, and length-scales $\mathscr{L}_i$ corresponding to the length ($i=1$), width ($i=2$), and depth ($i=3$) of the current. The current is assumed to be shallow, that is $\mathscr{L}_\alpha \gg \mathscr{L}_3$ for $\alpha \in \{1,2\}$. (Throughout, the subscripts $\alpha,\beta,\gamma$ will be used for indices limited to $1,2$.) This also aligns the coordinate system: $\pdv*{b}{x_\alpha}$ scales as $\mathscr{L}_3/\mathscr{L}_\alpha \ll 1$, the bed elevation is slowly varying. Considering the rotation of the coordinate system $\theta$, the bed can be quite steep provided that $\mathscr{L}_\alpha/\mathscr{L}_3 \gg \abs{\tan\theta}$ in the downslope direction. We assume the averaged velocities $\reavg{u_i}$ scale as $\mathscr{U}_i = \mathscr{L}_i/\mathscr{T}$, this can be viewed as a definition of the time scale. The scales of the turbulent transport terms are approximated by an eddy viscosity model. The details of the scaling analysis are given in \cref{app:deriv_channel_scales}. Eliminating the small terms we deduce that at leading order
\begin{subequations}\label{eqn:3Dsys_scales}
\allowdisplaybreaks[1]
\begin{align}
	\pdv{\reavg{u_j}}{x_j} &= 0,
	\label{eqn:3Dsys_scales_vol}
\\
	\pdv{\reavg{\phi}}{t} 
	+ \pdv{}{x_j} \ppar*{\reavg{u_j} \reavg{\phi}}
	+ \pdv{}{z} \ppar*{\tilde{w} \reavg{\phi} + J_3}
	&= 0,
	\label{eqn:3Dsys_scales_part}
\\
	\pdv{\reavg{u_\alpha}}{t} 
	\!+\! \pdv{}{x_{j}}\ppar*{\reavg{u_j}\reavg{u_\alpha}} 
	\!-\! \pdv{}{z} \ppar*{\! \tau_{3\alpha}^R + \nu \pdv{\reavg{u_\alpha}}{z}}
	\!+\! \pdv{p^T}{x_\alpha}
	&= R g_\alpha \reavg{\phi}
	\quad \text{for } \alpha \in \pbrc*{1,2},
	\label{eqn:3Dsys_scales_mom}
\\
	\pdv{p^T}{z} &= R g_3 \reavg{\phi},
	\label{eqn:3Dsys_scales_hydro}
\\
	\pdv{k}{t} 
	+ \pdv{}{x_j}\ppar*{\reavg{u_j}k}
	+ \pdv{T_3}{z}
	&= \tilde{\mathcal{P}} - \tilde{\epsilon}_K + R g_3 J_3,
	\label{eqn:3Dsys_scales_turb}
\end{align}
\end{subequations}
which represent conservation of volume, particles, $x$ \& $y$ momentum, $z$ momentum, and turbulent kinetic energy (TKE) respectively. For three dimensional variables (as used here) we use the convention that repeated indices should be summed over $\{1,2,3\}$. We denote the constant settling velocity of the particles by $\tilde{\vecb{u}}$ ($\tilde{w}<0$ for dense particles in a bottom current), 
the transport of particles by $\vecb{J}$, the Reynolds stress by $\vecb{\tau}^R$, the viscosity by $\nu$,  
the 
combined effect of pressure and TKE 
relative to the hydrostatic ambient by $p^T$, and the TKE by $k$ 
which is transported by $\vecb{T}$, produced by $\tilde{\mathcal{P}}$, and the dissipated by $\tilde{\epsilon}_K$. 
The hydrostatic approximation is common across the gravity current literature \citep[\eg][]{ar_Ellison_1959,ar_Parker_1986,ar_Bonnecaze_1993,bk_Ungarish_GCIv2}.

The form of the system \cref{eqn:3Dsys_scales} is generic, and can be derived by averaging a variety of models of dilute Boussinesq particle laden flow. Following \citet{ar_Parker_1986}, a simple approach is applicable to situations where the particles are smaller than the Kolmogorov length scale. From the perspective of the turbulent micro-scales the particles act as a concentration field $\phi$, similar to a concentration of salinity but moving at a speed $\vecb{u} + \tilde{\vecb{u}}$, where by diluteness the velocity of the mixture is the same as the velocity of the fluid, and  $k = \frac{1}{2} \reavg{\reflc{u_i}\reflc{u_i}}$. Averaging we obtain \cref{eqn:3Dsys_scales} (strictly the unsimplified system \cref{eqn:3Dsys}) with
\begin{align} \label{eqn:ReyStress_TurbTrans}
	\tau_{ij}^R &= - \reavg{\reflc{u_i}\reflc{u_j}},	&
	T_j &= \textfrac{1}{\rho}\reavg{\reflc{u_j}\reflc{p}} + \textfrac{1}{2}\reavg{\reflc{u_j}\reflc{u_i}\reflc{u_i}} - \nu \pdv{k}{x_j},	&
	\tilde{\mathcal{P}} &= \tau_{ji}^R \pdv{\reavg{u_i}}{x_j},
\\
	J_j &= \reavg{\reflc{u}_j \reflc{\phi}},
&
	\tilde{\epsilon}_K &= \nu \reavg{\pdv{\reflc{u_i}}{x_j}\pdv{\reflc{u_i}}{x_j}}.
\end{align}
The particles are moved by the coherent fluctuations of the concentration and velocity, and energy in these velocity fluctuations dissipated by viscosity.

For larger particles, we must use a more rigorous phase averaging approach. We use the results in \citet{bk_Drew_TMF}, and the details are provided in the supplementary material. We will state the non-dilute version of the definitions of variables and later impose the dilute assumption. The velocity field of the mixture is the weighted average of the velocity of the fluid and solid phases
\begin{align}
	\reavg{\vecb{u}} &= \frac{1}{\reavg{\rho}} \ppar*{ \reavg{\phi} \rho_f \reavg{\vecb{u}}_f + (1-\reavg{\phi}) \rho_s \reavg{\vecb{u}}_s }.
\end{align}
In this formalism, the concentration is only defined during the ensemble average so that $\reavg{\phi}=\phi$. The particles move relative to the the mixture for two reasons, settling $\tilde{\vecb{u}}$ and turbulent effects $\Delta \reavg{\vecb{u}}$, so that
\begin{align}
	\reavg{\vecb{u}}_s &= \reavg{\vecb{u}} + \tilde{\vecb{u}} + \Delta \reavg{\vecb{u}}.
\end{align}
The turbulent kinetic energy is defined as
\begin{align}
	k = \frac{\reavg{\rho u_j u_j}-\reavg{\rho} \reavg{u_j} \reavg{u_j}}{2\reavg{\rho}}.
\end{align}
In the dilute limit, $\reavg{\vecb{u}}$ tends to $\reavg{\vecb{u}}_f$, $k$ tends to $\frac{1}{2} \reavg{\reflc{u_i}\reflc{u_i}}$, and the expressions in \cref{eqn:ReyStress_TurbTrans} apply. However, the transport of particles and the viscous dissipation appear different
\begin{align}
	J_j &\eqdef \reavg{\phi} \Delta \reavg{u}_j,
&
	\tilde{\epsilon}_K &\eqdef \nu \reavg{\pdv{\reflc{u_i}}{x_j}\pdv{\reflc{u_i}}{x_j}} - R \reavg{\phi} g_i \tilde{u}_i.
\end{align}
The particles are transported by relative motion of the particle phase to the mixture, this relative motion a consequence of turbulence. The settling velocity is, by definition, that velocity at which the work done by gravity due to the settling motion is precisely balanced by the viscous dissipation in the fluid, and we take $\tilde{\epsilon}_K$ to represent all other viscous dissipation of TKE.

The boundary conditions for \cref{eqn:3Dsys_scales} (strictly the unsimplified system \cref{eqn:3Dsys}) are as follows. At the bed we impose no-slip, \ie
\begin{subequations} \label{eqn:3Dsys_BC}
\begin{align} \label{eqn:3Dsys_BC_bed}
    \reavg{u_j} = \reflc{u} = \reflc{v} &= 0
    &&\text{for}&
    z &= b(x,y).
\end{align}
Over a rigid bed we additionally have $w'=J_3=0$, and the treatment of the viscous boundary layer at the bed requires care (important in the analysis of \cref{sec:quasiequ,sec:suspension_efficiency}); for this reason, we include the full wall boundary layer in the model. For flows over an erosional bed we allow $w' \neq 0$, $J_3 \neq 0$, so that the erosion is captured by $J_3$. In the far field we require 
\begin{align} \label{eqn:3Dsys_BC_infty}
	\reflc{u_j} = J_j = \reavg{\phi} = p^T = \reflc{p} &= 0
&&\text{for}&
	z &=H(x,y,t).
\end{align}

In \citet{ar_Parker_1986,ar_Parker_1987}, $z=H$ is taken to be $z \to \infty$, and it is also assumed that $\reavg{u} = \reavg{v} = 0$ in the far field, so that all entrainment comes from the $\reavg{w}$ at $z \to \infty$. A similar approximation is used for jets, mixing-layers, wakes, and boundary-layers \citep[\eg][]{bk_Pope_TF}. The entrainment velocity is then defined, for a top-hat model, as 
\begin{align*}
    w_e = \pdv{h}{t} - \eval*{\reavg{w}}_{z \to \infty}.
\end{align*}
The function $h(x,y,t)$ is termed the depth of the current: for a top-hat model this is the location of the interface between the current and ambient. In gravity-currents, the approximation $z\to\infty$ works well under a deep quiescent ambient fluid which is our primary focus, but interpretation becomes problematic with a weakly counterflowing ambient. We use a more careful formulation with $z=H$ some interface beyond the strong influence of the gravity current, but still with $H-b \sim \mathscr{L}_3$. This change does not alter the depth-average model but aids interpretation. The equivalent condition in our formulation is
\begin{align} \label{eqn:3Dsys_BC_entrain}
    w_e = \pdv{}{t} \ppar*{\varsigma_h h} + \eval*{\ppar*{ \sum_{\alpha=1}^2 \reavg{u_\alpha}\pdv{H}{x_\alpha} - \reavg{w} }}_{z=H}.
\end{align}
\end{subequations}
We require that, at the elevation $z=H$, $\reavg{u_\alpha}$ ($\alpha \in \{1,2\}$) are small compared to their values within the current so that minimal momentum is transferred between the ambient and the current, $\eval{\reavg{u_\alpha}}_H \ll \mathscr{U}_\alpha$. In turbulent flow $w_e>0$, the flow entrains ambient fluid. The depth function $h$, the inclusion of the factor $\varsigma_h$, and other consequences of this definition are explored in \cref{sec:depth}. For now we simply note that typically $\varsigma_h \approx 1$.

We close this subsection by performing some manipulations of the system \cref{eqn:3Dsys_scales}. Firstly, the mean-flow kinetic energy (MKE) equation is obtained by multiplying \cref{eqn:3Dsys_scales_mom} by $\reavg{u_{\alpha}}$ and summing over $\alpha \in \{1,2\}$ and adding \cref{eqn:3Dsys_scales_hydro} times $\reavg{w}$,  giving
\begin{equation}\label{eqn:3Dsys_scales_mech}
	\pdv{e}{t} 
	+ \pdv{}{x_j}\ppar*{\reavg{u_j}e + \reavg{u_j}p^T} 
	- \pdv{}{z} \sum_{\alpha=1}^2\ppar*{\tau_{3\alpha}^R + \nu \pdv{\reavg{u_\alpha}}{z}} \reavg{u_\alpha}
	= - \tilde{\mathcal{P}} -\tilde{\epsilon}_M + R g_j \reavg{u_j} \reavg{\phi}.
\end{equation}
where
\begin{align}\label{eqn:3Dsys_MKE_defn}
	e &\eqdef \frac{1}{2} \sum_{\alpha=1}^2 \reavg{u_\alpha}^2,
&
	\tilde{\epsilon}_M &\eqdef \sum_{\alpha=1}^2 \nu \pdv{\reavg{u_\alpha}}{z}\pdv{\reavg{u_\alpha}}{z}.
\end{align}
Technically, $e$ is only the portion of the full kinetic energy resulting from the $x,y$ components of the mean flow; this expression arises naturally under the shallow assumption $\mathscr{L}_1, \mathscr{L}_2 \gg \mathscr{L}_3$ and turns out to be the useful quantity in this context. Secondly, the gravitational potential energy (GPE) that would be released if the excess mass of the current $R\reavg{\phi}$ was moved from elevation $z$ to $b$ is $R g (z-b) \reavg{\phi} \cos\theta$. The evolution equation for $(z-b) \reavg{\phi}$ is, by \cref{eqn:3Dsys_scales_part},
\begin{multline}\label{eqn:3Dsys_scales_grav}
	\pdv{}{t} (z-b) \reavg{\phi}
	+ \pdv{}{x_j} (z-b) \reavg{u_j} \reavg{\phi}
	+ \pdv{}{z} (z-b) \ppar*{\tilde{w} \reavg{\phi} + J_3}
\\
	= \pbrk*{\reavg{w} + \tilde{w}} \reavg{\phi} +  J_3
	- \sum_{\alpha=1}^2 \pdv{b}{x_\alpha} \reavg{u_\alpha} \reavg{\phi}.
\end{multline}
Again, technically this is not the full GPE, which is $R g_i x_i \reavg{\phi}$, but the portion of the GPE that turns out to be the useful quantity for shallow flows.
\subsection{Averaging over the depth} \label{sec:model_avg}

\begin{figure}
	\centering
	\input{Tables/ShapeFactorDefinition}
	\captionof{table}{Definitions and properties of the shape-factors. In the definitions we use the subscripts $\alpha,\beta,\gamma$ to indicate numerical values of $1$ or $2$. We state the equations in this section where the shape-factor appears, or else the first equation where it appears.  The values for top-hat flow are calculated using \cref{eqn:tophat}.}
	\label{tab:shape_factors}
	\vspace{5mm}
	\includegraphics{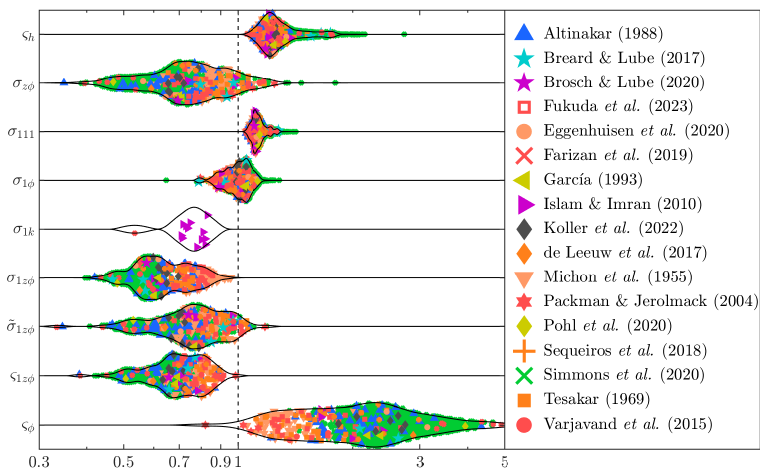}
	\captionof{figure}{Violin plots of the distribution of values that each shape-factor can take, computed from the dataset compiled by \citet{ar_Skevington_F006_TCFlowPower} (\citet{ar_Simmons_2020} is field data, all other is experimental). For each shape-factor (marked to the left) we plot the probability density for the distribution (computed using a kernel method) as a black line. In the computation, the data are weighted to account for the large number of samples from some sources. The data-points for each shape-factor are plotted at the horizontal location of their value, and given a random vertical displacement within the kernel.
	Here, $h$ is calculated by setting $\sigma_{11}=1$, and $\varsigma_h$ by \mbox{\cref{eqn:depth_definition}} with $\delta=10^{-2}$.}
	\label{fig:ShapeFactors}
	\nocite{phd_Altinakar_1988,ar_Breard_2017,ar_Brosch_2020,ar_Skevington_F006_TCFlowPower,ar_Eggenhuisen_2020,ar_Farizan_2019,ar_Garcia_1993,ar_Islam_2010,ar_Koller_2022,ar_Leeuw_2017,bk_Michon_1955,ar_Packman_2004,ar_Pohl_2020,ar_Sequerios_2018,ar_Simmons_2020,phd_Tesaker_1969,ar_Varjavand_2015}
\end{figure}

The equations in \cref{sec:model_3D} are integrated over the depth $b \leq z \leq H$. The resulting system of equations is written in terms of the depth-averaged variables
\begin{align}
	\Phi &= \frac{1}{h} \int_b^{H} \reavg{\phi} \dd{z},
&
	U_1 &= \frac{1}{h} \int_b^{H} \reavg{u_1} \dd{z},
&
	U_2 &= \frac{1}{h} \int_b^{H} \reavg{u_2} \dd{z},
&
	K &= \frac{1}{h} \int_b^{H} k \dd{z}.
\end{align}
Recall that in general $h \neq H-b$, and that under a deep quiescent ambient we may take $H \to \infty$ to simplify. 
For the concentration field $\reavg{\phi}$ (as an example) at a single point in $(x,y,t)$, the depth average $\Phi$ has eliminated all information about the variation of concentration with $z$. We carry this information forwards using a shape-function $\xi_\phi = \reavg{\phi}/\Phi$ which describes the variation of concentration relative to the depth average. It is informative to write the shape-function not as a function of $z$ but of $\zeta \eqdef (z-b)/h$. For some gravity currents, a careful choice of $h$ will enable the shape-functions to be invariant with respect to $(x,y,t)$, all profiles of $\reavg{\phi}$ being the same up to a rescaling in magnitude, $\Phi$, and a vertical stretch, $h$; self-similarity. Generalising to all variables the decomposition takes the form
\begin{subequations} \label{eqn:shape_avg_split}
\begin{align}
	\reavg{\phi}(x,y,z,t) 	&= \xi_\phi(x,y,\zeta,t) \cdot \Phi(x,y,t),
\\
	\reavg{u_1}(x,y,z,t) 	&= \xi_1(x,y,\zeta,t) \cdot U_1(x,y,t),
\\
	\reavg{u_2}(x,y,z,t) 	&= \xi_2(x,y,\zeta,t) \cdot U_2(x,y,t),
\\
	k(x,y,z,t) 				&= \xi_k(x,y,\zeta,t) \cdot K(x,y,t),
\end{align}
where $\xi_\phi$, $\xi_1$, $\xi_2$, and $\xi_k$ are shape-functions satisfying
\begin{align} \label{eqn:profile_normalisation}
	\int_0^{\zeta_H} \xi_{\omitdummy}(x,y,\zeta,t) \dd{\zeta} &= 1,
&
	\zeta_H &\eqdef \frac{H-b}{h}.
\end{align}
\end{subequations}
Consequently, $\Phi$, $U_1$, $U_2$, and $K$ are the depth-average values of density, velocity, and TKE, while $\xi_\phi$, $\xi_1$, $\xi_2$, and $\xi_k$ capture the shape of the current. In particular, the expressions defined in \cref{tab:shape_factors} are the features of the shape that influence the averaged properties of the current. The majority of the shape-factors are of the form
\begin{align}
	\sigma_{AB} &= \int_0^{\zeta_H} \xi_A \xi_B \dd{\zeta},
&&\text{or}&
	\sigma_{ABC} &= \int_0^{\zeta_H} \xi_A \xi_B \xi_C \dd{\zeta},
\end{align}
where we use $\xi_z = 2 \zeta$. The exceptions are indicated by the $\tilde{\sigma}_{\omitdummy}$ or $\varsigma_{\omitdummy}$ notation. Note that our choice of normalisation \cref{eqn:profile_normalisation} and shape-factors (\cref{tab:shape_factors,fig:ShapeFactors}) are different to those used by some authors, see \cref{app:ETshape}.

There are two special cases of flows we will consider, along with the general case. Firstly, when the shape of the current is in self-similar form, the shape-functions do not depend on $x$, $y$ or $t$, meaning the shape-factors in \cref{tab:shape_factors} are constants.  Secondly, for top-hat flow, the shape-functions take the form
\begin{align} \label{eqn:tophat}
	\xi_{\phi} &=
	\begin{cases}
		0, 								& 	1 < \zeta,			\\
		1,								&	0 < \zeta < 1,		\\
		\varsigma_\phi					& 	\zeta = 0,			
	\end{cases}
&
	\xi_1 = \xi_2 = \xi_k &=
	\begin{cases}
		0, 								& 	1 < \zeta,			\\
		1,								&	0 < \zeta < 1,
	\end{cases}
\end{align}
which yields the special values in \cref{tab:shape_factors}.

We next integrate the equations in \cref{sec:model_3D} over the depth $b \leq z \leq H$ using the results in \cref{app:depth_average} and applying the boundary conditions \cref{eqn:3Dsys_BC} \citep{ar_Ellison_1959,ar_Parker_1986}. The indices $\alpha,\beta,\gamma$ take a value of $1$ or $2$ and sums range over these values.
Conservation of fluid volume is, by \cref{eqn:3Dsys_scales_vol},
\begin{gather}\label{eqn:2Dsys_vol}
	\pdv{}{t} (\varsigma_h h) + \sum_\beta \pdv{}{x_\beta} 
	\Big(
		\underbrace{  h U_\beta   \vphantom{\big)}}_{\mathclap{\text{volume flux}}}
	\Big) 
	= S_h 
	= \underbrace{  w_e 			\vphantom{\big)}}_{\mathclap{\qquad\text{entrainment velocity}}}.
\end{gather}
Conservation of particle volume is, by \cref{eqn:3Dsys_scales_part},
\begin{gather}\label{eqn:2Dsys_part}
	\pdv{}{t}\ppar*{h \Phi} + \sum_\beta \pdv{}{x_\beta} 
	\Big(
		\underbrace{  \sigma_{\beta\phi} h U_\beta \Phi	\vphantom{\big)}}_{\mathclap{\text{particle flux}}}
	\Big)
	= S_{\Phi} = 
	- \underbrace{	\varsigma_{\phi} w_s \Phi \cos\theta	\vphantom{\big)}	}_{\mathclap{\text{deposition}}}
	+ \underbrace{	E_s.									\vphantom{\big)}	}_{\mathclap{\text{erosion}}}
\end{gather}
Conservation of momentum  is, by \cref{eqn:3Dsys_scales_mom}, using \cref{eqn:3Dsys_scales_hydro} to calculate the hydrostatic pressure,
\begin{subequations}\label{eqn:2Dsys_mom}
\begin{gather}\label{eqn:2Dsys_mom_DE}
	\pdv{}{t} \ppar*{h U_\alpha} 
	+ \sum_\beta \pdv{}{x_\beta} 
	\Big(
		  \underbrace{ \sigma_{\beta\alpha} h U_\beta U_\alpha 				\vphantom{\big)}}_{\mathclap{\text{momentum flux}}}
	\Big)
	+ \pdv{}{x_\alpha}
	\Big(
		\underbrace{ \textfrac{1}{2} \sigma_{z\phi} R g h^2 \Phi \cos\theta	\vphantom{\big)}}_{\mathclap{\text{pressure}}}
	\Big)
	= S_{\alpha},
\\
		\label{eqn:2Dsys_mom_S}
	S_{\alpha} =
	- \underbrace{ R g h \Phi \cos\theta \pdv{b}{x_\alpha} 	\vphantom{\bigg)}	}_{\mathclap{\text{pressure on bed slope}}} \qquad
	- \underbrace{	u_{\alpha\star}^2						\vphantom{\bigg)}	}_{\mathclap{\text{basal drag}}} \qquad
	+ \underbrace{	R g_\alpha h \Phi.						\vphantom{\bigg)}	}_{\mathclap{\text{downslope gravity}}}
\end{gather}
\end{subequations}
Conservation of MKE is, by \cref{eqn:3Dsys_scales_mech},
\begin{subequations}	\label{eqn:2Dsys_MKE}
\begin{multline}\label{eqn:2Dsys_MKE_DE}
	\sum_\beta \pdv{}{t} \ppar*{ \textfrac{1}{2} \sigma_{\beta\beta} h U_\beta^2 }
	+ \sum_{\beta\gamma} \pdv{}{x_\beta}  \Big(
		\underbrace{\textfrac{1}{2} \sigma_{\beta\gamma\gamma} h U_\beta U_\gamma^2	\vphantom{\big)}}_{\mathclap{\text{MKE flux}}}
	\Big)
\\
	+ \sum_\beta \pdv{}{x_\beta}  \Big(
		\underbrace{\textfrac{1}{2} \tilde{\sigma}_{\beta z\phi} U_\beta  R g h^2 \Phi \cos\theta	\vphantom{\big)}}_{\mathclap{\text{pressure work}}}
	\Big)
	= S_M
	= \tilde{S}_M 
	- \underbrace{h \mathcal{P} - h \mathcal{B}_M													\vphantom{\big)}}_{\mathclap{\text{energy transfer}}}
	,
\end{multline}
\begin{gather}\label{eqn:2Dsys_MKE_S}
	\tilde{S}_M 
	= 
	  \sum_\beta \underbrace{	\sigma_{\beta\phi} U_\beta R g_\beta h \Phi					\vphantom{\Big)}}_{\mathclap{\text{work by downslope gravity}}}	\qquad\qquad
	- \underbrace{h \epsilon_M  																	\vphantom{\Big)}}_{\mathclap{\text{MKE dissipation}}}
\end{gather}
\end{subequations}
Conservation of GPE is, by the product of \cref{eqn:3Dsys_scales_grav} and $R g \cos\theta$,
\begin{subequations}	\label{eqn:2Dsys_GPE}
\begin{multline}	\label{eqn:2Dsys_GPE_DE}
	\pdv{}{t}\ppar*{ \textfrac{1}{2} \sigma_{z\phi} R g h^2 \Phi \cos\theta}
	+ \sum_\beta \pdv{}{x_\beta} \Big(
		\underbrace{\textfrac{1}{2} \sigma_{\beta z\phi} U_\beta R g h^2 \Phi \cos\theta			\vphantom{\big)}}_{\mathclap{\text{GPE flux}}}
	\Big)	\\
	= S_G 
	= \tilde{S}_G 
	+ \underbrace{h \mathcal{B}_M + h \mathcal{B}_K													\vphantom{\big)}}_{\mathclap{\text{energy transfer}}}
	,
\end{multline}
\begin{equation}\label{eqn:2Dsys_GPE_S}
	\tilde{S}_G = 
	- \sum_\beta 
	  \underbrace{ \sigma_{\beta\phi} U_\beta R g h \Phi \cos\theta \pdv{b}{x_\beta} 				\vphantom{\bigg)}}_{\mathclap{\text{variation in datum}}}
	- \underbrace{ w_s R g h \Phi (\cos\theta)^2													\vphantom{\bigg)}}_{\mathclap{\text{energy loss to settling}}}
	.
\end{equation}\end{subequations}
(The variation in datum arises because we measure GPE relative to the local bed elevation in \cref{eqn:3Dsys_scales_grav}.) Conservation of TKE is, by \cref{eqn:3Dsys_scales_turb},
\begin{subequations}	\label{eqn:2Dsys_TKE}
\begin{gather}	\label{eqn:2Dsys_TKE_DE}
	\pdv{}{t} \ppar*{ h K }
	+ \sum_\beta \pdv{}{x_\beta} \Big(
		\underbrace{\sigma_{\beta k} h U_\beta K 													\vphantom{\big)}}_{\mathclap{\text{TKE flux}}}
	\Big)
	= S_K
	= \tilde{S}_K 
	+ \underbrace{h \mathcal{P} - h \mathcal{B}_K													\vphantom{\big)}}_{\mathclap{\text{energy transfer}}}
	,
\\
	\label{eqn:2Dsys_TKE_S}
	\tilde{S}_K = 
	- \underbrace{h \epsilon_K																		\vphantom{\big)}}_{\mathclap{\text{TKE dissipation}}}
	.
\end{gather}
\end{subequations}
Summing \cref{eqn:2Dsys_MKE,eqn:2Dsys_GPE,eqn:2Dsys_TKE} yields the equation for conservation of total energy, $E$,
\begin{subequations}\label{eqn:2Dsys_egy}
\begin{multline}\label{eqn:2Dsys_egy_DE}
	\pdv{}{t} \ppar*{h E}
	+ \sum_{\beta\gamma} \pdv{}{x_\beta}  \Big(
		\underbrace{\textfrac{1}{2} \sigma_{\beta\gamma\gamma} h U_\beta U_\gamma^2 \vphantom{\big)}}_{\mathclap{\text{MKE flux}}}
	\Big)
\\
	+ \sum_\beta \pdv{}{x_\beta}  \Big(
		  \underbrace{\sigma_{\beta k} h U_\beta  K 							\vphantom{\big)}}_{\mathclap{\text{TKE flux}}}
		+ \underbrace{\varsigma_{\beta z\phi} U_\beta  R g h^2 \Phi \cos\theta 	\vphantom{\big)}}_{\mathclap{\qquad \text{GPE flux and pressure work}}}
	\Big)
	= S_E,
\end{multline}
\begin{multline}\label{eqn:2Dsys_egy_S}
	S_E = 
	- \sum_\beta 
	  \underbrace{	\sigma_{\beta\phi} U_\beta R g h \Phi \cos\theta \pdv{b}{x_\beta}						\vphantom{\bigg)}	}_{\mathclap{\text{variation in GPE datum}}}
	+ \sum_\beta
	  \underbrace{	\sigma_{\beta\phi} U_\beta R g_\beta h \Phi												\vphantom{\bigg)}	}_{\mathclap{\text{work by downslope gravity}}}	\qquad\qquad
\\
	- \underbrace{	h \epsilon_T 																			\vphantom{\big)}	}_{\mathclap{\text{total dissipation}}} \qquad
	- \underbrace{	w_s R g h \Phi (\cos\theta)^2.															\vphantom{\big)}	}_{\mathclap{\text{GPE loss to settling}}}
\end{multline}
\end{subequations}
where the total energy (per unit mass) is defined as
\begin{equation}\label{eqn:2Dsys_egy_defn}
	E =
	\sum_\beta 
	\underbrace{\textfrac{1}{2} \sigma_{\beta\beta} U_\beta^2									\vphantom{\big)}}_{\mathclap{\text{MKE}}}
	+ \underbrace{K																				\vphantom{\big)}}_{\mathclap{\text{TKE}}}
	+ \underbrace{\textfrac{1}{2} \sigma_{z\phi} R g h \Phi \cos\theta							\vphantom{\big)}}_{\mathclap{\text{GPE}}}
	.
\end{equation}
While we will call $E$ the total energy going forward, it is technically only the portion of the total energy that is useful for describing shallow flows; see the discussion below \cref{eqn:3Dsys_MKE_defn,eqn:3Dsys_scales_grav}. Throughout our stating of the depth-averaged system, we have used the definitions of settling velocity, erosion, basal shear velocity, and depth-average total, mean-flow, and turbulent dissipation as
\begin{equation} \label{eqn:CHavg_bedvars}
\begin{gathered}
	\begin{aligned}
		w_s \cos\theta &= - \tilde{w},
	&
		E_s &= \eval[\Bigg]{ J_3 }_b,
	&
		u_{\alpha\star}^2 &= \eval[\Bigg]{ \nu \pdv{\reavg{u_\alpha}}{z} }_b,
	\end{aligned}
\\
	\begin{aligned}
		\epsilon_T &= \epsilon_M + \epsilon_K,
	&
		h \epsilon_M &= \int_b^{H} \tilde{\epsilon}_M \dd{z},
	&
		h \epsilon_K &= \int_b^{H} \tilde{\epsilon}_K \dd{z} - \eval[\Big]{T_3}_b.
	\end{aligned}
\end{gathered}\end{equation}
The terms marked as `energy transfer'  are the depth-averaged TKE production, turbulent buoyancy-flux, and mean-flow buoyancy-flux, defined as
\begin{subequations}\label{eqn:energytrans}
\begin{align} 
	\mathcal{P} 	
	&\eqdef \frac{1}{h} \int_b^{H} \tilde{\mathcal{P}} \dd{z},	
\\
	\mathcal{B}_K 	
	&\eqdef \frac{R g \cos\theta}{h} \int_b^{H} J_3 \dd{z}.
\\
	\label{eqn:energytrans_meanflowprod}
	\mathcal{B}_M 	
	&\eqdef \frac{R g \cos\theta}{h} \int_b^{H} \reavg{w}\reavg{\phi} \dd{z},
\end{align}\end{subequations}
respectively. We understand $\mathcal{P}$ as the rate of conversion of MKE to TKE, $\mathcal{B}_K$ as the rate of conversion of TKE to GPE, and $\mathcal{B}_M$ as the rate of conversion of MKE to GPE.

We briefly note that in \cref{fig:ShapeFactors} and other figures going forward, the data considered is for unidirectional currents ($U_2=\partial/\partial y=0$) and the depth is defined following \citet{ar_Ellison_1959} and \citet{ar_Parker_1986} by setting $\sigma_{11} = 1$; that is take
\begin{equation} \label{eqn:depth_ET}
	h = \frac**{ \ppar*{ \int_b^{H} \reavg{u} \dd{z} }^2 }{ \int_b^{H} \reavg{u}^2 \dd{z} }.
\end{equation}
In our analysis $h$ is arbitrary, and will be discussed thoroughly in \cref{sec:depth}.

\subsection{Interpretation as a classical volumetric model} \label{sec:model_interp}

As discussed in the introduction, the purpose of this derivation is to produce a system of equations for the same unknown functions as \cite{ar_Parker_1986}, but allowing for the imposition of shape-functions selected by the modeller to improve accuracy. We have also incorporated arbitrary slowly-varying topography $b$, again this can be specified by the modeller, along with generalising to 3-dimensional flows (that is, we include variation in $y$). The equations governing volume \cref{eqn:2Dsys_vol}, particles \cref{eqn:2Dsys_part}, and momentum \cref{eqn:2Dsys_mom} are precisely these generalisations, and the case $\varsigma_h=1$ has been used previously \citep{ar_Dorrell_2014,ar_Sher_2015,ar_Negretti_2017}. For the energetics, we have derived a set of three equations for the separate contributions from MKE \cref{eqn:2Dsys_MKE}, GPE \cref{eqn:2Dsys_GPE}, and TKE \cref{eqn:2Dsys_TKE}, along with their sum which describes the evolution of the total energy \cref{eqn:2Dsys_egy}. \cite{ar_Parker_1986} present their model with an equation for the evolution of TKE, and so we may naively think that \cref{eqn:2Dsys_TKE} is the appropriate equation. However, the key consideration is not the component of energy modelled, but the closures required to complete the model. As was found by \cite{ar_Parker_1986} in their top-hat model, the energy transfer terms are intimately related to other model closures, and we must therefore eliminate them from the model. That is not to say these terms are not important \citep{ar_Odier_2014}, but rather that they cannot be specified independently. The only equation which does not include energy transfer terms is the equation for total energy \cref{eqn:2Dsys_egy}. This forms the model: the functions $h$, $\Phi$, $U_\alpha$, $K$ are solved for using the system of equations \cref{eqn:2Dsys_vol,eqn:2Dsys_part,eqn:2Dsys_mom,eqn:2Dsys_egy}. To close the system the modeller must specify the topography, $b$, and shape of the concentration, velocity, and turbulence fields through the shape-functions in \cref{eqn:shape_avg_split} which give rise to the shape-factors in the model through the expressions in \cref{tab:shape_factors}. Additionally, the modeller must specify the parameters and closures present in the model from \cite{ar_Parker_1986}, including entrainment of ambient fluid at speed $w_e$, erosion of the bed at rate $E_s$, drag from the bed with shear velocity $u_{\alpha\star}$, and viscous dissipation of energy at rate $\epsilon_T$.

\section{Consistency requirements for energy transfer} \label{sec:production}

The system of governing equations \cref{eqn:2Dsys_vol,eqn:2Dsys_part,eqn:2Dsys_mom,eqn:2Dsys_egy} lack explicit inclusion of the energy transfer terms defined in \cref{eqn:energytrans}, and yet they form a closed system of equations for the unknown functions $h$, $\Phi$, $U_\alpha$, and $K$.
Consequently, we can eliminate the time evolution in the additional energetic equations \cref{eqn:2Dsys_MKE,eqn:2Dsys_GPE,eqn:2Dsys_TKE} to obtain expressions for $\mathcal{P}$, $\mathcal{B}_M$, and $\mathcal{B}_K$ in terms of the source terms and the spatial gradients of the unknown functions. This allows us to explore  the bulk energetics of gravity currents.  The expressions obtained are, unfortunately, rather complicated, but they give the full implications of the equations in \cref{sec:model_avg}. We endeavour to give some interpretation of the expressions here, and give simplified expressions in \cref{sec:quasiequ}. Verification of these manipulations using the computer algebra software Maple is provided as supplemental information. We begin with a derivation of the energy transfer terms in \cref{sec:production_dervation}, and discuss their use for the development of energetically consistent model closures in \cref{sec:production_discussion}.

\subsection{Derivation of energetic consistency requirements} \label{sec:production_dervation}

We begin by rearranging the equations for MKE \cref{eqn:2Dsys_MKE_DE} and GPE \cref{eqn:2Dsys_GPE_DE} so that the time derivatives can be easily substituted from the governing equations for volume, particles, and momentum \cref{eqn:2Dsys_vol,eqn:2Dsys_part,eqn:2Dsys_mom}:
\begin{subequations} \label{eqn:source_energies}
\begin{multline} \label{eqn:source_MKE_from_GE}
	S_M =
	- \ppar*{\sum_\beta \frac{1}{2} \sigma_{\beta\beta}  U_\beta^2} \pdv{h}{t}
	+ \sum_\beta \sigma_{\beta\beta} U_\beta \pdv{}{t}(h U_\beta)
	+ \sum_\beta \frac{1}{2} h U_\beta^2 \pdv{\sigma_{\beta\beta}}{t}
\\
	+ \sum_\beta \pdv{}{x_\beta} \ppar*{ \sum_\gamma \frac{1}{2} \sigma_{\beta\gamma\gamma} h U_\beta U_\gamma^2 +   \frac{1}{2} \tilde{\sigma}_{\beta z\phi} U_\beta R g h^2 \Phi \cos\theta },
\end{multline}
\begin{multline} \label{eqn:source_GPE_from_GE}
	S_G =
	  \frac{1}{2} \sigma_{z\phi} R g h \Phi \cos\theta \pdv{h}{t}
	+ \frac{1}{2} \sigma_{z\phi} R g h \cos\theta \pdv{}{t} \ppar*{h\Phi}
	+ \frac{1}{2} R g h^2 \Phi \cos\theta \pdv{\sigma_{z\phi}}{t}
\\
	+ \sum_\beta  \pdv{}{x_\beta} \ppar*{ \frac{1}{2} \sigma_{\beta z \phi}  U_\beta R g h^2 \Phi \cos\theta }.
\end{multline}
\end{subequations}
Due to the structure of the source terms $S_M$, $S_G$, and $S_K$, it is not possible to rearrange them into expressions for $\mathcal{P}$, $\mathcal{B}_K$, and $\mathcal{B}_M$. Instead, two of these can be expressed in terms of the remaining one. We choose to express $\mathcal{P}$ and $\mathcal{B}_K$ in terms of $\mathcal{B}_M$, which yields 
\begin{align} \label{eqn:production_rearranged}
	h \mathcal{P} &= - h \mathcal{B}_M - \ppar*{ S_M - \tilde{S}_M },
&
	h \mathcal{B}_K &= - h \mathcal{B}_M + \ppar*{ S_G - \tilde{S}_G },
\end{align}
with the second equality in the TKE equation \cref{eqn:2Dsys_TKE_DE} then stating simply $S_{E} + S_K + S_G = \tilde{S}_{E} + \tilde{S}_K + \tilde{S}_G$. The reason for this choice is that $\mathcal{B}_M$ is a property of the mean-flow, and can therefore be deduced from the equations describing the mean-flow, whereas the others are properties of the turbulence requiring closure. A consequence of this choice is that $h \mathcal{P}$ will be derived as the implied loss of MKE, while $h \mathcal{B}_K$ is derived as the implied gain of GPE, and this will be seen in the expressions to follow. To deduce $\mathcal{B}_M$, we employ the continuity equation
\begin{align} \label{eqn:meanflow_uplift_derivation}
	\pdv{\reavg{u}}{x} + \pdv{\reavg{v}}{y} + \pdv{\reavg{w}}{z} &= 0,
&&\text{thus}&
	\reavg{w} &= - \int_b^{z} \eval*{ \pdv{\reavg{u}}{x} + \pdv{\reavg{v}}{y} }_{z'} \dd{z'}.
\end{align}
Substituting into the definition of $\mathcal{B}_M$ \cref{eqn:energytrans_meanflowprod} and decomposing into shape-functions \cref{eqn:shape_avg_split} to rewrite  in terms of shape-factors, we obtain the expression for $h \mathcal{B}_M$ given below. To obtain expressions for the other turbulent energy transfers, we begin with the expressions in \cref{eqn:production_rearranged}, into which we substitute our expression for $h \mathcal{B}_M$, the rearranged energy equations \cref{eqn:source_energies}, and eliminate time derivatives using the governing equations \cref{eqn:2Dsys_vol,eqn:2Dsys_part,eqn:2Dsys_mom}. These manipulations are algebraically challenging, and verification of the results using the computer algebra package Maple are provided as supplemental information.  Firstly, the depth integrated rate of conversion of MKE to TKE is  
\small
\begin{multline} \label{eqn:egytransfer_full_turb}
	h \mathcal{P} =
	\frac{1}{2} \ppar*{\sum_\beta \frac{\sigma_{\beta\beta}}{\varsigma_h}  U_\beta^2} w_e
	+ \sum_\beta \sigma_{\beta\beta} U_\beta u_{\beta\star}^2 
	- h \epsilon_M
\\
	+ \sum_\beta (\sigma_{\beta\phi}-\sigma_{\beta\beta}) U_\beta R h \Phi \ppar*{ g_\beta - g \cos\theta \pdv{b}{x_\beta} }
\\
	+ \frac{1}{2} \sum_{\beta\gamma} \ppar*{ 
		- \frac{\sigma_{\gamma\gamma}}{\varsigma_h}  
		+ 2 \sigma_{\gamma\gamma} \sigma_{\beta\gamma}
		- \sigma_{\beta\gamma\gamma}
	} U_\gamma^2 U_\beta \pdv{h}{x_\beta}
	+ \frac{1}{2} \sum_{\beta} \ppar*{ 
		- \frac{\sigma_{\beta\beta}}{\varsigma_h}  
		+ 4\sigma_{\beta\beta}^2
		- 3\sigma_{\beta\beta\beta}
	} h U_\beta^2 \pdv{U_\beta}{x_\beta}
\\
	+ \frac{1}{2} \sum_{\beta\neq\gamma} \pgrp*{ 
	\ppar*{ 
		- \frac{\sigma_{\gamma\gamma}}{\varsigma_h}  
		+ 2 \sigma_{\gamma\gamma} \sigma_{\beta\gamma}
		- \sigma_{\beta\gamma\gamma} 
	} h U_\gamma^2 \pdv{U_\beta}{x_\beta}
	+ 2 \ppar*{ 
		\sigma_{\gamma\gamma} \sigma_{\beta\gamma}
		- \sigma_{\beta\gamma\gamma}
	} h U_\beta U_\gamma \pdv{U_\gamma}{x_\beta}
	}
\\
	+ \sum_{\beta} \ppar*{ 
		  \sigma_{\beta\beta} \sigma_{z\phi}
		- \varsigma_{\beta z\phi}
	}  U_\beta R g h \Phi \cos\theta \pdv{h}{x_\beta}
	+ \frac{1}{2} \sum_{\beta} \ppar*{ 
		  \sigma_{\beta\beta} \sigma_{z\phi}
		- \tilde{\sigma}_{\beta z\phi}
	} U_\beta R g h^2 \cos\theta \pdv{\Phi}{x_\beta}
\\
	+ \frac{1}{2} \sum_\beta \pgrp*{ 
		- \frac{1}{\varsigma_h} \pdv{}{t} \ppar*{\varsigma_h \sigma_{\beta\beta}}
		+ \sum_\gamma \ppar*{
			2 \sigma_{\beta\beta} \pdv{\sigma_{\beta\gamma}}{x_\gamma}
			- \pdv{\sigma_{\beta\beta\gamma}}{x_\gamma}
		} U_\gamma
	} h U_\beta^2
\\
	+ \frac{1}{2} \sum_\beta \ppar*{ 
		\sigma_{\beta\beta} \pdv{\sigma_{z\phi}}{x_\beta} 
		- \pdv{\tilde{\sigma}_{\beta z \phi}}{x_\beta} 
		+ \tilde{\sigma}_{D\beta z \phi}
	} U_\beta R g h^2 \Phi \cos\theta
	.
\end{multline}
\normalsize
The terms in the first line are the energy required to accelerate entrained fluid, the work done by drag, and the MKE lost to viscous effects. The second line results from an imbalance between work done by downslope gravity on the mean velocity and that from velocity/density profiles, along with the imbalance of the implied work by varying GPE datum and the vertical transport of material. The first term in the third line is the result of imbalances in the energy changes associated with varying depth from a changing volume of fluid, work done by depth-average momentum, and the depth resolved transport of MKE; the second term and the fourth line similarly with the effect of varying velocity. The fifth line results from an imbalance of work done by the depth-average pressure/GPE and depth resolved transport of these quantities. The sixth line includes the effective variation in MKE from the shape-functions varying temporally and spatially, and similarly the variation in MKE flux. The seventh line includes the imbalance between depth-average and depth resolved pressure work from varying shape-functions, along with the vertical transport generated by varying $\xi_\beta$. Next, the depth integrated rate of conversion of TKE to GPE is 
\small
\begin{multline} \label{eqn:egytransfer_full_buoy}
	h \mathcal{B}_K = \frac{1}{2} R g h \cos\theta \Bigg\lgroup
	  \frac{\sigma_{z\phi}}{\varsigma_h} \Phi w_e
	+ \pbrk*{ 2 - \sigma_{z\phi} \varsigma_\phi } w_s \Phi \cos\theta 
	+ \sigma_{z\phi} E_s
\\
	+ \sum_\beta \ppar*{
		- \frac{\sigma_{z\phi}}{\varsigma_h}
		- \sigma_{\beta \phi}\sigma_{z\phi}
		+ 2\varsigma_{\beta z \phi}
	} \Phi \pdv{}{x_\beta} (h U_\beta)
	+ \sum_\beta \ppar*{
		- \sigma_{\beta \phi}\sigma_{z\phi} 
		+ \sigma_{\beta z \phi}
	} U_\beta h  \pdv{\Phi}{x_\beta}
\\
	+ \pbrk*{
		\varsigma_h \pdv{}{t} \ppar*{ \frac{\sigma_{z\phi}}{\varsigma_h}}
		+ \sum_{\beta} \ppar*{
			- \sigma_{z\phi} \pdv{\sigma_{\beta\phi}}{x_\beta} 
			+ \pdv{\sigma_{\beta z \phi}}{x_\beta} 
			+ \tilde{\sigma}_{D\beta z \phi}
		} U_\beta
	} h \Phi 
	\Bigg\rgroup
	.
\end{multline}
\normalsize
The terms in the first line are the GPE generated by entrainment, the energy expended holding particles in suspension less the amount deposited, and the energy required to erode. The first term in the second line is the imbalance in GPE from varying volume flux as a consequence of changing depth, depth-average transport of GPE, and depth resolved transport of both GPE and pressure; the second term similar with varying concentration. The third line includes the effect of varying shape-factors, including the direct change of GPE, the imbalance between depth-average and depth resolved GPE flux, along with the vertical transport generated by varying $\xi_\beta$. Finally, the depth-average rate of conversion of MKE to GPE is
\begin{equation} \label{eqn:egytransfer_full_mean}
	\small
	h \mathcal{B}_M 
	= \frac{1}{2} R g h \Phi \cos\theta \sum_\beta \Bigg(
		  2 \sigma_{\beta\phi} U_\beta \pdv{b}{x_\beta}
		- \tilde{\sigma}_{\beta z\phi} h \pdv{U_\beta}{x_\beta}
		+ \ppar*{\sigma_{\beta z\phi}-\tilde{\sigma}_{\beta z\phi}} U_\beta \pdv{h}{x_\beta}
		- \tilde{\sigma}_{D\beta z\phi} h U
	\Bigg).
\end{equation}
where the terms are the increase in GPE from flow generated by varying bed elevation, the work done by pressure due to velocity gradients, the imbalance in the change of GPE and work done by pressure due to changing depth, and the vertical transport generated by varying $\xi_\beta$.

\begin{figure}
	\centering
	\includegraphics{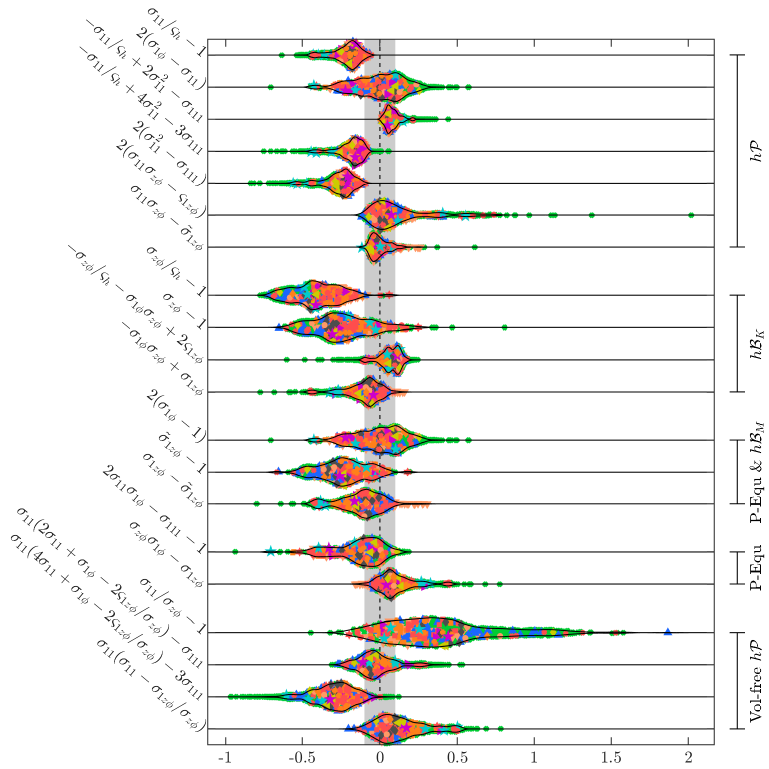}
	\caption{The difference between the coefficients of the energy transfer terms  and the top-hat approximation to these terms \cref{eqn:egytransfer_tophat}, plotted using the same format as \cref{fig:ShapeFactors}. Note that the majority of coefficients vanish in a top-hat model, in which case the coefficients are plotted without modification. There are sections dedicated to: $h\mathcal{P}$ \cref{eqn:egytransfer_full_turb}; $h \mathcal{B}_K$ \cref{eqn:egytransfer_full_buoy}; $h \mathcal{B}_M$ \cref{eqn:egytransfer_full_mean}; the pseudo-equilibrium simplification \cref{eqn:quasiequ_egytransfer}; and the volume-free production \cref{eqn:egytransfer_volfree_production}, as indicated to the right.}
	\label{fig:EnergyCoeff}
\end{figure}

Using the values of the shape-factors in \cref{tab:shape_factors}, top-hat flows simplify to
\begin{subequations}\label{eqn:egytransfer_tophat}
\begin{align}
	\label{eqn:egytransfer_tophat_P}
	h \mathcal{P} &= 
	U_1 u_{1\star}^2 + U_2 u_{2\star}^2
	+ \textfrac{1}{2} (U_1^2 + U_2^2) w_e - h \epsilon_M,
\\
	\label{eqn:egytransfer_tophat_BK}
	h \mathcal{B}_K &= 
	\textfrac{1}{2} R g h \cos\theta \ppar[\big]{ 
		\Phi w_e 
		+ \pbrk*{ 2 - \varsigma_\phi } w_s \Phi \cos\theta
		+ E_s
	},
\\
	h \mathcal{B}_M &= 
	\textfrac{1}{2} R g h \Phi \cos\theta \ppar*{ 2 U \pdv{b}{x} + 2 V \pdv{b}{y} - h \pdv{U}{x} - h \pdv{V}{y} }.
\end{align}\end{subequations}
For two-dimensional flow ($U_2=\pdv*{b}{y}=0$) with negligible mean-flow dissipation in the boundary layer ($\epsilon_M=0$), the first two expressions match those from \cite{ar_Parker_1986}, and the third is consistent though never explicitly stated. 

However, using the values from \cref{fig:ShapeFactors} we find that the additional terms from a complete analysis \cref{eqn:egytransfer_full_turb,eqn:egytransfer_full_buoy,eqn:egytransfer_full_mean} are not negligible. The difference between the coefficients and the top-hat approximation are plotted in \cref{fig:EnergyCoeff}. This can be understood as representing each of $h \mathcal{P}$, $h \mathcal{B}_K$, and $h \mathcal{B}_M$ as the sum of the top-hat model and a correction based on the shape of the current, and a coefficient of $\pm 1$ means the correction from including shape-factors is as large as the contribution from the top-hat model. Of course, we are only comparing the size of the coefficients, and the size of the term also depends on what this coefficient multiplies, but this analysis gives an indication of which terms are important. The region where the terms are of magnitude less than $10\%$ of the top-hat terms is indicated by a grey band, and while a few of the coefficients do lie in this region where they can (arguably) be neglected, many of them lie substantially outside of this region . Thus, we cannot neglect the effect of shape on the energetics.
\subsection{Using energetic consistency requirements} \label{sec:production_discussion}

The governing equations we have derived includes balances for volume \cref{eqn:2Dsys_vol}, particles \cref{eqn:2Dsys_part}, momentum \cref{eqn:2Dsys_mom}, and total energy \cref{eqn:2Dsys_egy}. This system is similar to that proposed by \cite{ar_Parker_1986}, but crucially allows for the specification of flow shape in a model that captures energetics. Shape-factors are known to have a substantial influence on prediction in models which do not capture energetics \citep{ar_Dorrell_2014}. This modelling framework has the potential to be much more accurate because it allows for the inclusion of additional physics. However, these additional physics appear in the model as closures specified by the modeller, and there is a possibility that the selected closures are unsuitable in a subtle way. Suppose that we take a selection of developed closures for shape-functions \citep{ar_Abad_2011,ar_Islam_2010}, entrainment \citep{ar_Ellison_1959,ar_Parker_1986,ar_Cenedese_2010}, and erosion \citep{ar_Parker_1986,ar_Dorrell_2018,ar_Guo_2020}, in addition to making some reasonable choice of drag coefficient, and constructing some expression for the total dissipation (while method for this construction is presented in \citet{ar_Parker_1986}, it would be better to have an empirically verified closure). Then we can simulate a current using the governing equations, and post-process the results to obtain the energy transfer that implicitly occurred using \cref{eqn:egytransfer_full_turb,eqn:egytransfer_full_buoy,eqn:egytransfer_full_mean}. It is extremely unlikely that independently developed closures will somehow yield the energy transfer seen in real currents.

Consequently, the purpose of \cref{eqn:egytransfer_full_turb,eqn:egytransfer_full_buoy,eqn:egytransfer_full_mean} is as consistency requirements to be used during the development of closures. When a set of closures for the governing equations have been developed using data from experiment or simulation, a final verification can be performed by using the developed closures to predict the energy transfer, which can be compared to the measured energy transfer. This final verification, if successful, demonstrates that the set of closures are energetically consistent and will produce the correct evolution of TKE. While the TKE may not always be the subject of interest, having a model which is tightly constrained to real physics gives confidence in the model in general.

For these consistency requirements to be useful in closure development, two things are required. Firstly, the balances must be verified. Provided we are dealing with a shallow Boussinesq gravity current under a deep unstratified ambient, this is principally verification that there have been no algebraic mistakes in the derivation. Secondly, some simplification of the consistency requirements must be made in order to make them useable for model development. The supplied Maple script can be used to apply simplifications to specific circumstances. The next section provides an example of simplifying the system to a specific simulation configuration, and verifying the accuracy of both the full and simplified form of the energetic transfer expressions.

\section{Comparison of model to high resolution simulations} \label{sec:quasiequ}

\subsection{Two-dimensional pseudo-equilibrium flow} \label{sec:quasiequ_simulations}

In this subsection we consider two-dimensional steady flow down a smooth slope ($\pdv*{}{t} = \pdv*{}{y} = U_2 = b = 0$, $\theta \neq 0$, use notation $U \equiv U_1$). Under such conditions we may expect the fluid to reach a \emph{pseudo-equilibrium} configuration after flowing a sufficient distance, where the shape enters self-similar form (shape-factors constant) and the following dimensionless parameters become independent of $x$,
\begin{align} \label{eqn:quasiequ_defn_dimless}
	\frac{U}{\sqrt{R g h \cos\theta}},
&&
	\frac{w_s\cos\theta}{\sqrt{K}},
&&
	\frac{U^2}{2 K}.
\end{align}
Up to constant coefficients, these are the Froude number (ratio of flow speed to  velocity scale), the Rouse number (ratio of particle settling to upwards turbulent diffusion), and the ratio of MKE to TKE respectively. These conditions imply 
\begin{align} \label{eqn:quasiequ_defn_simp}
	\dv{}{x} \ppar*{h \Phi} = \dv{U}{x} = \dv{K}{x} = 0
\end{align}
as used by \citet{ar_Parker_1986}. 
Simplifying the governing system \cref{eqn:2Dsys_vol,eqn:2Dsys_part,eqn:2Dsys_mom,eqn:2Dsys_egy} using the pseudo-equilibrium conditions \cref{eqn:quasiequ_defn_simp}
\begin{subequations}\label{eqn:quasiequ_condn}
\allowdisplaybreaks[1]
\begin{gather}
	\label{eqn:quasiequ_condn_vol}
	\dv{h}{x} = \frac{w_e}{U},
\\
	\label{eqn:quasiequ_condn_part}
	0
	=
	\varsigma_{\phi} w_s \Phi \cos\theta
	- E_s,
\\			
	\label{eqn:quasiequ_condn_mom}
	\ppar*{ \sigma_{11} U^2 + \textfrac{1}{2} \sigma_{z\phi} R g h \Phi \cos\theta } \frac{w_e}{U}
	= R g h \Phi \sin\theta
	- u_{1\star}^2,
\\			
	\label{eqn:quasiequ_condn_egy}
	\begin{multlined}[b][\multlinedwidth]
		\ppar*{ \textfrac{1}{2} \sigma_{111} U^2 + \sigma_{1k} K + \varsigma_{1z\phi} R g h \Phi \cos\theta } w_e	\\
		=
		\sigma_{1\phi} U R g h \Phi \sin\theta - h \epsilon_T - w_s R g h \Phi (\cos\theta)^2.
	\end{multlined}
\end{gather}
\end{subequations}

\begin{figure}
	\centering
	\includegraphics{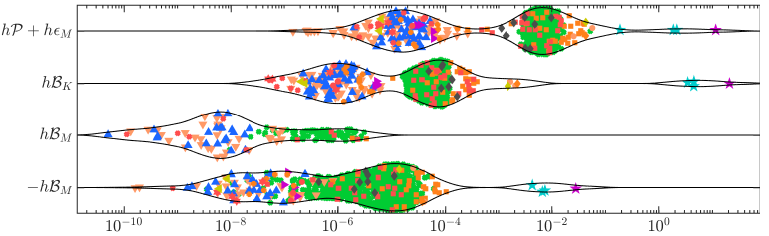}
	\caption{The depth integrated energy transfer calculated using the pseudo-equilibrium balance \cref{eqn:quasiequ_egytransfer}, plotted using the same format as \cref{fig:ShapeFactors}. The value of $h \mathcal{B}_M$ can be positive or negative, so it is split over two lines.}
	\label{fig:EnergyTransfer}
\end{figure}

To obtain consistency requirements for the energetics in pseudo-equilibrium we substitute \cref{eqn:quasiequ_defn_simp,eqn:quasiequ_condn} into the full expressions for $h \mathcal{P}$, $h \mathcal{B}_K$ and $h \mathcal{B}_M$ \cref{eqn:egytransfer_full_turb,eqn:egytransfer_full_buoy,eqn:egytransfer_full_mean} to obtain
\begin{subequations}\label{eqn:quasiequ_egytransfer}
\begin{align}
	h \mathcal{P} &=
	\sigma_{1\phi} U u_{1\star}^2
	+ \ppar*{ 2 \sigma_{11} \sigma_{1\phi} - \sigma_{111} } \textfrac{1}{2} U^2 w_e
\notag \\ & \hspace{4cm}
	+ \ppar*{ \sigma_{z\phi} \sigma_{1\phi} - \sigma_{1z\phi} } \textfrac{1}{2} R g h \Phi \cos\theta w_e
	- h \epsilon_M,
	\label{eqn:quasiequ_egytransfer_prod}
\\
	h \mathcal{B}_K &= 
	\textfrac{1}{2} R g h \Phi \cos\theta \ppar[\big]{
		\tilde{\sigma}_{1z\phi} w_e
		+ 2 w_s \cos\theta
	},
	\label{eqn:quasiequ_egytransfer_buoy}
\\
	h \mathcal{B}_M 
	&= \textfrac{1}{2} R g h \Phi \cos\theta \ppar*{ \sigma_{1z\phi} - \tilde{\sigma}_{1z\phi} } w_e.
	\label{eqn:quasiequ_egytransfer_Mbuoy}
\end{align}
\end{subequations}
This simplification puts $h \mathcal{P}$ and $h \mathcal{B}_K$ in a format where they are easily compared to top-hat expressions in \cref{eqn:egytransfer_tophat}, revealing an alteration in the magnitude of the terms and the introduction of some additional terms, see \cref{fig:EnergyCoeff}. Compared to \cref{eqn:egytransfer_tophat} the turbulent production $h \mathcal{P}$ typically has a smaller contribution from the energy needed to accelerate entrained fluid up to speed, but a new contribution from the energy required to uplift the mass, while the turbulent production induced by entrainment is reduced, and similarly the production by the mean-flow $h \mathcal{B}_M$ has a negative contribution from the energy consumed by the entrainment uplift. The magnitude of the pseudo-equilibrium transfer in real currents is plotted in \cref{fig:EnergyTransfer}, showing that $h \mathcal{B}_M$ is small for this regime (but the non-equilibrium terms may be large). It is difficult to compare the size of $h \mathcal{B}_K$ and $h \mathcal{P}$ without knowing the mean flow dissipation $h \epsilon_M$. Note that the currents used to plot \cref{fig:EnergyTransfer} are not necessarily in pseudo-equilibrium balance, see \cref{sec:suspension_efficiency}.

\begin{figure}
	\centering
	\includegraphics{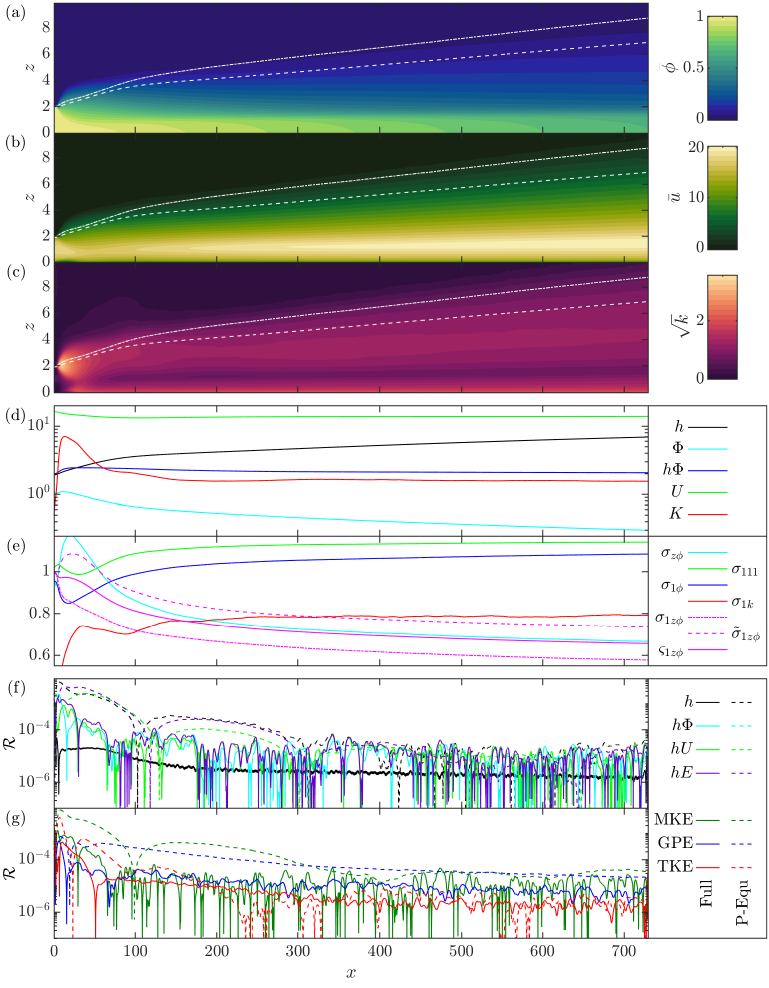}
	\caption{Plots computed from the DNS data of \citet{ar_Zuniga_2024}. (a-c) The time averaged fields of concentration, velocity, and TKE, the depths $h$ \cref{eqn:depth_ET} and $\tilde{h}$ \cref{eqn:depth_definition} shown in dashed and dash-dot white respectively. (d) The spatial variation of the depth-average quantities. (e) The spatial variation of the shape-factors. (f) The residual in the full depth-averaged equations \cref{eqn:2Dsys_vol,eqn:2Dsys_part,eqn:2Dsys_mom,eqn:2Dsys_egy} or the pseudo-equilibrium equations \cref{eqn:quasiequ_condn}, computed from the simulation data using \cref{eqn:model_error}. In the legend the equation is indicated using the corresponding conserved quantity. (g) The residual in the energy equations \cref{eqn:2Dsys_egy,eqn:quasiequ_condn_egy} split over contributions from MKE, GPE and TKE in \cref{eqn:2Dsys_MKE,eqn:2Dsys_GPE,eqn:2Dsys_TKE}, dividing by full energy flux rather than the flux for the specific equation.}
	\label{fig:Bala_GeneralProperties}
\end{figure}

To investigate the pseudo-equilibrium dynamics we employ data from a direct numerical simulation originally published in \citet{ar_Zuniga_2024}. (For similar simulations see \citet{ar_Salinas_2019a,ar_Salinas_2019b,ar_Salinas_2020,ar_Salinas_2021,ar_Salinas_2022,ar_Salinas_2023}, and for an experimental configuration see \citet{ar_Odier_2014}.) Here, a conservative ($w_s=E_s=0$) gravity current flowed down a slope of $\theta = 3\degree$ over $0<x<730$ ($b=0$), having been fed into the domain at $\Rey = 5650$ at $x=0$, $0<z<2$, the units of their simulation such that $Rg = 1/\sin\theta$. The flow was simulated until it became statistically steady in time, and then time averaged statistics were computed. The resulting flow is shown in \cref{fig:Bala_GeneralProperties}(a-c). We compute the depth-averaged variables and shape-factors directly through integration over $0 = b \leq z \leq H = 60$, the full height of the simulation domain, see \cref{fig:Bala_GeneralProperties}(d,e). On $x>300$ the depth-averages satisfy well the pseudo-equilibrium conditions \cref{eqn:quasiequ_defn_simp}, $K$ being the last to become steady. However, the concentration shape $\xi_\phi$ continues to slowly evolve over the entire length of the simulation as can be seen through the shape-factors $\sigma_{z\phi}$, $\sigma_{1z\phi}$ and $\tilde{\sigma}_{1z\phi}$. 

Before going any further, we validate the depth-average model \cref{eqn:2Dsys_vol,eqn:2Dsys_part,eqn:2Dsys_mom,eqn:2Dsys_egy}. In steady-state each equation is of the form $\dv*{\mathcal{F}}{x} = \mathcal{S}$ for some flux $\mathcal{F}$ and source $\mathcal{S}$, and we calculate the residual in each as
\begin{align} \label{eqn:model_error}
	\mathcal{R} = \frac{1}{\mathcal{F}} \abs*{ \dv{\mathcal{F}}{x}  - \mathcal{S} }
\end{align}
were the flux and source are computed directly from the simulation data. Consequently, $\mathcal{R}^{-1}$ is the distance over which cumulative  error would produce an $\order{1}$ change to the flux if we were to integrate \cref{eqn:model_error} with the given source.\footnote{In fact, since the residuals alternate sign, the change to the flux may be seen as a random walk and thus  residuals accumulate over a distance of $\mathcal{R}^{-2}$} By \cref{fig:Bala_GeneralProperties}(f), once the current has established ($x>100$) the length-scale of  residual accumulation is around $10^5$, vastly longer than the simulation domain. Splitting the  residual in the energy equation over its components (\cref{fig:Bala_GeneralProperties}(g)) we see a similarly good agreement. The larger  residual on $x<100$ is a consequence of the assumption that horizontal scales are much larger than vertical by which we omitted terms in \cref{eqn:3Dsys_scales}. We also plot the residual in the pseudo-equilibrium balance, ensuring the computed residual is simply the direct simplification of \cref{eqn:model_error} employing the assumptions. Once the flow has reached the balance ($x>300$) the residual in the system (\cref{fig:Bala_GeneralProperties}(f)) is no larger than for the full equation, showing that this is an accurate simplification, however the residual in the split energy equations (\cref{fig:Bala_GeneralProperties}(g)) is slightly larger with an accumulation length-scale of $10^4$, which we will find is a consequence of the varying shape-factors.

\begin{figure}
	\centering
	\includegraphics{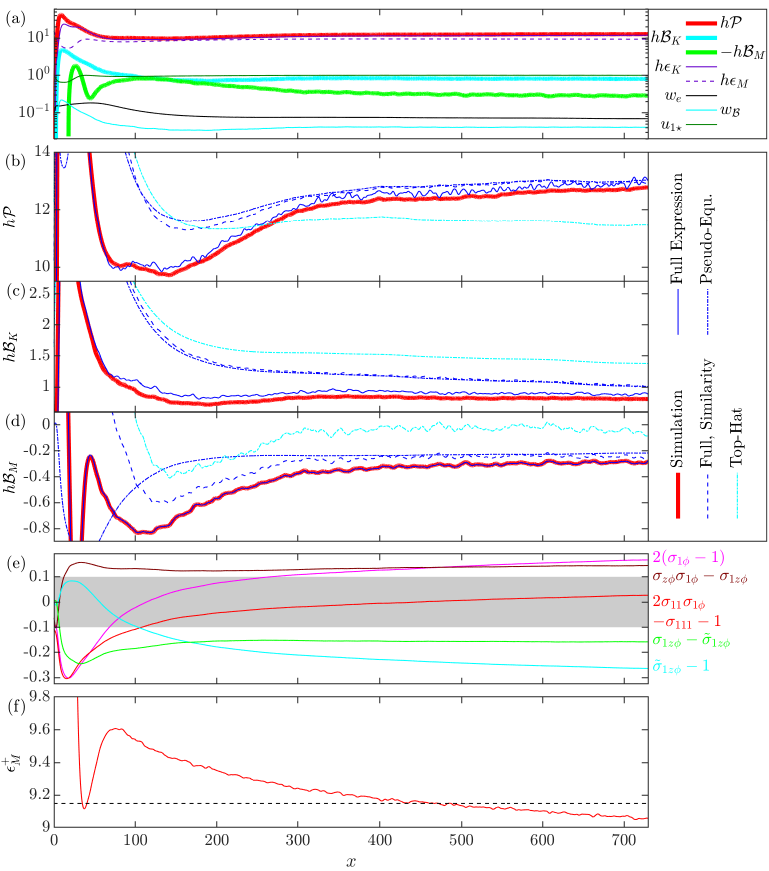}
	\caption{Further plots computed from the DNS data of \citet{ar_Zuniga_2024}.  (a) The properties of turbulence, entrainment and drag computed directly from DNS data. (b-d) Comparing the values of $h\mathcal{P}$, $h\mathcal{B}_K$ and $h\mathcal{B}_M$ direct from simulation to the expressions in the full model \cref{eqn:egytransfer_full_turb,eqn:egytransfer_full_buoy,eqn:egytransfer_full_mean}, along with the simplifications of self-similar flow (\cref{eqn:egytransfer_full_turb,eqn:egytransfer_full_buoy,eqn:egytransfer_full_mean} neglecting derivatives of shape-factors), the pseudo-equilibrium balance \cref{eqn:quasiequ_egytransfer}, and top-hat flow \cref{eqn:egytransfer_tophat}. (e) The coefficients of the additional terms in \cref{eqn:quasiequ_egytransfer} with respect to  top-hat flow. (f) The dimensionless mean-flow dissipation (red) and the approximation using \citet{ar_Reichardt_1951} (black dashed).}
	\label{fig:Bala_Energetics}
\end{figure}

The properties of the turbulence and other source terms used in the residual analysis are computed directly from the simulation data using \cref{eqn:3Dsys_BC_entrain,eqn:CHavg_bedvars,eqn:energytrans}, and are plotted in \cref{fig:Bala_Energetics}(a). Crucially, we find that the viscous dissipation of the mean-flow $h \epsilon_M$ is almost equal to the turbulent production $h \mathcal{P}$ Whether $h \mathcal{P} \simeq h \epsilon_M$ in general is discussed later.

The values of $h\mathcal{P}$, $h\mathcal{B}_K$ and $h\mathcal{B}_M$ are plotted in \cref{fig:Bala_Energetics}(b-d). We see that the values of the full model expressions from \cref{eqn:egytransfer_full_turb,eqn:egytransfer_full_buoy,eqn:egytransfer_full_mean} agree very well with the exact values from the simulations across all $x$ (including small $x$ where the curves go out of the figure, the small amount of noise at large $x$ arises from the $x$ derivatives). Reducing to self-similar form by neglecting the $x$ derivatives of shape-functions in the consistency requirements \cref{eqn:egytransfer_full_turb,eqn:egytransfer_full_buoy,eqn:egytransfer_full_mean} introduces a large error in the proximal region $x<300$ (see \cref{fig:Bala_Energetics}(b-d)), but in the distal region $x>300$ where the flow is equilibrated the error is small and provides a reasonable approximation. Simplifying further to the pseudo-equilibrium expressions \cref{eqn:quasiequ_egytransfer} does not increase the error in the distal regions, the assumptions on the depth-averaged variables \cref{eqn:quasiequ_defn_simp} being satisfied to a much greater degree of accuracy than the assumption that shape-functions are independent of $x$. The top-hat approximation increases the error more substantially. Plotting the coefficients of the difference of the top-hat approximation and the pseudo-equilibrium approximation in \cref{fig:Bala_Energetics}(e) (the same coefficients as in \cref{fig:EnergyCoeff}) we find that they are moderately large, which is the cause of the error.

\subsection{Mean flow dissipation} \label{sec:quasiequ_meandiss}

Given the comparison in \cref{fig:Bala_Energetics}(b-d), the greatest error in approximating the turbulent processes arises from neglecting the mean-flow dissipation $h \epsilon_M$ because $h \mathcal{P} \simeq h \epsilon_M$. We ask, is this an unusual property of this flow or something we should expect to see in general? The mean-flow dissipation is a property of near bed flow. Working in dimensionless wall variables $u^+ = \reavg{u}/u_{1\star}$ and $z^+ = z u_{1\star}/\nu$ we see that
\begin{align}  \label{eq:meandiss_wallvar}
	h \epsilon_M 
	= \int_0^{H} \nu \ppar*{\pdv{\reavg{u}}{z}}^2 \dd{z}
	= u_{1\star}^3 \int_0^{H u_{1\star}/\nu} \ppar*{\pdv{u^+}{z^+}}^2 \dd{z^+}.
\end{align}
While internal shear layers also technically contribute to this dissipation, these produce a change in velocity of order $U$ and thus to have a leading order contribution the the mean flow dissipation they must have a thickness of order $\nu U^2/u_{1\star}^3$ so that the Reynolds number of the shear layer is of order $U^3/u_{1\star}^3$. In environmental currents the Reynolds number of shear layers is much larger, even in the presence of jet sharpening \citep{ar_Dorrell_2019}, and so the contribution to the mean flow dissipation is only from the viscous sub-layer between the bed and the logarithmic layer. This motivates a definition of the dimensionless mean-flow dissipation as
\begin{align}
	\epsilon_M^+ \eqdef \int_0^{H u_{1\star}/\nu} \ppar*{\pdv{u^+}{z^+}}^2 \dd{z^+} = \frac{h \epsilon_M}{u_{1\star}^3}
\end{align}
As a rough estimate, we may expect that there is a viscous sub-layer with $u^+ = z^+$ over $0<z^+<10$, and a turbulent log-layer where $u^+ = \kappa^{-1} \ln z^+ + \text{const}$ on $z^+>10$ where $\kappa = 0.41$ is the von Kármán constant, resulting in the approximation $\epsilon_M^+ \approx 10 + {1}/{10 \kappa^2} = 10.6 ~ (3~\sf)$ showing that even the log-layer does not contribute significantly to the value. The estimate can be improved using the matched asymptotic of \citet{ar_Reichardt_1951} \citep[see][]{ar_Kadivar_2021} which  yields a value of $\epsilon_M^+ = 9.15 ~ (3~\sf)$.
We plot $\epsilon_M^+$ in \cref{fig:Bala_Energetics}(f) and find that it lies very close to this estimate, but there is some slow downstream evolution which is similar in kind to the evolution of the shape-factors in \cref{fig:Bala_GeneralProperties}(e). 

\begin{figure}
	\centering
	\includegraphics{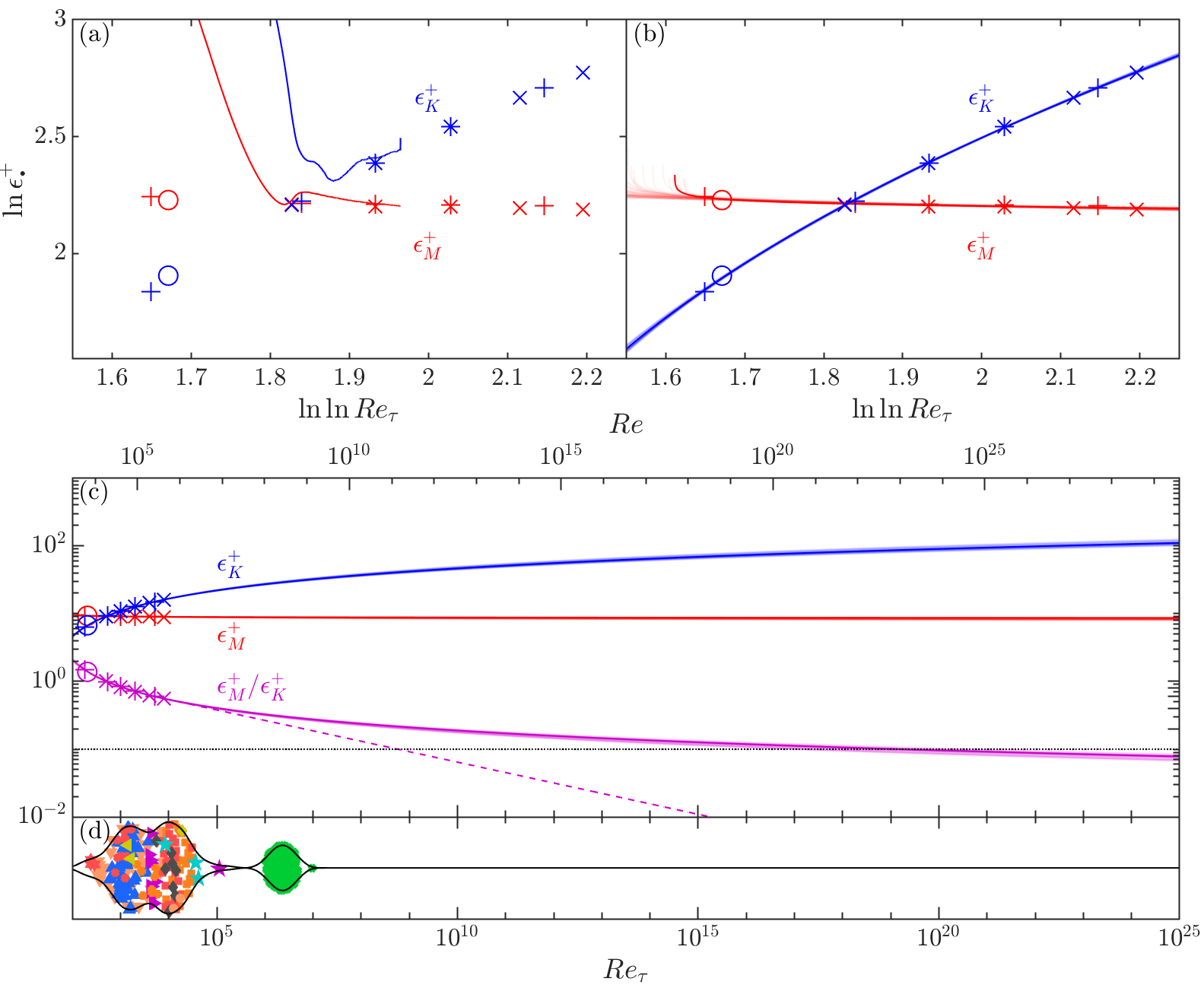}
	\caption{Extrapolation of the dissipation to high Reynolds numbers, using three datasets of channel flow simulations: $+$ from \citet{ar_Lee_2015}, $\times$ from \citet{ar_Kaneda_2021}, and $\circ$ from \citet{ar_Orlandi_2019}. $\epsilon_M^+$ is red and $\epsilon_K^+$ is blue. (a) The channel flow data and the gravity current data from \citet{ar_Zuniga_2024} (solid). (b) Curves of best-fit for the channel flow dissipation. Plotted faintly are samples from a probability distribution over curves, showing the uncertainty of the best-fit curve. (c) Extrapolation of the dissipation best-fit, and ratio of the extrapolations (purple). We also show a power law extrapolation of the best-fit ratio through the final data-point.  Two abscissa are given for (c) showing $\Rey_\tau$ and $\Rey$. (d) The Reynolds number for particulate gravity currents, plotted using the same format as \cref{fig:ShapeFactors}.}
	
	\label{fig:ChannelDissipation_smooth}
\end{figure}

While it is true that $\epsilon_M^+$ is finite at high Reynolds numbers, the dimensionless turbulent dissipation, $\epsilon_K^+ \eqdef h \epsilon_K / u_{\star1}^3$, grows at high Reynolds number, so that $\epsilon_M^+/\epsilon_K^+ \to 0$ as $\Rey \to \infty$. But how large does $\Rey$ need to be for $\epsilon_K^+$ to dominate? To answer, we use data from channel flow between two boundaries (sometimes called plane Poiseuille flow) from \citet{ar_Lee_2015} and \citet{ar_Kaneda_2021}. For this analysis we will employ a Reynolds number $\Rey_\tau$ where the length scale is the half height of the channel and the velocity scale is $u_{\star1}$. \Cref{fig:ChannelDissipation_smooth}(a) shows that, sufficiently far downstream of the release, the gravity current of \citet{ar_Zuniga_2024} has the same dissipations as channel flow (length scale in $\Rey_\tau$ is flow depth \cref{eqn:depth_ET}). From the data, the dissipations have curves of best-fit given by
\begin{align} \label{eqn:dissipation_smooth_fit}
	\epsilon_M^+ &= 9.17 \cdot (\ln \Rey_\tau - 5.01 )^{-0.0148},
&
	\epsilon_K^+ &= 3.62 \cdot (\ln \Rey_\tau - 3.28 )^{0.856},
\end{align}
with coefficients of determination ($R^2$) of $0.8867$ and $0.9997$ respectively (\cref{fig:ChannelDissipation_smooth}(b)). To account for the uncertainty in parameters we use a Bayesian approach. The uncertainty of the data is estimated as the root-mean-square difference between the points and the respective best-fit line. We use a uniform (improper) prior for the parameters, so the Bayesian approach is equivalent to a likelihood approach. Sample curves from the resulting posterior distribution are plotted faintly in \cref{fig:ChannelDissipation_smooth}(b,c) showing almost no scatter. Extrapolations of these curves, and their ratio, are plotted in \cref{fig:ChannelDissipation_smooth}(c), which has axes for both $\Rey_\tau$ and $\Rey$. The latter is calculated with the channel-average velocity as the velocity scale (equivalent to depth-average velocity for gravity currents), the two Reynolds numbers related by $\Rey_\tau \simeq 0.09 \Rey^{0.88}$ \citep{bk_Pope_TF}. To account for error in the functional form in the extrapolation we also show a power-law (straight line in log scale) extrapolation of the best fit curve from the final data-point (dashed). We say that the mean flow dissipation is small when it is below $10\%$ of the turbulent dissipation, and negligible when below $1\%$. The small threshold is reached somewhere in the interval $10^{18} \lesssim \Rey_\tau \lesssim 10^{21}$ ($10^9$ for power-law extrapolation).
The threshold is substantially above the largest values ever measured in gravity currents, $\Rey_\tau \lesssim 10^7$ (\cref{fig:ChannelDissipation_smooth}(d), green points from \citealt{ar_Simmons_2020}). 
Thus, mean-flow dissipation is relevant at the Reynolds numbers of all geophysical currents.

\begin{figure}
	\centering
	\includegraphics{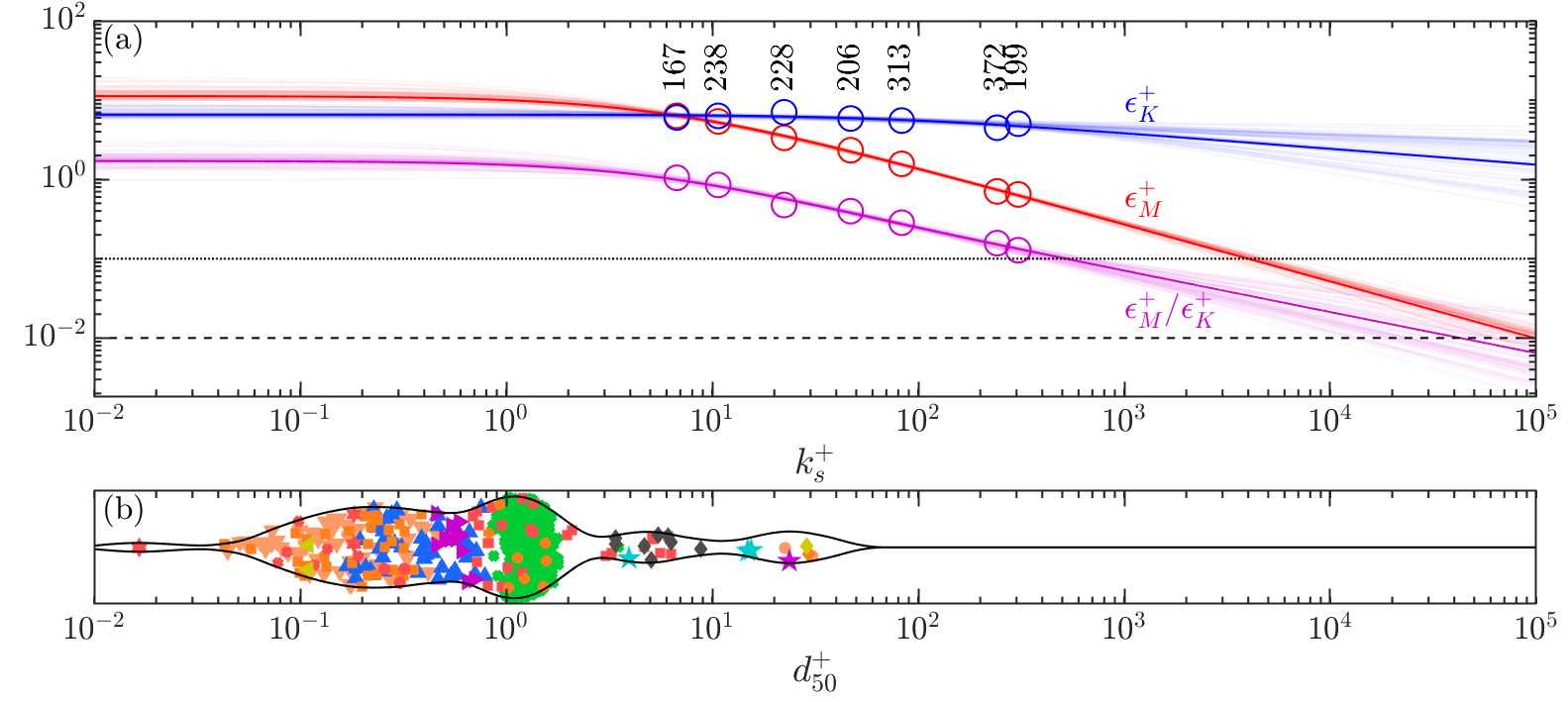}
	\caption{(a) Extrapolation of the dissipation to high roughness, $\circ$ is data from \citet{ar_Orlandi_2019}. $\epsilon_M^+$ is red, $\epsilon_K^+$ is blue, and their ratio is purple, with $\Rey_\tau$ given above. Curves of best-fit are shown, and plotted faintly are samples from a probability distribution over curves, showing the uncertainty of the best-fit curves. (b) The mean particle diameter for particulate gravity currents, plotted using the same format as \cref{fig:ShapeFactors}.}
	\label{fig:ChannelDissipation_rough}
\end{figure}

The data from channel flow considered so far is all from smooth channels, and particulate gravity currents always flow over a rough bed, the roughness appearing at the particle scale and (for environmental currents) at the sales of benthic fauna and bed-forms \citep{ar_Olu_2017,ar_Sen_2017,ar_AzpirozZabala_2024}. 
A large body of research exists into flow over rough beds, see the recent review of \cite{ar_Kadivar_2021}. 
There are a great many parameters which can be used to characterise the bed roughness as documented by \citet{ar_Thakkar_2017}. 
For research on how roughness effects the log-law region see \citet{ar_Shringarpure_2012,ar_Bilgin_2023}, and for the additional layers of the flow which are present in flow over roughness see \citet{ar_Nikora_2004,ar_Mazzuoli_2017,ar_Forooghi_2018}. We use data from \citet{ar_Orlandi_2019}, representing the roughness using the equivalent sand roughness $k_s^+$ which is calculated from the log-layer by fitting $u^+ = (1/\kappa) \log(y^+/k_s^+) + 8$ \citep[\eg][]{bk_Schlichting_BLT}. The best fit curves in \cref{fig:ChannelDissipation_rough}(a) are
\begin{align} \label{eqn:dissipation_rough_fit}
	\epsilon_M^+ &= 38.9 \cdot (k_s^+ + 5.62 )^{-0.719},
	&
	\epsilon_K^+ &= 15.5 \cdot (k_s^+ + 70.3 )^{-0.201},
\end{align}
with coefficients of determination $0.9978$ and $0.7126$. For the mean-flow dissipation to be small we require $k_s^+ \gtrsim 500$ and to be negligible $k_s^+ \gtrsim 10^4$. We expect these thresholds to be only weakly dependent on $\Rey_\tau$, because $\epsilon_K^+$ is largely independent of $k_s^+$ and only varying by a factor of $\sim 10$ over the $\Rey_\tau$ of gravity currents, while $\epsilon_M^+$ is independent of $\Rey_\tau$. To reach $k_s^+ \gtrsim 10^4$ requires bedforms $\gtrsim 10^4 d_{50}^+$ high (\cref{fig:ChannelDissipation_rough}(b), green points from \citealt{ar_Simmons_2020}), which may occur in natural settings but not in experiments. Mean-flow dissipation is therefore important for experimental flows and moderately sized turbidity currents, but not for large currents over bedforms.

\subsection{Summary of comparison to simulations}

There are two distinct alterations to the energetics of \citet{ar_Parker_1986} that have been discussed in this section: flow shape and  mean-flow dissipation. Regarding flow shape, there is a clear hierarchy of approaches shown in \cref{fig:Bala_Energetics}: the top-hat model is the least accurate, then the approaches which include flow shape but not its variation, and most accurate is the model which includes the variations of flow shape. When fully including flow shape and providing closures from the simulation, the model approaches the accuracy of the full simulation, showing that the energetics can, in principle, be accurately captured by a depth-average model. This accuracy improvement is relevant for predictive models (\cref{sec:model_interp}) because the flow shape is incorporated into the models. Conversely, the mean-flow dissipation is only relevant for assessing the consistency requirements (\cref{sec:production}) which are important for model development and interpretation. It is not included in the predictive model outside of the total dissipation, which can be approximated separately. Nonetheless, it can make a leading order contribution to the implied energetics which is important for the use of consistency requirements in closure development (\cref{sec:production_discussion}), and the understanding of particle suspension (\cref{sec:suspension_efficiency}).

\section{Depth, entrainment, and detrainment} \label{sec:depth}

\subsection{Gravity current depth} \label{sec:depth_defn}

To this point we have not discussed the depth $h(x,y,t)$; the model presented is independent of the definition. A wide variety of measures of depth can be used \citep[\eg][]{ar_Salinas_2023}, and the symmetry group in \cref{app:depth_group} can be used to transform between different measures. For two-dimensional flow ($U_2=\pdv*{}{y}=0$) \citet{ar_Ellison_1959} measure the depth by enforcing $\sigma_{11} = 1$, equivalent to \cref{eqn:depth_ET}. To be consistent with previous work we have used \cref{eqn:depth_ET} in all figures, and state explicitly whenever it is used with most expressions being for general depth. The advantage of \cref{eqn:depth_ET} is that it enables self-similar flow to be captured whenever it occurs, if instead depth is taken to be some arbitrary function then even if the flow profiles are self-similar the shape-functions in \cref{eqn:shape_avg_split} will still depend on $(x,y,t)$. As a general point, whatever definition of depth is used in a depth-averaged model, it must enable self-similar flow to be captured.

There is another measure of depth in the system: $\tilde{h} \eqdef \varsigma_h h$ appears in the definition of entrainment \cref{eqn:3Dsys_BC_entrain} which gives rise to the volume equation \cref{eqn:2Dsys_vol}. By \cref{eqn:3Dsys_BC_entrain} the entrainment across $z=b+\tilde{h}$ should be the same as across $z=H$, so $b+\tilde{h}$ can be taken to be any elevation above the region in which the velocity field is influenced by the presence of the current. However, the surface $z=b+\tilde{h}$ moves with the velocity field, and we wish for the shape-factor $\varsigma_h=\tilde{h}/h$ to be a constant for self-similar flow. This means that it must be on the edge of where the velocity field is influenced by the current, that is
\begin{align}  \label{eqn:depth_definition}
	\int_{b}^{b+\tilde{h}} \sqrt{\reavg{u}^2+\reavg{v}^2} \dd{z} &= (1-\delta) \int_b^{H} \sqrt{\reavg{u}^2+\reavg{v}^2} \dd{z}
&&\text{for}&
	0 < \delta &\ll 1.
\end{align}

Our definition of $\tilde{h}$ is comparable to that used in some experimental investigations of entrainment, in that it seeks a level at which the bed-parallel velocity has become small \citep{ar_Odier_2014,ar_Maggi_2023}. In these, the entrainment rate is defined for steady flow in terms of the velocity of fluid flow across this surface, $w_e = - w(z=\tilde{h})$. However, in a depth-average model, the conditions which arise most naturally are
\begin{equation}\begin{aligned}
	w_e &= \dv{}{x} (Uh)
	&&&&\text{for spatially evolving steady flow, and}
\\
	w_e &= \dv{\tilde{h}}{t} 
	&&&&\text{for temporally evolving uniform flow.}
\end{aligned}\end{equation}
The condition for steady flow is well known \citep{ar_Ellison_1959,ar_Negretti_2017,ar_Martin_2019,ar_Maggi_2023}, the uniform case less so. The definition of $\tilde{h}$ is important because it controls the time evolution of depth through conservation of volume \cref{eqn:2Dsys_vol}: steady currents should have the same entrainment rate as uniform currents when they have the same flow properties.
Further investigation is required to establish whether \cref{eqn:depth_definition} gives the same entrainment in these two scenarios. Throughout, we use \cref{eqn:depth_definition} with $\delta = 0.01$.

\subsection{Relating entrainment to turbulence} \label{sec:depth_entrinment}

The discussion above reveals the ad-hoc nature of the volume equation \cref{eqn:2Dsys_vol}. The interface $z = \tilde{h}$ is not a physical property, but rather is constructed through heuristic arguments, with different researchers coming up with different constructions. The entrainment rate $w_e$ then measures the flow rate across our constructed interface. This formulation is a historical artefact: the original depth-averaged model was presented by \citet{ar_SaintVenant_1871} for open-channel flow, and in that context there is a clear interface between air and water determining $z = \tilde{h}$. The source term is no longer entrainment, but could be interpreted as heavy rain inflating the volume of the river or canal. There is no such clarity for gravity currents: sometimes the upper interface is moderately sharp, but often it is highly diffuse and the attempt to represent it as a mathematical surface is artificial.

Likewise, the physical interpretation of entrainment $w_e$ is not clear: it is some measure of the progressive mixing and dilution of the current, but one dependent on the imposed interface. The physical origin of entrainment is known to be turbulence, in particular it is the buoyancy-flux which leads to the fluid becoming more dilute and disperse \citep{ar_Strang_2001,ar_Arneborg_2007,ar_Wells_2010,ar_Odier_2014}. There is a natural velocity which arises from these turbulent fluctuations, which we will term the turbulent buoyancy-velocity,
\begin{equation} \label{eqn:buoyancy_velocity}
	w_{\mathcal{B}} 
	\eqdef \frac**{2 \int_b^H J_3 \dd{z}}{\int_b^H \reavg{\phi} \dd{z}}
	= \frac{2 \mathcal{B}_K}{R g \Phi \cos\theta}
\end{equation}

For a top-hat model (\cref{tab:shape_factors}) of a compositional current ($w_s=E_s=0$) the consistency requirement for $\mathcal{B}_K$ \cref{eqn:egytransfer_tophat_BK} gives $w_e = w_{\mathcal{B}}$, consistent with \citet{ar_Wells_2010}. For a particle-driven current, \cref{eqn:egytransfer_tophat_BK} becomes
\begin{equation}\label{eqn:entrainment_consistency}
	w_{\mathcal{B}} = w_e + \pbrk*{ 2 - \varsigma_\phi } w_s \cos\theta + \frac{E_s}{\Phi}.
\end{equation}
This can be used to eliminate $w_e$ from the governing equations, after which the top-hat volume equation becomes
\begin{gather}\label{eqn:2Dsys_vol_detrainment}
	\pdv{h}{t} + \sum_\beta \pdv{}{x_\beta} (h U_\beta)
	= w_e
	= w_{\mathcal{B}} - \pbrk*{ 2 - \varsigma_\phi } w_s \cos\theta - \frac{E_s}{\Phi}
\end{gather}
Different approaches have been taken to close the entrainment in model of particle-driven currents. Suppose that $\hat{w}_e$ is a closure of entrainment in compositional currents. Historical models either neglect entrainment outright \citep{ar_Bonnecaze_1993}, or assume that closures can be transferred directly from compositional to particle-driven currents as $w_e = \hat{w}_e$ \citep{ar_Parker_1986}. However, there is a range of physical processes in particle-driven currents that are not present in compositional currents, and so it is not obvious that the aggregate effect of these processes (entrainment) is transferred so easily. Instead, it is reasonable to suppose that the effects of turbulence are most similar, the large scale vortices performing the mixing will view the particles as a concentration field. Thus $w_{\mathcal{B}}$ can be approximated as the same in the two classes of current. That is, if $\hat{w}_{\mathcal{B}}$ is a closure for buoyancy-velocity in compositional currents (top-hat gives $\hat{w}_{\mathcal{B}} = \hat{w}_e$) then $w_{\mathcal{B}} \simeq \hat{w}_{\mathcal{B}}$. Thus
\begin{equation} \label{eqn:tophat_entrainment_closure}
	w_e
	\simeq \hat{w}_e - \pbrk*{ 2 - \varsigma_\phi } w_s \cos\theta - \frac{E_s}{\Phi},
\end{equation}
we expect particle-driven currents to be modified relative to compositional currents by some detrainment, driven by settling and a reduction of available energy due to erosion. In the limit of no turbulence and a horizontal bed ($w_{\mathcal{B}} = E_s = \theta = 0$) with vertical continuity near the bed ($\varsigma_\phi=1$) this becomes the standard expression for dilute laminar detrainment $w_e \simeq - w_s$ (in this case $w_e = - w_s$ at early times, the approximation is exact despite the top-hat criterion not always being satisfied as time advances, \citealt{ar_Dorrell_2010}). For bypass flows where erosion balances deposition ($E_s = \varsigma_\phi \Phi w_s \cos\theta$) we find that $w_e \simeq \hat{w}_e - 2 w_s \cos\theta$. 

The approach \cref{eqn:tophat_entrainment_closure} is comparable to the approach of \citet{ar_Toniolo_2006} who linearly interpolated between the dynamics of compositional currents and laminar deposition to model their experiments. \Citet{ar_Pittaluga_2018} effectively use $w_e=\hat{w}_e - w_s \cos\theta$, which is subsequently adopted by \citet{ar_Ma_2024}. The energetic interpretation of detrainment presented here gives a clear guide of how it is best implemented.

To understand the impact of detrainment on the turbulence, we first reconstruct the TKE equation from \citet{ar_Parker_1986} by substituting the top-hat consistency requirements from \cref{eqn:quasiequ_egytransfer} into our TKE equation \cref{eqn:2Dsys_TKE}, yielding
\begin{multline}
	\pdv{}{t} \ppar*{ h K }
	+ \sum_\beta \pdv{}{x_\beta}(h U_\beta K)
	= \sum_\alpha \ppar*{ U_\alpha u_{\alpha\star}^2 + \textfrac{1}{2} U_\alpha^2 w_e }
\\
	- \textfrac{1}{2} R g h \cos\theta \ppar*{ \Phi w_e + \underbrace{\pbrk*{ 2 - \varsigma_\phi } w_s \Phi \cos\theta} + E_s }
	- h \epsilon_T.
\end{multline}
The term with an underbrace has a rather strange interpretation. If $\varsigma_\phi<2$ it is reasonable; the TKE is expended holding particles against gravity. However, for $\varsigma_\phi=2$ particle suspension costs no energy, and for $\varsigma_\phi>2$ TKE is generated by particle settling ($1 \lesssim \varsigma_\phi \lesssim 5$ by \cref{fig:ShapeFactors}). This term has long been recognised by Parker as greatly troubling.\footnote{Private correspondence} Rewriting using $w_\mathcal{B}$ we arrive at
\begin{multline} \label{eqn:tophat_TKE_detrainment}
	\pdv{}{t} \ppar*{ h K }
	+ \sum_\beta \pdv{}{x_\beta}(h U_\beta K)
	= \sum_\alpha \ppar*{ U_\alpha u_{\alpha\star}^2 + \overbrace{\textfrac{1}{2} U_\alpha^2 w_e }}
\\
	- \textfrac{1}{2} R g h \Phi \cos\theta w_{\mathcal{B}}
	- h \epsilon_T.
\end{multline}
with $w_e$ computed using \cref{eqn:entrainment_consistency}. This reformulation, and reinterpretation as $w_\mathcal{B}$ being the specified closure and not $w_e$, removes the troubling term. Energy is now simply expended generating buoyancy-velocity according to some closure for $w_\mathcal{B}$, as would be the case in a compositional current. The expression with the overbrace is the MKE converted to TKE during the process of accelerating entrained fluid. Provided that the entire term is positive (\ie the net entrainment is positive, $w_e>0$) there is no problem with interpretation. However, in a situation where the particle dynamics do not just reduce the entrainment ($w_e < w_\mathcal{B}$) but actually create detrainment ($w_e < 0$) there is a loss of TKE in the model. This is because the top-hat model does not account for the momentum loss to the ambient fluid in this case. This can be resolved by adding another source term to the momentum equation \cref{eqn:2Dsys_mom} which is $0$ for $w_e \geq 0$ and $w_e U_\alpha$ when $w_e<0$, and similarly a term to the MKE and total energy equations \cref{eqn:2Dsys_MKE,eqn:2Dsys_egy} which is $w_e (U_1^2+U_2^2)/2$ when $w_e<0$. Tracking through the influence of these extra terms, the overbraced term in \cref{eqn:tophat_TKE_detrainment} is zero when $w_e<0$.

This remedy for $w_e<0$ in the TKE equation is specific to the case of the top-hat model, and is not required in the model with shape-functions. Instead, the detrainment of fluid adjusts the velocity shape-function to be raised relative to the concentration profile, capturing the real physics at play. The principle goal of this manuscript is to include flow shape in the modelling framework, which is important to the flow generally and not just for detrainment. The pseudo-equilibrium case in \cref{sec:quasiequ} is quite straightforward, the consistency requirement \cref{eqn:quasiequ_egytransfer_buoy} yields
\begin{equation} \label{eqn:egytransfer_volfree_pseudoequ_entrain}
	w_e = \frac{w_\mathcal{B} - 2 w_s \cos\theta}{\tilde{\sigma}_{1z\phi}}.
\end{equation}
Note that typically $\tilde{\sigma}_{1z\phi}<1$ (\cref{fig:ShapeFactors}) and thus for compositional currents $w_e>w_\mathcal{B}$. The expression for $w_e$ can be substituted into the pseudo-equilibrium governing equations \cref{eqn:quasiequ_condn} to obtain the balances written in terms of the buoyancy-velocity $w_\mathcal{B}$. Perhaps more importantly, we can obtain the consistency requirements for turbulent-production and mean-flow buoyancy-flux by substitution into \cref{eqn:quasiequ_egytransfer_prod,eqn:quasiequ_egytransfer_Mbuoy}, which are

\begin{subequations}\begin{align}
	h \mathcal{P} &=
	\sigma_{1\phi} U u_{1\star}^2
	+ \frac{ 2 \sigma_{11} \sigma_{1\phi} - \sigma_{111} }{\tilde{\sigma}_{1z\phi}} \textfrac{1}{2} U^2 \ppar*{ w_\mathcal{B} - 2 w_s \cos\theta }
\notag \\ & \hspace{2cm}
	+ \frac{ \sigma_{z\phi} \sigma_{1\phi} - \sigma_{1z\phi} }{\tilde{\sigma}_{1z\phi}} \textfrac{1}{2} R g h \Phi \cos\theta \ppar*{ w_\mathcal{B} - 2 w_s \cos\theta }
	- h \epsilon_M,
\\
	h \mathcal{B}_M 
	&= \textfrac{1}{2} R g h \Phi \cos\theta \ppar*{ \frac{\sigma_{1z\phi}}{\tilde{\sigma}_{1z\phi}} - 1 } \ppar*{ w_\mathcal{B} - 2 w_s \cos\theta }.
\end{align}\end{subequations}
We can also use  \cref{eqn:egytransfer_volfree_pseudoequ_entrain} to get an approximation for entrainment in pseudo-equilibrium particulate currents:
\begin{align}\label{eqn:pseudoequ_volfree_entrainment}
	\tilde{\sigma}_{1z\phi} w_e + 2 w_s \cos\theta = w_\mathcal{B} &\simeq \hat{w}_\mathcal{B} = \tilde{\sigma}_{1z\phi} \hat{w}_e,
	&&\text{thus}&
	w_e &\simeq \hat{w}_e - \frac{2 w_s \cos\theta}{\tilde{\sigma}_{1z\phi}}.
\end{align}

\subsection{A volume-free energetic model of gravity-currents} \label{sec:depth_volfree}

The approach used up to now is to rearrange a consistency requirement for buoyancy-flux $h \mathcal{B}_K$ to express entrainment $w_e$ in terms of the buoyancy-velocity $w_\mathcal{B}$, which we expect to have the same closures in particle driven and compositional currents. We then substitute for $w_e$ wherever we see it. To apply this to the general case we would need to rearrange \cref{eqn:egytransfer_full_buoy} and substitute the result into the volume equation \cref{eqn:2Dsys_vol}. This is algebraically complex; here we present an equivalent but simpler approach (the equivalence is shown in the supplemental Maple document). The consistency requirement \cref{eqn:egytransfer_full_buoy} is derived from the GPE equation \cref{eqn:2Dsys_GPE} (see \cref{eqn:production_rearranged}), as
\begin{multline}	\label{eqn:2Dsys_GPE_volfree}
	\pdv{}{t}\ppar*{ \textfrac{1}{2} \sigma_{z\phi} R g h^2 \Phi \cos\theta}
	+ \sum_\beta \pdv{}{x_\beta} \ppar*{ \textfrac{1}{2} \sigma_{\beta z\phi} U_\beta R g h^2 \Phi \cos\theta }	\\
	= \tilde{S}_G 
	+ \frac{1}{2} R g h \Phi \cos\theta w_\mathcal{B}
	+ h \mathcal{B}_M
	.
\end{multline}
$\tilde{S}_G$ is defined in \cref{eqn:2Dsys_GPE_S} and contains the effects of bed variation and particle settling. $h \mathcal{B}_M$ can be found in \cref{eqn:egytransfer_full_mean} derived as the uplift of concentration by the mean flow, and is a consequence of the incompressibility condition \cref{eqn:3Dsys_scales_vol}.

\begin{figure}
	\centering
	\includegraphics{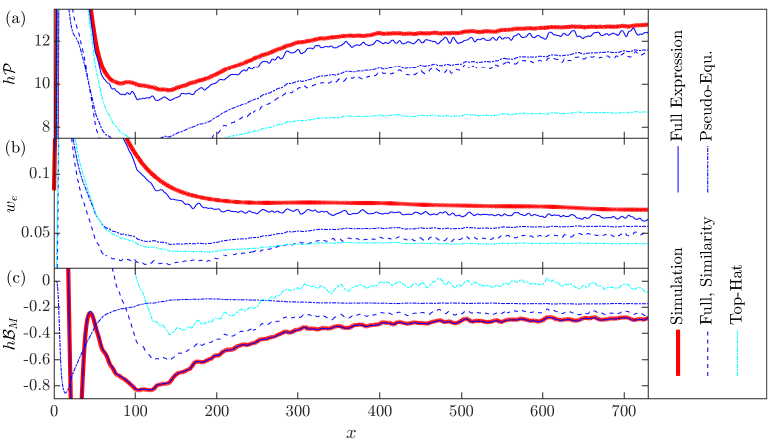}
	\caption{Equivalent plots to those in \cref{fig:Bala_Energetics} but for the case of the volume-free energetic model (\cref{sec:depth_volfree}).}
	\label{fig:Bala_Energetics_VolFree}
\end{figure}

We can now offer an alternative to the classical volumetric framework for modelling gravity currents, which was outlined in \cref{sec:model_interp}. The governing equations are those for particles \cref{eqn:2Dsys_part}, momentum \cref{eqn:2Dsys_mom}, GPE \cref{eqn:2Dsys_GPE_volfree}, and total energy \cref{eqn:2Dsys_egy}. This model is closed by imposition of shape-functions along with buoyancy-velocity $w_\mathcal{B}$; erosion $E_s$; drag $u_{\alpha\star}$; and dissipation $\epsilon_T$. Consistency requirements for this system of equations can be derived using the volume equation \cref{eqn:2Dsys_vol} to obtain an expression for entrainment $w_e$, and either the MKE equation \cref{eqn:2Dsys_MKE} or the TKE equation \cref{eqn:2Dsys_TKE} to obtain a closure for turbulent production $h \mathcal{P}$. This is equivalent to rearranging the expressions in \cref{sec:production}, see there for interpretation of terms. Substituting the definition of $w_\mathcal{B}$ from \cref{eqn:buoyancy_velocity} into the consistency requirement for $h \mathcal{B}_K$ \cref{eqn:egytransfer_full_buoy} and rearranging for $w_e$ we obtain
\small
\begin{multline} \label{eqn:egytransfer_volfree_entrain}
	w_e = \frac{\varsigma_h}{\sigma_{z\phi}} \Bigg\lgroup
	  w_\mathcal{B}
	- \pbrk*{ 2 - \sigma_{z\phi} \varsigma_\phi } w_s \cos\theta 
	- \sigma_{z\phi} \frac{E_s}{\Phi}
\\
	+ \sum_\beta \ppar*{
		  \frac{\sigma_{z\phi}}{\varsigma_h}
		+ \sigma_{\beta \phi}\sigma_{z\phi}
		- 2\varsigma_{\beta z \phi}
	} \pdv{}{x_\beta} (h U_\beta)
	+ \sum_\beta \ppar*{
		  \sigma_{\beta \phi}\sigma_{z\phi} 
		- \sigma_{\beta z \phi}
	} \frac{U_\beta h}{\Phi} \pdv{\Phi}{x_\beta}
\\
	+ \pbrk*{
		- \varsigma_h \pdv{}{t} \ppar*{ \frac{\sigma_{z\phi}}{\varsigma_h}}
		+ \sum_{\beta} \ppar*{
			  \sigma_{z\phi} \pdv{\sigma_{\beta\phi}}{x_\beta} 
			- \pdv{\sigma_{\beta z \phi}}{x_\beta} 
			- \tilde{\sigma}_{D\beta z \phi}
		} U_\beta
	} h
	\Bigg\rgroup
	.
\end{multline}
\normalsize
The two formulations of entrainment: $w_e$ from velocity of ambient fluid fluid into the current and $w_\mathcal{B}$ from the turbulent mixing, are thus related as is expected but has not previously been shown. (Note that the appropriate way to approximate a current by a top-hat model when using a closure for buoyancy-velocity is to define depth by $\sigma_{z\phi}=1$, so $h$ is twice the elevation of the centre of excess mass, \citet{ar_Arneborg_2007,ar_Anjum_2013}.) The expression for $w_e$, \cref{eqn:egytransfer_volfree_entrain}, can be substituted into the consistency requirement for $h \mathcal{P}$ \cref{eqn:egytransfer_full_turb} yielding
\small
\begin{multline} \label{eqn:egytransfer_volfree_production}
	h \mathcal{P} =
	\frac{1}{2} \ppar*{\sum_\beta \frac{\sigma_{\beta\beta}}{\sigma_{z\phi}}  U_\beta^2} \ppar*{ 
	      w_\mathcal{B}
		- \pbrk*{ 2 - \sigma_{z\phi} \varsigma_\phi } w_s \cos\theta 
		- \sigma_{z\phi} \frac{E_s}{\Phi}
	}
	+ \sum_\beta \sigma_{\beta\beta} U_\beta u_{\beta\star}^2 
	- h \epsilon_M
\\
	+ \sum_\beta (\sigma_{\beta\phi}-\sigma_{\beta\beta}) U_\beta R h \Phi \ppar*{ g_\beta - g \cos\theta \pdv{b}{x_\beta} }
\\
	+ \frac{1}{2} \sum_{\beta\gamma} \ppar*{ 
		\sigma_{\gamma\gamma} \pbrk*{ 2 \sigma_{\beta\gamma} + \sigma_{\beta\phi} - 2 \frac{\varsigma_{\beta z \phi}}{\sigma_{z\phi}} }
		- \sigma_{\beta\gamma\gamma}
	} U_\gamma^2 U_\beta \pdv{h}{x_\beta}
\\
	+ \frac{1}{2} \sum_{\beta} \ppar*{
		\sigma_{\beta\beta} \pbrk*{ 4\sigma_{\beta\beta} + \sigma_{\beta\phi} - 2 \frac{\varsigma_{\beta z \phi}}{\sigma_{z\phi}} }
		- 3\sigma_{\beta\beta\beta}
	} h U_\beta^2 \pdv{U_\beta}{x_\beta}
\\
	+ \frac{1}{2} \sum_{\beta\neq\gamma} \pgrp*{ 
	\ppar*{ 
		\sigma_{\gamma\gamma} \pbrk*{ 2\sigma_{\beta\gamma} + \sigma_{\beta\phi} - 2 \frac{\varsigma_{\beta z \phi}}{\sigma_{z\phi}} }
		- \sigma_{\beta\gamma\gamma} 
	} h U_\gamma^2 \pdv{U_\beta}{x_\beta}
	+ 2 \ppar*{ 
		\sigma_{\gamma\gamma} \sigma_{\beta\gamma}
		- \sigma_{\beta\gamma\gamma}
	} h U_\beta U_\gamma \pdv{U_\gamma}{x_\beta}
	}
\\
	+ \frac{1}{2} \sum_{\beta\gamma} \sigma_{\gamma\gamma} \ppar*{ 
		  \sigma_{\beta\phi}
		- \frac{\sigma_{\beta z\phi}}{\sigma_{z\phi}}
	} \frac{h U_\gamma^2 U_\beta}{\Phi} \pdv{\Phi}{x_\beta}
\\
	+ \sum_{\beta} \ppar*{ 
		  \sigma_{\beta\beta} \sigma_{z\phi}
		- \varsigma_{\beta z\phi}
	}  U_\beta R g h \Phi \cos\theta \pdv{h}{x_\beta}
	+ \frac{1}{2} \sum_{\beta} \ppar*{ 
		  \sigma_{\beta\beta} \sigma_{z\phi}
		- \tilde{\sigma}_{\beta z\phi}
	} U_\beta R g h^2 \cos\theta \pdv{\Phi}{x_\beta}
\\
	\!+\! \frac{1}{2} \! \sum_\beta \!\! \pgrp*{ 
		\!\!- \frac{1}{\sigma_{z\phi}} \! \pdv{}{t} \! \ppar*{ \sigma_{z\phi} \sigma_{\beta\beta} \!}
		\!+\!\! \sum_\gamma \! \ppar*{
			\!\!  \sigma_{\beta\beta} \pdv{}{x_\gamma} \! \pbrk*{ 2\sigma_{\beta\gamma} \!\!+\! \sigma_{\gamma\phi} \! }
			\!-\! \pdv{\sigma_{\beta\beta\gamma}}{x_\gamma}
			\!-\! \frac{\sigma_{\beta\beta}}{{\sigma_{z\phi}}} \! \pbrk*{\! \pdv{\sigma_{\gamma z \phi}}{x_\gamma} \!+\! \tilde{\sigma}_{D\gamma z \phi} \!}
		\! } U_\gamma
	\!}\! h U_\beta^2
\\
	+ \frac{1}{2} \sum_\beta \ppar*{ 
		\sigma_{\beta\beta} \pdv{\sigma_{z\phi}}{x_\beta} 
		- \pdv{\tilde{\sigma}_{\beta z \phi}}{x_\beta} 
		+ \tilde{\sigma}_{D\beta z \phi}
	} U_\beta R g h^2 \Phi \cos\theta
	.
\end{multline}
\normalsize
The consistency requirement for mean-flow uplift of particles \cref{eqn:egytransfer_full_mean} is not affected, and is used to close \cref{eqn:2Dsys_GPE_volfree}. The accuracy of these expressions is verified in \cref{fig:Bala_Energetics_VolFree} in the same way as the consistency expressions for the classical volumetric model were verified in \cref{fig:Bala_Energetics}, \cref{sec:quasiequ}. The coefficients in \cref{eqn:egytransfer_volfree_entrain,eqn:egytransfer_volfree_production} calculated for real currents are included in \cref{fig:EnergyCoeff}.

This new formulation does not contain an equation for volume, and eliminates the need for an arbitrary imposition of a interfacial surface. This model is closed directly by a measure of the turbulence, the buoyancy-velocity \cref{eqn:buoyancy_velocity}. The only time when the interfacial surface $z=\tilde{h}$ needs to be specified is when using the consistency relationship for entrainment \cref{eqn:egytransfer_volfree_entrain}, which depends on $\varsigma_h = \tilde{h}/h$. This model is volume-free in that it does not explicitly include a volume equation, and instead the evolution is governed by the energetics. The way that incompressibility \cref{eqn:3Dsys_scales_vol} enters the formulation is through the derivation of mean-flow uplift of concentration \cref{eqn:meanflow_uplift_derivation}, and in deducing the implied entrainment rate \cref{eqn:egytransfer_volfree_entrain}.

A key observation is that, even for compositional currents, the entrainment $w_e$ is no longer equal to, or even proportional to, the buoyancy-velocity $w_\mathcal{B}$ \cref{eqn:egytransfer_volfree_entrain}. Consequently, in the general case, we require new closures for $w_\mathcal{B}$. Early results on this have been documented by \citet{ar_Wells_2010}, but further work is required to parametrise the stirring and mixing of the density field. See \cite{ar_Caulfield_2021} for a review of this research area.

\section{Novel implications for particle auto-suspension} \label{sec:suspension_efficiency}

Here, we investigate how the energetics of a gravity current influence the particle load it carries. We begin our discussion by illustrating how the Knapp-Bagnold criterion \citep{ar_Knapp_1938,ar_Bagnold_1962} can be derived from our system of equations. This criterion identifies when the increased downslope gravitational work provided by suspended particles provides the requisite energy to keep the particles suspended, a condition referred to as auto-suspension. From the TKE equation \cref{eqn:2Dsys_TKE}, to not deplete the supply of turbulent energy we require the turbulent production exceeds the buoyancy-flux. Using the pseudo-equilibrium balance, which was shown in \cref{sec:quasiequ} to give a good description of a slowly evolving current on a flat bed, the TKE equation \cref{eqn:2Dsys_TKE} simplifies to
\begin{align}  \label{eqn:KB_quasiequ}
	h \mathcal{P} = h \mathcal{B}_K + h \epsilon_K + \sigma_{1k} K w_e.
\end{align}
This implies the Knapp-Bagnold energetic principle
\begin{equation} \label{eqn:KB_principle}
	h \mathcal{P} > h \mathcal{B}_K.
\end{equation}
Consequently, neglecting entrainment and mean-flow dissipation,
\begin{align}
	\sigma_{1\phi} U u_{1\star}^2
	>
	w_s R g h \Phi (\cos\theta)^2
\end{align}
Approximating $\sigma_{1\phi} \approx 1$ and $(\cos\theta)^2 \approx 1$, and using \cref{eqn:quasiequ_condn_mom} to rewrite drag in terms of bed slope, we arrive at
\begin{align}
	U \sin\theta > w_s,
\end{align}
the standard representation of the Knapp-Bagnold criterion. This can either be interpreted as the maximum particle size that can be transported by a given current (upper bound on $w_s$), or the minimum speed of the gravity current to be sustainable over long distances (lower bound on $U$). For gravity currents, this simple analysis ignores a lot of effects, and we  expand on it using the understanding of gravity current energetics developed.

For dilute mono- and poly-disperse fluvial systems, it has been observed that \citep{ibk_Garcia_2008,ar_Maren_2009,ar_Dorrell_2018,ar_Skevington_F006_TCFlowPower}
\begin{align} \label{eq:suspension_efficiency}
	h \mathcal{B}_K = \mathcal{E}_{\Phi} \cdot h \mathcal{P};
\end{align}
$\mathcal{E}_{\Phi}$, related to the flux Richardson number, represents the efficiency of suspending sediment, and typically satisfies $0 < \mathcal{E}_{\Phi} < 1$.  For gravity currents \citep{ar_Skevington_F006_TCFlowPower} we treat \cref{eq:suspension_efficiency} as the definition of the efficiency and explore when its value may be predictable. From the quasi-equilibrium balance \cref{eqn:KB_quasiequ}, the final term can be neglected because TKE, $K$, is small compared to the MKE, $\propto \textfrac{1}{2} U^2$, see \cref{eqn:quasiequ_egytransfer_prod}. Substituting in \cref{eq:suspension_efficiency} we obtain

\begin{align}
	\frac{h \mathcal{B}_K}{h \epsilon_K} = \frac{\mathcal{E}_{\Phi}}{1 - \mathcal{E}_{\Phi}} = \Gamma,
\end{align}
where $\Gamma$ is a mixing coefficient. The analysis of \citet{ar_Osborn_1980} gives $\Gamma<1/5$ so that $\mathcal{E}_{\Phi}<1/6$. 
More recently larger values of $\Gamma$ have been reported \citep{ar_Maffioli_2016,ar_Mashayek_2021}, we take $\mathcal{E}_{\Phi}=1/6$ as a reasonable estimate.

\begin{figure}
	\centering
	\includegraphics{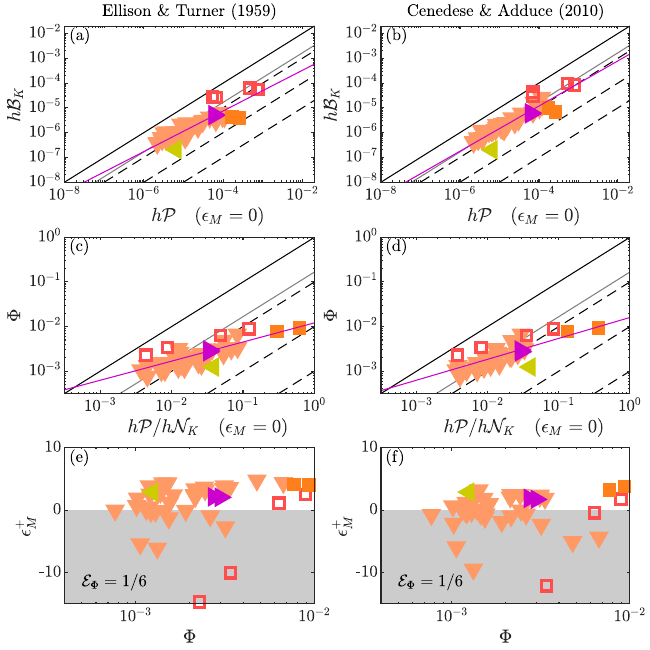}
	\caption{Energetics of particle suspension in a classical volumetric model using the entrainment models of \citet{ar_Ellison_1959} (a,c,e) and \citet{ar_Cenedese_2010} (b,d,f), for symbols see legend in \cref{fig:ShapeFactors}. (a,b) Turbulent-production (neglecting mean-flow dissipation) against buoyancy-flux, rearranged in (c,d) to show the concentration. In both (b,c) lines of constant efficiency are shown in black ($\mathcal{E}_\Phi=1$), grey ($\mathcal{E}_\Phi = 1/6$), and dashed black ($\mathcal{E}_\Phi \in \pbrc*{10^{-1},10^{-2},10^{-3}}$), and a best fit line is shown in purple. (e,f) The dimensionless mean-flow dissipation against the concentration, the region outside of the bounds \cref{eq:dissipation_bounds} shaded grey.}
	\label{fig:KB_parkerstyle}
\end{figure}

To be able to examine the suspension efficiency of the dataset used in \cref{fig:ShapeFactors} we require values of $h \mathcal{P}$ and $h \mathcal{B}_K$. Ideally, we would use direct measurements, which we do not have for most of the data, or we would use the full expressions \cref{eqn:egytransfer_full_turb,eqn:egytransfer_full_buoy,eqn:egytransfer_full_mean} but these require derivatives of the depth-averaged quantities and shape-factors. Instead, we use the pseudo-equilibrium expressions \cref{eqn:quasiequ_egytransfer} which were shown to provide a good level of accuracy provided the current is slowly evolving (\cref{fig:Bala_Energetics}). We make the weak requirement that the flow has a non-zero erosion rate ($E_s>0$) by the condition of \citet{ar_Guo_2020} \citep[see][]{ar_Skevington_F006_TCFlowPower} and is dilute, $\Phi \leq 10^{-2}$. We also require that the residual \cref{eqn:model_error} in the pseudo-equilibrium momentum balance \cref{eqn:quasiequ_condn_mom} is such that error accumulates over a distance of more than $10^{3/2} h$, that is $h \mathcal{R} < 10^{-3/2}$ where we use the entrainment closure from \citet{ar_Cenedese_2010}. We include only experimental data, the field measurements reported by \citet{ar_Simmons_2020} are for flow in a sinuous channel and levee overspill and are not in pseudo-equilibrium. 

For the moment we neglect the mean-flow dissipation (we will discuss this later), and examine the balance \cref{eq:suspension_efficiency} in \cref{fig:KB_parkerstyle}(a,b). We use the entrainment relationships from \citet{ar_Ellison_1959} and \citet{ar_Cenedese_2010} to show how little different empirical closures effect the results, and in \cref{fig:KB_parkerstyle} a classical volumetric model is assumed (\cref{sec:model_interp}) where entrainment models from compositional currents are used directly (\cref{sec:depth_entrinment}). We observe an approximately linear trend (\cref{fig:KB_parkerstyle}(a,b)), indicating that \cref{eq:suspension_efficiency} is perhaps a reasonable approximation, although there is over an order of magnitude of variation in the efficiency $\mathcal{E}_\Phi$. It is possible to rewrite \cref{eq:suspension_efficiency} to give a prediction of the sediment concentration:
\begin{align} \label{eq:suspension_efficiency_part}
	\Phi = \mathcal{E}_{\Phi} \cdot \frac{h \mathcal{P}}{h \mathcal{N}_K} .
\end{align}
Here, $\mathcal{N}_K$ is the normalised buoyancy-flux, that is the buoyancy-flux per unit sediment concentration. In pseudo-equilibrium this is, by \cref{eqn:quasiequ_egytransfer_buoy},
\begin{equation} \label{eqn:normalised_buoyprod}
	h \mathcal{N}_K 
	\eqdef \frac{h \mathcal{B}_K}{\Phi} 
	= \textfrac{1}{2} R g h \cos\theta \ppar[\big]{
		\tilde{\sigma}_{uz\phi} w_e
		+ 2 w_s \cos\theta
	}.
\end{equation}
Plotting the balance \cref{eq:suspension_efficiency_part} in \cref{fig:KB_parkerstyle}(c,d) we observe that the scatter collapses around a non-linear trend (these plots are very similar to figure 3(c) in  \citealt{ar_Skevington_F006_TCFlowPower}).  This implies that some process correlated with the particle concentration is controlling the mixing efficiency. This should not be the case, and indicates a problem with the analysis.

So far we have neglected $h \epsilon_M$, the mean-flow dissipation, which was shown in \cref{sec:quasiequ_meandiss} to be of leading order for the experimental flows in the dataset, with bounds   
\begin{align} \label{eq:dissipation_bounds}
	0 < \epsilon_M^+ \lessapprox 9.15.
\end{align}
Narrowing down this value further requires both detailed information about the structure of the bed roughness and how this relates to dissipation. Consequently, we engage in a plausibility analysis: could the results be explained by mean-flow dissipation in the permissible range?

The effect of increasing $h \epsilon_M$ moves points in \cref{fig:KB_parkerstyle}(c,d) to the left, and we observe what happens when we move all the points to a line of equal efficiency and back calculate the dissipation. Moving all the points to the line $\mathcal{E}_\Phi = 1/6$  results in \cref{fig:KB_parkerstyle}(e,f). Many of the points require $\epsilon_M^+<0$, which we interpret as the real current either having more production or less buoyancy flux than captured by the model.

\begin{figure}
	\centering
	\includegraphics{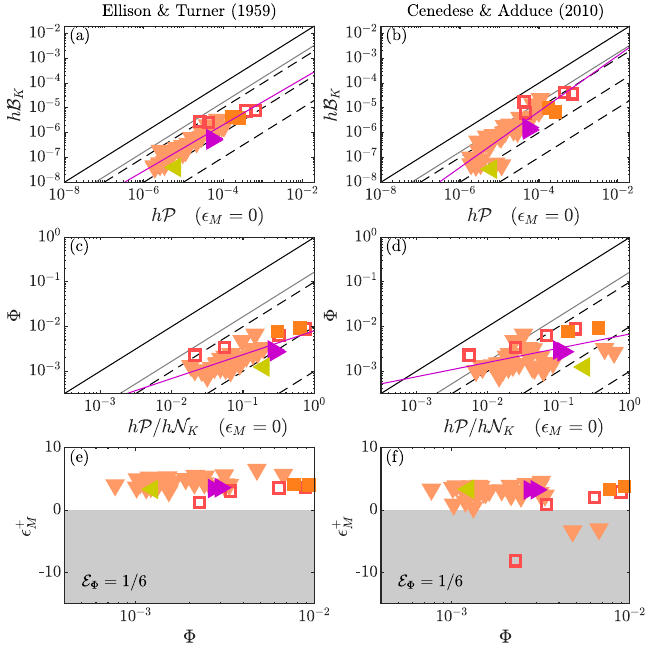}
	\caption{The same as \cref{fig:KB_parkerstyle} but for a volume-free energetic model.}
	\label{fig:KB_volfree}
\end{figure}

To this point we have been considering a classical volumetric model (\cref{sec:model_interp}), where a closure for entrainment developed for compositional currents, $\hat{w}_e$, may be directly used for a particle driven current, $w_e \simeq \hat{w}_e$. In the volume-free energetic model (\cref{sec:depth_volfree}), it is the turbulent buoyancy-velocity \cref{eqn:buoyancy_velocity} for which a closure for a compositional current, $\hat{w}_\mathcal{B}$, may be used in a particle driven current, $w_\mathcal{B} \simeq \hat{w}_\mathcal{B}$. We do not have such a closure, but have shown how to construct something equivalent from entrainment models in \cref{eqn:pseudoequ_volfree_entrainment}.
We repeat the analysis of energetics in particle suspension, see \cref{fig:KB_volfree}. Now, there is a difference between the closures of \citet{ar_Ellison_1959} and \citet{ar_Cenedese_2010}, which arises because $h \mathcal{B}_K$ is determined solely by the entrainment closure and is not dependent on the settling velocity. For both closures, there is a broad range of efficiencies when mean-flow dissipation is neglected. Exploring what values $\epsilon_M^+$ could take (\cref{fig:KB_volfree}(e-f)) there is reasonably narrow range of values for  satisfying the bounds \cref{eq:dissipation_bounds}, except for three points in \cref{fig:KB_volfree}(f) (perhaps these flows are rapidly depositing sediment, or the entrainment model is in error).

Our plausibility analysis is thus a success, the departure from constant efficiency  \cref{eq:suspension_efficiency_part} observed by \cite{ar_Skevington_F006_TCFlowPower} and reproduced in \cref{fig:KB_parkerstyle}(c) can be corrected using two effects: mean-flow dissipation, and the settling velocity reducing the entrainment rate. Compared to \cite{ar_Parker_1986}, both of these effects reduce the implied rate of turbulent production by \cref{eqn:quasiequ_egytransfer_prod}. The reduced entrainment rate, by design, exactly cancels the settling term in \cref{eqn:quasiequ_egytransfer_buoy} meaning that the implied buoyancy-flux is no longer given an artificial boost to account for the need to suspend particles. The reduction in both of these turbulent terms allows, in principle, for the suspension efficiency \cref{eq:suspension_efficiency} to be independent of particle concentration as expected.

\section{Future work} \label{sec:future}

We here overview additional work required to realise the potential of this new modelling framework. To enable the model to be used predictively, accurate closures are required for the shape, entrainment, erosion, drag, and dissipation. A large experimental dataset exists for the shape of the velocity and concentration profiles, but comparatively little data exists for TKE (\cref{fig:ShapeFactors}). The shape depends on the Froude number \citep{ar_Abad_2011} along with a wider set of dimensionless parameters \cref{eqn:quasiequ_defn_dimless}, and establishing this dependence is key to understanding the evolution of currents. The total dissipation also needs to be properly quantified, \cite{ar_Parker_1986} provides an approximate form based on heuristic arguments but an empirically verified closure would be preferable. Closures for the other parameters exist, but have been developed separately; ideally, the full set of closures should be developed in accordance with the consistency relationships (\cref{sec:production}) which would ensure that the energetics predicted by the model are accurate. At lab scale or over smooth beds, the use of the consistency requirement for turbulent production requires an understanding of the mean-flow dissipation and how this relates to roughness (\cref{sec:quasiequ_meandiss}). Regarding entrainment, it is worth further exploring whether an energetic approach would be more accurate than the prevailing volumetric approach (\cref{sec:depth}). The transfer of closures from compositional to particulate currents needs to be verified against high-resolution velocity and concentration fields across different settling velocities. The model and closures capture the body of the current, and require boundary conditions for the front to be developed, similar those by \citet{ar_Benjamin_1968} and \citet{ar_Ungarish_2018} but with the profiles in the current specifiable. The goal is to develop a set of closures sufficiently accurate to capture the extremely long run-out of real-world currents \citep{ar_AzpirozZabala_2017} and the development of sedimentary deposits over geological timescales \citep{ar_Wahab_2022}.
\section{Summary and conclusion} \label{sec:conclusion}

In this work we have constructed a novel depth-averaged modelling framework that for the first time accurately captures the bulk dynamics of a gravity current, allowing for arbitrarily shaped profiles of concentration, velocity, and turbulent kinetic energy (TKE) to be implemented (\cref{sec:model}). Prior to this work, profiles of velocity and concentration have been used in depth-average models of gravity current that do not attempt to capture energetics, and separately research has been conducted on the turbulence and mixing processes. Our work connects these two research efforts, enabling models which accurately capture energetics in a depth-average framework (\cref{sec:quasiequ}). These energetic balances inform the dynamics of particle suspension, and by providing a robust mathematical framework within which to understand the energetics we enable a new and deeper understanding of particulate gravity currents (\cref{sec:suspension_efficiency}).

With the modelling framework we propose two classes of predictive model for the flow depth, concentration, velocity and TKE, which require different closures. Both require specification of the shape of the velocity, concentration, and TKE profiles, along with expressions for the erosion of particles, basal drag, and viscous dissipation of energy. The first class of model, which we term a classical volumetric model (\cref{sec:model_interp}), requires specification of some interface over which entrainment occurs along with the entrainment rate (\cref{sec:depth_defn}). The resulting model consists of equations for volume, concentration, momentum, and total energy. Models which make use of volume are by far the most common class of model in the gravity current community, and go back to \citet{ar_Ellison_1959} and \citet{ar_Parker_1986}. Our model is a direct generalisation of the classical model derived by \cite{ar_Parker_1986}, and differs from it only by the possibility to specify arbitrary profiles through shape-factors. The top-hat version of our model where all shape-factors are unity precisely recovers the model of \cite{ar_Parker_1986}. In the optimal case where the shape-factors are taken directly from DNS simulation, the model is able to reproduce the results of DNS with almost no error and is substantially more accurate than the top-hat version (\cref{sec:quasiequ}). |This gives confidence that, with high-quality closures, the new model can produce accurate predictions of the flow evolution including the energetics.

This approach has some shortcomings in its capturing of entrainment. The construction of an interface over which there is entrainment of ambient fluid is artificial, the upper region of a gravity current is highly diffuse and there is no surface separating the current from the ambient (\cref{sec:depth_entrinment}). Moreover, it is commonplace to using an entrainment closure from compositional currents in a particulate current, which results in the implied uplift of particles by turbulence (the turbulent buoyancy-flux) being an increasing function of particle settling-velocity (\cref{sec:production_dervation}). The buoyancy-flux should only be dependent on the distribution of particles and the strength of the turbulence, and artificially inflating it to provide additional support for the particles is not only erroneous but also confuses attempts to understand particle suspension (\cref{sec:suspension_efficiency}). Instead, the available work of the buoyancy-flux is split between upholding the particles against settling and entraining ambient fluid.

For these reasons, we propose a second class of model wherein it is the buoyancy-flux which is closed for directly, not the entrainment, which removes the need for an interface over which the entrainment occurs. Careful consideration shows that the appropriate formulation of the model is as equations for concentration, momentum, gravitational potential energy, and total energy, and we term this class volume-free energetic models (\cref{sec:depth_volfree}). This class corresponds much more closely to a physical understanding of gravity currents, which do have budgets of excess mass, momentum, and energy, but typically not a clear region to define a volume. Models which require a closure of the buoyancy-flux directly are far less common, but do exist \citep{ar_Arneborg_2007,ar_Wells_2010}.  It is likely that the processes which generate buoyancy-flux in compositional currents are the same as those in particulate currents, which would allow for the same closures to be used for both. This assumption was used to help understand the dynamics of particle suspension in \cref{sec:suspension_efficiency}.

Both classes of predictive model have some effects which are directly closed for, and others which are implied by the model as indirect predictions. For these indirect predictions to be accurate, model closures must be validated to ensure they give the correct indirect predictions through what we have termed consistency requirements (\cref{sec:production_discussion}).  In both classes the turbulent-production, which transfers energy from the mean-flow to the turbulence, is implied from the loss of mean-flow energy (\cref{sec:production_dervation,sec:depth_volfree}). There is also loss of mean-flow energy directly to viscosity in the boundary layer by the bed, and we show that this makes a leading order contribution at Reynolds numbers up-to and beyond geophysical scales (\cref{sec:quasiequ_meandiss}). For the case of bed-roughness, which is present for almost all particulate currents and many compositional currents, structures that are very large ($\sim 10^4$ wall units) are required before the mean flow dissipation is negligible. However, an understanding of mean-flow dissipation is lacking in the literature at large, which makes using the consistency requirements challenging. Large scale particulate currents over bedforms, along with gravity currents propagating along a strong density interface rather than a solid boundary (a type of intrusion), will not have such a strong mean-flow dissipation and so the consistency requirement may be more easily employed. What is much more straightforward is in the use of the consistency requirement for buoyancy-flux in classical volumetric models (\cref{sec:production_dervation}), which becomes a consistency requirement for entrainment in volume-free energetic models (\cref{sec:depth_volfree}). For particulate currents, the implied entrainment incorporates the particle detainment by settling, which has been discussed by several authors without theoretical justification \citep{ar_Toniolo_2006,ar_Pittaluga_2018,ar_Ma_2024}; here we provide that justification. These consistency requirements not only provide insight into entrainment, but in future can be used to constrain closures to produce energetically consistent models in the developed depth-average framework. This would enable the energetics to be captured in system-scale models, giving accurate prediction of hazardous environmental flows at geophysical scales.

\begin{myappendices}
	\section{The scales of the flow within the current}  \label{app:deriv_channel_scales}

The full system of Boussinesq RANS equations is
\begin{subequations}\label{eqn:3Dsys}
\allowdisplaybreaks[1]
\begin{align}
	\pdv{\reavg{u_j}}{x_j} &= 0,
	\label{eqn:3Dsys_vol}
\\
	\pdv{\reavg{\phi}}{t} + \pdv{}{x_j} \ppar*{\pbrk*{\reavg{u_j} + \tilde{u}_j} \reavg{\phi} + J_j } &= 0,
	\label{eqn:3Dsys_part}
\\
	\pdv{\reavg{u_i}}{t} + \pdv{}{x_j}\ppar*{\reavg{u_j}\reavg{u_i} - \tau_{ji}^R - \nu \pdv{\reavg{u_i}}{x_j}} + \frac{1}{\rho_f} \pdv{\reavg{p}}{x_i} &= R g_i \reavg{\phi},
	\label{eqn:3Dsys_mom}
\\
	\pdv{k}{t} + \pdv{}{x_j}\ppar*{\reavg{u_j}k + T_j} &= P - \tilde{\epsilon}_K + R g_j J_j.
	\label{eqn:3Dsys_turb}
\end{align}\end{subequations}
Here, we document the scale analysis used to simplify \cref{eqn:3Dsys} to \cref{eqn:3Dsys_scales}. Throughout this section we will make claims of processes that occur in the flow, and use these to balance terms. We do not claim that these processes are always occurring, but simply that they are processes we wish to capture, and they set the largest scale the terms can take anywhere in the current given a particular slope.

As stated in the main text, we employ a time-scale $\mathscr{T}$ , and length-scales $\mathscr{L}_i$ ($\pbrc*{ \mathscr{L}_1,\mathscr{L}_2 } \gg \mathscr{L}_3$), and assume the Reynolds averaged velocities scale as $\mathscr{U}_i = \mathscr{L}_i/\mathscr{T}$. We denote the scale of the Reynolds averaged pressure by $\mathscr{P}$, the TKE scale by $\mathscr{K}$, the TKE dissipation scale by $\mathscr{E}$, and the scale of the Reynolds averaged concentration as $\varphi$. Without loss of generality we assume that the $x,y$ plane is orientated so that $y$ is horizontal and $g_2=0$. We  define the gravitational scales to be $\mathscr{G}_i = \abs{R g_i \varphi}$ so that $\mathscr{G}_1 = \mathscr{G}_3 \abs{\tan\theta}$. All scales should be understood as the scale of the depth average (\cref{app:depth_average}) of the given quantity. Note that it is possible to perform the depth average first and then the analysis of scales, which is preferable from the perspective of formal justification. However, this approach increases the complexity of the analysis substantially, which is why the order of presentation here has been chosen.

We first consider the momentum equation \cref{eqn:3Dsys_mom}. To examine the scales of the system we require scales for the components of the Reynolds stress. For the purpose of constructing scales only we employ the eddy viscosity approximation, that is
\begin{align} \label{eqn:deviatoric_stress}
	\tau_{ij}^R &= - \textfrac{2}{3} k \delta_{ij} + \tau_{ij}^D,
	&\text{where}&&
	\tau_{ij}^D &= \nu_t \ppar*{ \pdv{\reavg{u_i}}{x_j} + \pdv{\reavg{u_j}}{x_i} }
\end{align}
is the deviatoric Reynolds stress, and the eddy viscosity $\nu_t(\vecb{x},t)$ has scale $\mathscr{N}$. We also employ the scales
\begin{align}
	\norm{\tilde{\vecb{u}}} &\lesssim \frac{\mathscr{L}_3}{\mathscr{T}},
	&
	J_i &\sim \frac{ \mathscr{N} \varphi }{ \mathscr{L}_i },
	&&\text{and}&
	T_i &\sim \frac{ \mathscr{N} \mathscr{K} }{ \mathscr{L}_i }.
\end{align}
Now \cref{eqn:3Dsys_mom} becomes (outside of the viscous boundary layer so we can neglect viscosity)
\begin{align}
	\underbrace{ \pdv{\reavg{u_i}}{t} + \pdv{}{x_j}\ppar*{\reavg{u_j}\reavg{u_i}}   \vphantom{\Bigg)}}_{\textstyle \mathscr{L}_i/\mathscr{T}^2} &= 
	- \underbrace{ \frac{1}{\rho_f} \pdv{\reavg{p}}{x_i} 					        \vphantom{\Bigg)}}_{\textstyle \mathscr{P}/\rho_f\mathscr{L}_i}
	- \underbrace{ \frac{2}{3} \pdv{k}{x_i} 							            \vphantom{\Bigg)}}_{\textstyle \mathscr{K}/\mathscr{L}_i}
	+ \underbrace{ \pdv{}{x_j} \ppar*{ \nu_t \pdv{\reavg{u_i}}{x_j} }	        	\vphantom{\Bigg)}}_{\textstyle \mathscr{N}\mathscr{L}_i/\mathscr{L}_j^2\mathscr{T}}
	+ \underbrace{ \pdv{\nu_t}{x_j} \pdv{\reavg{u_j}}{x_i} 				            \vphantom{\Bigg)}}_{\textstyle \mathscr{N}/\mathscr{L}_i\mathscr{T}}
	+ \underbrace{ R g_i \reavg{\phi}						                        \vphantom{\Bigg)}}_{\textstyle \mathscr{G}_i},
\end{align}
where the scales of the flow within the current are given beneath the brace under each term. In the downslope direction ($i=1$) the driving force, scale $\mathscr{D}$, is provided by the larger of the pressure+TKE gradient and the longitudinal component of gravity, \ie
\begin{align}
	\mathscr{D} = \max \ppar*{ \frac{\textfrac{1}{\rho_f} \mathscr{P} + \mathscr{K}}{\mathscr{L}_1} , \mathscr{G}_1 }.
\end{align}
The driving force accelerates the flow until the turbulent viscous effects are sufficiently strong, causing all three effects to appear at leading order
\begin{align}
	\frac{\mathscr{L}_1}{\mathscr{T}^2} &= \mathscr{D} = \frac{\mathscr{N}\mathscr{L}_1}{\mathscr{L}_3^2\mathscr{T}},
	&\text{thus}&&
	\mathscr{D} &= \frac{\mathscr{L}_1}{\mathscr{T}^2},
	&
	\mathscr{N} &= \frac{\mathscr{L}_3^2}{\mathscr{T}}.
\end{align}
In the bed-normal ($i=3$) direction, the pressure+TKE gradient is generated by the effects of gravity,
\begin{align}
	\frac{\textfrac{1}{\rho_f}\mathscr{P} + \mathscr{K}}{\mathscr{L}_3} &= \mathscr{G}_3  \gg \frac{\mathscr{L}_3}{\mathscr{T}^2}.
\end{align}
How this balance interacts with the downslope balance depends on the slope angle. On a very shallow slope
\begin{equation}\begin{gathered}
	\begin{aligned}
		\abs{\tan\theta} &\leq \frac{\mathscr{L}_3}{\mathscr{L}_1}
	&\text{we have}&&
		\mathscr{G}_1 &\leq \frac{\textfrac{1}{\rho_f} \mathscr{P} + \mathscr{K}}{\mathscr{L}_1},
	&\text{thus}&&
		\mathscr{D} &= \frac{\textfrac{1}{\rho_f} \mathscr{P} + \mathscr{K}}{\mathscr{L}_1}
	\end{aligned}
\\
	\begin{aligned}
	&\text{so that}&
		\mathscr{G}_3 \mathscr{L}_3 &= \textfrac{1}{\rho_f}\mathscr{P} + \mathscr{K} = \mathscr{D} \mathscr{L_1} = \frac{\mathscr{L}_1^2}{\mathscr{T}^2},
	\end{aligned}
\\
	\begin{aligned}	
	&\text{and}&
		\mathscr{G}_3 &\gg \frac{\mathscr{L}_3}{\mathscr{T}^2}
	&\text{implies}&&
		\mathscr{L}_1 &\gg \mathscr{L}_3;
	\end{aligned}
\end{gathered}\end{equation}
the current is driven by longitudinal pressure gradients. On a moderate to steep slope
\begin{equation}\begin{gathered}
	\begin{aligned}
		\abs{\tan\theta} &\geq \frac{\mathscr{L}_3}{\mathscr{L}_1}
	&\text{we have}&&
		\mathscr{G}_1 &\geq \frac{\textfrac{1}{\rho_f} \mathscr{P} + \mathscr{K}}{\mathscr{L}_1},
	&\text{thus}&&
		\mathscr{D} &= \mathscr{G}_1
	\end{aligned}
\\
	\begin{aligned}
	&\text{so that}&
		\textfrac{1}{\rho_f}\mathscr{P} + \mathscr{K} &= \mathscr{G}_3 \mathscr{L}_3 = \frac{\mathscr{G}_1 \mathscr{L}_3}{\abs{\tan\theta}} = \frac{\mathscr{D} \mathscr{L}_3}{\abs{\tan\theta}} = \frac{\mathscr{L}_1 \mathscr{L}_3}{\mathscr{T}^2 \abs{\tan\theta}},
	\end{aligned}
\\
	\begin{aligned}	
	&\text{and}&
		\mathscr{G}_3 &\gg \frac{\mathscr{L}_3}{\mathscr{T}^2}
	&\text{implies}&&
		\mathscr{L}_1 &\gg \mathscr{L}_3 \abs{\tan\theta};
	\end{aligned}
\end{gathered}\end{equation}
the current is driven directly by the longitudinal component of gravity. Combining the two cases, the scales have bounds
\begin{align}
	\mathscr{G}_1 &\leq \frac{\mathscr{L}_1}{\mathscr{T}^2},
	&
	\mathscr{G}_3 &\leq \frac{\mathscr{L}_1^2}{\mathscr{T}^2 \mathscr{L}_3},
	&
	\textfrac{1}{\rho_f} \mathscr{P} + \mathscr{K} &\leq \frac{\mathscr{L}_1^2}{\mathscr{T}^2}.
\end{align}
where the first is equality on moderate to steep slopes, and the latter two are equality on shallow slopes. Analysing the turbulent production, the dominant contribution is
\begin{gather}
	P \simeq \nu_t \ppar*{ \pdv{\reavg{u}}{z} }^2 +  \nu_t \ppar*{ \pdv{\reavg{v}}{z} }^2
	\sim \mathscr{N} \frac{\mathscr{U}_1^2 + \mathscr{U}_2^2}{\mathscr{L}_3^2} 
	= \frac{\mathscr{L}_1^2 + \mathscr{L}_2^2}{\mathscr{T}^3},
\shortintertext{thus by \cref{eqn:3Dsys_turb}}
	\mathscr{K} = \frac{\mathscr{L}_1^2 + \mathscr{L}_2^2}{\mathscr{T}^2},
	\qquad
	\mathscr{E} = \frac{\mathscr{L}_1^2 + \mathscr{L}_2^2}{\mathscr{T}^3}.
\end{gather}
Using the developed scales, simplifications can be made to the system of equations \cref{eqn:3Dsys} by neglecting terms order $\mathscr{L}_3/\mathscr{L}_1$ or $\mathscr{L}_2/\mathscr{L}_1$ smaller than the largest, yielding \cref{eqn:3Dsys_scales}. This analysis preserves terms which are large in the wall boundary layer which is dominated by $z$ derivatives, up to the fact that we need to re-include the viscous stress.

	\section{Comparison to Ellison-Turner variables and shape-factors} \label{app:ETshape}


The choice of shape-factors used in the main text differs from the choice made by \citet{ar_Ellison_1959}, and developed by \citet{ar_Parker_1986,ar_Parker_1987}. There, the fluxes were simplified by defining
\begin{subequations}
\allowdisplaybreaks[1]
\begin{align}
	\reavg{u}(x,y,z,t) 		&= \ElTu{\xi_u}(x,y,\zeta,t) \cdot \ElTu{U}(x,t),
\\
	\reavg{\phi}(x,y,z,t) 	&= \ElTu{\xi_\phi}(x,y,\zeta,t) \cdot \ElTu{\Phi}(x,t),
\\
	k(x,y,z,t) 				&= \ElTu{\xi_k}(x,y,\zeta,t) \cdot \ElTu{K}(x,t),
\end{align}\end{subequations}
where the quantities $\ElTu{\xi_u}$, $\ElTu{\xi_\phi}$, $\ElTu{\xi_k}$ satisfy
\begin{subequations} \label{eqn:ET_shape_normalise}
\begin{align}
	\int_0^{\zeta_H} \ElTu{\xi_u} \dd{\zeta} &= 1,
&
	\int_0^{\zeta_H} \ElTu{\xi_u^2} \dd{\zeta} &= 1,
\\
	\int_0^{\zeta_H} \ElTu{\xi_u} \ElTu{\xi_\phi} \dd{\zeta} &= 1,
&
	\int_0^{\zeta_H} \ElTu{\xi_u} \ElTu{\xi_k} \dd{\zeta} &= 1.
\end{align}\end{subequations}
Above, we have identified variables specific to the Ellison-Turner scaling by an ET subscript. The definitions have been modified to be compatible with our  lack of self-similar assumption so that the variables $\ElTu{\xi_{\omitdummy}}$ depend on $x$, $y$, and $t$ as well as $\zeta$, and our finite range of integration in the vertical direction. 

\begin{table}
	\centering
	$
\renewcommand*{\arraystretch}{1.3}
\begin{array}{ l | r@{}l | c *{1}{ | r@{}l }}
		& \multicolumn{2}{c|}{\text{Definition}}	& \text{Simplified}		& \multicolumn{2}{c}{\text{Equiv.}}	\\
\hline
	a_0 &	&	&&
			\varsigma_{h} &									 \\
	a_1	&   \int_0^{\zeta_H} & \ElTu{\xi_\phi} \dd{\zeta}		&& 
			1&/\sigma_{1\phi}						\\
	a_2	& 2 \int_0^{\zeta_H} & \int_{\zeta_1}^{\zeta_H} \eval*{\ElTu{\xi_\phi}}_{\zeta_2} \dd{\zeta_2} \dd{\zeta_1}
		& 2 \int_0^{\zeta_H}   \zeta \ElTu{\xi_\phi} \dd{\zeta}	& 
			\sigma_{z\phi} &/\sigma_{1\phi}	\\
	a_3 &   \int_0^{\zeta_H} & \ElTu{\xi_u^3} \dd{\zeta}		&& 
			\sigma_{111}&/\sigma_{11}^3						\\
	a_4 & 2 \int_0^{\zeta_H} & \int_{\zeta_1}^{\zeta_H} \eval*{\ElTu{\xi_u}}_{\zeta_1} \eval*{\ElTu{\xi_\phi}}_{\zeta_2} \dd{\zeta_2} \dd{\zeta_1}	& a_7
													\\
	a_5 &   \int_0^{\zeta_H} & \int_{\zeta_1}^{\zeta_H} \zeta_1 \eval*{\ppar*{\pdv**{\ElTu{\xi_u}}{\zeta}}}_{\zeta_1} \eval*{\ElTu{\xi_\phi}}_{\zeta_2} \dd{\zeta_2} \dd{\zeta_1}	& \textfrac{1}{2} (a_8-a_7)	&
		& 													\\
	a_6 & 2 \int_0^{\zeta_H} & \int_{\zeta_1}^{\zeta_H} \eval*{\ElTu{\xi_u}}_{\zeta_2} \eval*{\ElTu{\xi_\phi}}_{\zeta_2} \dd{\zeta_2} \dd{\zeta_1} & a_8 &
			&												\\
	a_7 & 2 \int_0^{\zeta_H} & \int_0^{\zeta_1} \eval*{\ElTu{\xi_u}}_{\zeta_2} \eval*{\ElTu{\xi_\phi}}_{\zeta_1} \dd{\zeta_2} \dd{\zeta_1}	&& 
			\tilde{\sigma}_{1z\phi}&/\sigma_{11}\sigma_{1\phi}	\\
	a_8 & 2 \int_0^{\zeta_H} & \zeta \ElTu{\xi_\phi} \ElTu{\xi_u} \dd{\zeta}	&& 
			\sigma_{1z\phi}&/\sigma_{11}\sigma_{1\phi}  	\\
	a_9	&	\int_0^{\zeta_H} & \ElTu{\xi_k} \dd{\zeta}			&& 
			1&/\sigma_{1k} 							\\
	r_0 & 	 				& \ElTu{\xi_\phi} \big|_{\zeta = 0} 	&&
			\varsigma_\phi&/\sigma_{1\phi}					
\end{array}
$
	\caption{The shape-factors used by \cite{ar_Parker_1987}, with $a_0$ being additional here. The first column is the symbols used for the shape-factors. The second is their definition (modified here to account for possible lateral variation). The third is simplified expressions for the shape-factors. In the fourth we express these shape-factors in terms of the ones defined in \cref{tab:shape_factors}.}
	\label{tab:shape_factors_old}
\end{table}

These shape-functions may be used to compute shape-factors, and those measured by \citet{ar_Parker_1987} and \citet{ar_Islam_2010} are defined in \cref{tab:shape_factors_old}. In this table, several of the shape-factors have simplified expressions listed, and in each case this has been achieved by switching the order of integration of $\zeta_1$ and $\zeta_2$. For $a_5$, we subsequently integrate the product $\zeta \dv*{\ElTu{\xi_u}}{\zeta}$ by parts. The simplifications for $a_2$, $a_4$, $a_5$ and $a_6$ do not rely on the finite extent of the integrals and we may take the limit $\zeta_H \to \infty$; nor on the shape-functions satisfying \cref{eqn:ET_shape_normalise}. Thus, these equalities are satisfied by any shape-function, including those measured in experiment. What is concerning, is that some experimental results do not satisfy the equalities proved above. In the measurements of \cite{ar_Parker_1987} we have 
\begin{align}
	\frac{a_4}{a_7} &= 1.00,
&
	\frac{a_5}{\textfrac{1}{2} (a_8-a_7)} &= 0.86,
&
	\frac{a_6}{a_8}	&= 1.16,
\intertext{while in those of \cite{ar_Islam_2010}}
	\frac{a_4}{a_7} &= 0.87,
&
	\frac{a_5}{\textfrac{1}{2} (a_8-a_7)} &= 1.08,
&
	\frac{a_6}{a_8}	&= 0.71.
\end{align}
If the integrals were evaluated exactly then all the ratios would be $1$. We expect that the discrepancy comes from under-resolved numerical integration, though differences of up to $30\%$ do suggest significant problems.

To close this section, we demonstrate how to convert between the Ellison-Turner variables (\cref{tab:shape_factors_old}) and those used here (\cref{tab:shape_factors}). Observe that
\begin{align}
	\ElTu{U}^2 &=  U_1^2,
&
	\ElTu{U} \ElTu{\Phi} &=  \sigma_{1\phi} U_1 \Phi,
&
	\ElTu{U} \ElTu{K} &= \sigma_{1k} U_1 K.
\end{align}
Consequently
\begin{equation}\label{eqn:ET_variables}
\begin{aligned}
	\ElTu{U} 			&= U,
&
	\ElTu{\Phi} 		&= \sigma_{1\phi} \Phi,
&
	\ElTu{K} 			&= \sigma_{1k} K,
\\
	\ElTu{\xi_u}		&= \xi_1,
&
	\ElTu{\xi_\phi}		&= \frac{1}{\sigma_{1\phi}} \xi_\phi,
&
	\ElTu{\xi_k}		&= \frac{1}{\sigma_{1k}} \xi_k.
\end{aligned}
\end{equation}
Substitution of \cref{eqn:ET_variables} into the definitions in \cref{tab:shape_factors_old} yields the equivalent expressions listed, which can be inverted to obtain
\begin{equation}
\begin{aligned}
	\sigma_{z\phi} 		&= \frac{a_2}{a_1},
&
	\sigma_{11}			&= 1,
&
	\sigma_{111}		&= a_3,
&
	\sigma_{1\phi}		&= \frac{1}{a_1},
\\
	\sigma_{1k}			&= \frac{1}{a_9},
&
	\sigma_{1z\phi}		&= \frac{a_8}{a_1},
&
	\tilde{\sigma}_{1z\phi}	&= \frac{a_7}{a_1},
&
	\varsigma_\phi 		&= \frac{r_0}{a_1}.
\end{aligned}\end{equation}
These can in turn be substituted into \cref{eqn:2Dsys_vol,eqn:2Dsys_part,eqn:2Dsys_mom,eqn:2Dsys_TKE} and taking the steady-state over a flat bed with constant $a_i$ obtain the system from \cite{ar_Parker_1987}.
	\section{The depth-average operator} \label{app:depth_average}

To average the system of equations \cref{eqn:3Dsys_scales} over the depth we introduce the depth-averaging operator, $\spavg{\omitdummy}{}{}$, defined as
\begin{equation} \label{eqn:CHavg_defn}
	\spavg{f}{}{}(x,y,t) \eqdef \frac{1}{h(x,y,t)} \int_{b(x,y)}^{H(x,y,t)} f(x,y,z,t)  \dd{z}.
\end{equation}
Using the Leibniz rule, the depth-average of derivatives transforms as 
\begin{multline} \label{eqn:Davg_deriv}
	h \spavg{\pdv{f_t}{t} + \pdv{f_x}{x} + \pdv{f_y}{y} + \pdv{f_z}{z}}{}{} 
	\\ 
	= \pdv{}{t} \ppar*{ h \spavg{f_t}{}{} } + \pdv{}{x} \ppar*{ h \spavg{f_x}{}{} }  + \pdv{}{y} \ppar*{ h \spavg{f_y}{}{} } + \eval[\Big]{f_x}_{b} \pdv{b}{x} + \eval[\Big]{f_y}_{b} \pdv{b}{y} - \eval[\Big]{f_z}_{b}
	\\ - \eval[\Big]{f_t}_{H} \pdv{H}{t} - \eval[\Big]{f_x}_{H} \pdv{H}{x} - \eval[\Big]{f_y}_{H} \pdv{H}{y} + \eval[\Big]{f_z}_{H}.
\end{multline}
 To depth integrate an equation we apply the operator $h \spavg{\omitdummy}{}{}$ using \cref{eqn:Davg_deriv} to the system \cref{eqn:3Dsys_scales} and apply the boundary conditions \cref{eqn:3Dsys_BC}. We substitute for the depth average variables as
\begin{align}
	\Phi &= \spavg{\reavg{\phi}}{}{},
&
	U_\alpha &= \spavg{\reavg{u_\alpha}}{}{},
&
	K &= \spavg{k \vphantom{\reavg{v}}}{}{}.
\end{align}

	\section{Depth-rescaling symmetry group} \label{app:depth_group}

For any $c(x,y,t)>0$, equations 
\cref{eqn:2Dsys_vol} to \cref{eqn:2Dsys_egy} 
are invariant under
\begin{subequations}	\label{eqn:deph_symmetry}
\begin{align}
	h 				&\mapsto c 				h,				&
	\Phi 			&\mapsto \frac{1}{c} 	\Phi,			&
	U_\alpha 		&\mapsto \frac{1}{c} 	U_\alpha,		&
	K 				&\mapsto \frac{1}{c} 	K,				\\
	\zeta			&\mapsto \frac{1}{c}	\zeta,			&
	\xi_\phi		&\mapsto c 				\xi_\phi, 		&
	\xi_\alpha		&\mapsto c 				\xi_\alpha,		&
	\xi_k			&\mapsto c 				\xi_k,
\end{align}\end{subequations}
with $b$ and $H$ unchanging, which results in the changes to the shape-factors 
\begin{subequations}
\begin{gather}
	\begin{aligned}
		\varsigma_h 					& \mapsto c^{-1} \varsigma_h,					&
		\sigma_{z\phi}					& \mapsto 	c^{-1} \sigma_{z\phi}, 				&
		\sigma_{\alpha\beta}			& \mapsto 	c \sigma_{\alpha\beta},				
	\end{aligned}
\\
	\begin{aligned}
		\sigma_{\alpha\beta\gamma}		& \mapsto 	c^2 \sigma_{\alpha\beta\gamma},		&
		\sigma_{\alpha\phi}				& \mapsto 	c \sigma_{\alpha\phi},				&
		\sigma_{\alpha k}				& \mapsto 	c \sigma_{\alpha k},				
	\end{aligned}
\\
	\begin{aligned}
		\sigma_{\alpha z\phi}			& \mapsto 	\sigma_{\alpha z\phi},				&
		\tilde{\sigma}_{\alpha z\phi}	& \mapsto 	\tilde{\sigma}_{\alpha z\phi},		&
		\varsigma_{\alpha z\phi}		& \mapsto 	\varsigma_{\alpha z\phi},			&
		\tilde{\sigma}_{D\alpha z\phi}	& \mapsto 	\tilde{\sigma}_{D\alpha z\phi},	
	\end{aligned}
\\
	\varsigma_{\phi}				  \mapsto	c \varsigma_\phi.
\end{gather}\end{subequations}
This symmetry can be used to transform between different measures of depth.

\end{myappendices}

\bibliographystyle{jfm}
\bibliography{Bibliography/Books,Bibliography/Mine,Bibliography/ShallowWater,Bibliography/GravityCurrents,Bibliography/ChannelHydraulics,Bibliography/TurbidityCurrents,Bibliography/SalinityCurrents,Bibliography/StratifiedTurbulence,Bibliography/SlowManifoldTheory,Bibliography/Sedimentology,Bibliography/AnalysisHyperbolic,Bibliography/WallTurbulence,Bibliography/SedimentTransport,Bibliography/CarbonCaptureStorage,Bibliography/Lakes}

\begin{thebibliography}{107}
\expandafter\ifx\csname natexlab\endcsname\relax\def\natexlab#1{#1}\fi
\def\au#1{#1} \def\ed#1{#1} \def\yr#1{#1}\def\at#1{#1}\def\jt#1{\textit{#1}}
  \def\bt#1{#1}\def\bvol#1{\textbf{#1}} \def\vol#1{#1} \def\pg#1{#1}
  \def\publ#1{#1}\def\arxiv#1{#1}\def\org#1{#1}\def\st#1{\textit{#1}}

\bibitem[Abad {\em et~al.\/}(2011)Abad, Sequeiros, Spinewine, Pirmez, Garcia \&
  Parker]{ar_Abad_2011}
{\sc \au{Abad, J.D.}, \au{Sequeiros, O.E.}, \au{Spinewine, B.}, \au{Pirmez,
  C.}, \au{Garcia, M.H.} \& \au{Parker, G.}} \yr{2011}  \at{Secondary current
  of saline underflow in a highly meandering channel: Experiments and theory}.
  \jt{Journal of Sedimentary Research}  \bvol{81}~(11),  \pg{787--813}.

\bibitem[Altinakar(1988)]{phd_Altinakar_1988}
{\sc \au{Altinakar, M.~S.}} \yr{1988}  \at{Weakly depositing turbidity currents
  on small slopes}. PhD thesis, {\'E}cole Polytechnique F{\'e}d{\'e}rale,
  Lausanne Switzerland.

\bibitem[Anjum {\em et~al.\/}(2013)Anjum, Mcelwaine \&
  Caulfield]{ar_Anjum_2013}
{\sc \au{Anjum, H.~J.}, \au{Mcelwaine, J.~N.} \& \au{Caulfield, C.~P.}}
  \yr{2013}  \at{The instantaneous froude number and depth of unsteady gravity
  currents}.  \jt{Journal of Hydraulic Research}  \bvol{51}~(4),
  \pg{432--445}.

\bibitem[Arneborg {\em et~al.\/}(2007)Arneborg, Fiekas, Umlauf \&
  Burchard]{ar_Arneborg_2007}
{\sc \au{Arneborg, L.}, \au{Fiekas, V.}, \au{Umlauf, L.} \& \au{Burchard, H.}}
  \yr{2007}  \at{Gravity current dynamics and entrainment—a process study
  based on observations in the arkona basin}.  \jt{Journal of Physical
  Oceanography}  \bvol{37}~(8),  \pg{2094–2113}.

\bibitem[Azpiroz-Zabala {\em et~al.\/}(2017)Azpiroz-Zabala, Cartigny, Talling,
  Parsons, Sumner, Clare, Simmons, Cooper \& Pope]{ar_AzpirozZabala_2017}
{\sc \au{Azpiroz-Zabala, M.}, \au{Cartigny, M. J.~B.}, \au{Talling, P.~J.},
  \au{Parsons, D.~R.}, \au{Sumner, E.~J.}, \au{Clare, M.~A.}, \au{Simmons,
  S.~M.}, \au{Cooper, C.} \& \au{Pope, E.~L.}} \yr{2017}  \at{Newly recognized
  turbidity current structure can explain prolonged flushing of submarine
  canyons}.  \jt{Science Advances}  \bvol{3}~(10),  \pg{e1700200}.

\bibitem[Azpiroz-Zabala {\em et~al.\/}(2024)Azpiroz-Zabala, Sumner, Cartigny,
  Peakall, Clare, Darby, Parsons, Dorrell, Özsoy, Tezcan, Wynn \&
  Johnson]{ar_AzpirozZabala_2024}
{\sc \au{Azpiroz-Zabala, M.}, \au{Sumner, E.~J.}, \au{Cartigny, M. J.~B.},
  \au{Peakall, J.}, \au{Clare, M.~A.}, \au{Darby, S.~E.}, \au{Parsons, D.~R.},
  \au{Dorrell, R.~M.}, \au{Özsoy, E.}, \au{Tezcan, D.}, \au{Wynn, R.~B.} \&
  \au{Johnson, J.}} \yr{2024}  \at{Benthic biology influences sedimentation in
  submarine channel bends: Coupling of biology, sedimentation and flow}.
  \jt{The Depositional Record}  \bvol{10}~(1),  \pg{159--175}.

\bibitem[Bagnold(1962)]{ar_Bagnold_1962}
{\sc \au{Bagnold, R.~A.}} \yr{1962}  \at{Auto-suspension of transported
  sediment; turbidity currents}.  \jt{Proceedings of the Royal Society, Series
  A, Mathematical and Physical Sciences}  \bvol{265}~(1322),  \pg{315--319}.

\bibitem[Bagnold(1966)]{ar_Bagnold_1966}
{\sc \au{Bagnold, R.~A.}} \yr{1966}  \at{An approach to the sediment transport
  problem from general physics}.  \jt{Geoloical Survey Professional Paper}
  \bvol{422}~(1),  \pg{I1--I37}.

\bibitem[Benjamin(1968)]{ar_Benjamin_1968}
{\sc \au{Benjamin, T.~B.}} \yr{1968}  \at{Gravity currents and related
  phenomena}.  \jt{Journal of Fluid Mechanics}  \bvol{31}~(2),  \pg{209--248}.

\bibitem[Bilgin \& Cantwell(2023)]{ar_Bilgin_2023}
{\sc \au{Bilgin, E.} \& \au{Cantwell, B~J.}} \yr{2023}  \at{Application of the
  universal velocity profile to rough-wall pipe flow}.  \jt{Physics of Fluids}
  \bvol{35},  \pg{055135}.

\bibitem[Bolla~Pittaluga {\em et~al.\/}(2018)Bolla~Pittaluga, Frascati \&
  Falivene]{ar_Pittaluga_2018}
{\sc \au{Bolla~Pittaluga, M.}, \au{Frascati, A.} \& \au{Falivene, O.}}
  \yr{2018}  \at{A gradually varied approach to model turbidity currents in
  submarine channels}.  \jt{Journal of Geophysical Research: Earth Surface}
  \bvol{123}~(1),  \pg{80--96}.

\bibitem[Bonnecaze {\em et~al.\/}(1993)Bonnecaze, Huppert \&
  Lister]{ar_Bonnecaze_1993}
{\sc \au{Bonnecaze, R.~T.}, \au{Huppert, H.~E.} \& \au{Lister, J.~R.}}
  \yr{1993}  \at{Particle-driven gravity currents}.  \jt{Journal of Fluid
  Mechanics}  \bvol{250},  \pg{339--369}.

\bibitem[Breard \& Lube(2017)]{ar_Breard_2017}
{\sc \au{Breard, E. C.~P.} \& \au{Lube, G.}} \yr{2017}  \at{Inside pyroclastic
  density currents – uncovering the enigmatic flow structure and transport
  behaviour in large-scale experiments}.  \jt{Earth and Planetary Science
  Letters}  \bvol{458},  \pg{22--36}.

\bibitem[Brosch \& Lube(2020)]{ar_Brosch_2020}
{\sc \au{Brosch, E.} \& \au{Lube, G.}} \yr{2020}  \at{Spatiotemporal sediment
  transport and deposition processes in experimental dilute pyroclastic density
  currents}.  \jt{Journal of Volcanology and Geothermal Research}  \bvol{401},
  \pg{106946}.

\bibitem[Cantero {\em et~al.\/}(2012)Cantero, Shringarpure \&
  Balachandar]{ar_Cantero_2012}
{\sc \au{Cantero, M.~I.}, \au{Shringarpure, M.} \& \au{Balachandar, S.}}
  \yr{2012}  \at{Towards a universal criteria for turbulence suppression in
  dilute turbidity currents with non-cohesive sediments}.  \jt{Geophysical
  Research Letters}  \bvol{39},  \pg{L14603}.

\bibitem[Carter {\em et~al.\/}(2015)Carter, Gavey, Talling \&
  Liu]{ar_Carter_2015}
{\sc \au{Carter, L.}, \au{Gavey, R.}, \au{Talling, P.~J.} \& \au{Liu, J.~T.}}
  \yr{2015}  \at{Insights into submarine geohazards from breaks in subsea
  telecommunication cables}.  \jt{Oceanography}  \bvol{27}~(2),  \pg{58--67}.

\bibitem[Caulfield(2021)]{ar_Caulfield_2021}
{\sc \au{Caulfield, C.~P.}} \yr{2021}  \at{Layering, instabilities, and mixing
  in turbulent stratified flows}.  \jt{Annual Review of Fluid Mechanics}
  \bvol{53},  \pg{113--45}.

\bibitem[Cenedese \& Adduce(2010)]{ar_Cenedese_2010}
{\sc \au{Cenedese, C.} \& \au{Adduce, C.}} \yr{2010}  \at{A new
  parameterization for entrainment in overflows}.  \jt{Journal of Physical
  Oceanography}  \bvol{40}~(8),  \pg{1835–1850}.

\bibitem[Curray {\em et~al.\/}(2002)Curray, Emmel \& Moore]{ar_Curray_2002}
{\sc \au{Curray, J.~R.}, \au{Emmel, F.~J.} \& \au{Moore, D.~G.}} \yr{2002}
  \at{The bengal fan: morphology, geometry, stratigraphy, history and
  processes}.  \jt{Marine and Petroleum Geology}  \bvol{19}~(10),
  \pg{1191--1223}.

\bibitem[Dorrell \& Hogg(2010)]{ar_Dorrell_2010}
{\sc \au{Dorrell, R.} \& \au{Hogg, A.~J.}} \yr{2010}  \at{Sedimentation of
  bidisperse suspensions}.  \jt{International Journal of Multiphase Flow}
  \bvol{36}~(6),  \pg{481--490}.

\bibitem[Dorrell {\em et~al.\/}(2018)Dorrell, Amy, Peakall \&
  McCaffrey]{ar_Dorrell_2018}
{\sc \au{Dorrell, R.~M.}, \au{Amy, L.~A.}, \au{Peakall, J.} \& \au{McCaffrey,
  W.~D.}} \yr{2018}  \at{Particle size distribution controls the threshold
  between net sediment erosion and deposition in suspended load dominated
  flows}.  \jt{Geophysical Research Letters}  \bvol{45}~(3),  \pg{1443--1452}.

\bibitem[Dorrell {\em et~al.\/}(2014)Dorrell, Darby, Peakall, Sumner, Parsons,
  \& Wynn]{ar_Dorrell_2014}
{\sc \au{Dorrell, R.~M.}, \au{Darby, S.~E.}, \au{Peakall, J.}, \au{Sumner,
  E.~J.}, \au{Parsons, D.~R.},  \& \au{Wynn, R.~B.}} \yr{2014}  \at{The
  critical role of stratification in submarine channels: implications for
  channelization and long runout of flows}.  \jt{Journal of Geophysical
  Research: Oceans}  \bvol{119}.

\bibitem[Dorrell {\em et~al.\/}(2013)Dorrell, Hogg \&
  Pritchard]{ar_Dorrell_2013}
{\sc \au{Dorrell, R.~M.}, \au{Hogg, A.~J.} \& \au{Pritchard, D.}} \yr{2013}
  \at{Polydisperse suspensions: Erosion, deposition, and flow capacity}.
  \jt{Journal of Geophysical Research: Earth Surface}  \bvol{118}.

\bibitem[Dorrell {\em et~al.\/}(2019)Dorrell, Peakall, Darby, Parsons, Johnson,
  Sumner, Wynn, Özsoy \& Tezcan]{ar_Dorrell_2019}
{\sc \au{Dorrell, R.~M.}, \au{Peakall, J.}, \au{Darby, S.~E.}, \au{Parsons,
  D.~R.}, \au{Johnson, J.}, \au{Sumner, E.~J.}, \au{Wynn, R.~B.}, \au{Özsoy,
  E.} \& \au{Tezcan, D.}} \yr{2019}  \at{Self-sharpening induces jet-like
  structure in seafloor gravity currents}.  \jt{Nature Communications}
  \bvol{10},  \pg{1381}.

\bibitem[Drew \& Passman(1999)]{bk_Drew_TMF}
{\sc \au{Drew, D.~A.} \& \au{Passman, S.~L.}} \yr{1999} {\em Theory of
  Multicomponent Fluids\/}, 1st edn. {\em AMS\/} 135.  \publ{Springer}.

\bibitem[Eggenhuisen {\em et~al.\/}(2017)Eggenhuisen, Cartigny \&
  de~Leeuw]{ar_Eggenhuisen_2017}
{\sc \au{Eggenhuisen, J.~T.}, \au{Cartigny, M. J.~B.} \& \au{de~Leeuw, J.}}
  \yr{2017}  \at{Physical theory for near-bed turbulent particle suspension
  capacity}.  \jt{Earth Surface Dynamics}  \bvol{5},  \pg{269--281}.

\bibitem[Eggenhuisen {\em et~al.\/}(2020)Eggenhuisen, Tilston, Leeuw, Pohl \&
  Cartigny]{ar_Eggenhuisen_2020}
{\sc \au{Eggenhuisen, J.~T.}, \au{Tilston, M.~C.}, \au{Leeuw, J.}, \au{Pohl,
  F.} \& \au{Cartigny, M.~J.}} \yr{2020}  \at{Turbulent diffusion modelling of
  sediment in turbidity currents: An experimental validation of the rouse
  approach}.  \jt{The Depositional Record}  \bvol{6}~(1),  \pg{203--216}.

\bibitem[Ellison \& Turner(1959)]{ar_Ellison_1959}
{\sc \au{Ellison, T.~H.} \& \au{Turner, J.~S.}} \yr{1959}  \at{Turbulent
  entrainment in stratified flows}.  \jt{Journal of Fluid Mechanics}
  \bvol{6}~(3),  \pg{423--448}.

\bibitem[Farizan {\em et~al.\/}(2019)Farizan, Yaghoubi, Firoozabadi \&
  Afshin]{ar_Farizan_2019}
{\sc \au{Farizan, A.}, \au{Yaghoubi, S.}, \au{Firoozabadi, B.} \& \au{Afshin,
  H.}} \yr{2019}  \at{Effect of an obstacle on the depositional behaviour of
  turbidity currents}.  \jt{Journal of Hydraulic Research}  \bvol{57}~(1),
  \pg{75--89}.

\bibitem[Forooghi {\em et~al.\/}(2018)Forooghi, Stroh, Schlatter \&
  Frohnapfel]{ar_Forooghi_2018}
{\sc \au{Forooghi, P.}, \au{Stroh, A.}, \au{Schlatter, P.} \& \au{Frohnapfel,
  B.}} \yr{2018}  \at{Direct numerical simulation of flow over dissimilar,
  randomly distributed roughness elements: A systematic study on the effect of
  surface morphology on turbulence}.  \jt{Physical Review Fluids}  \bvol{3},
  \pg{044605}.

\bibitem[Fukuda {\em et~al.\/}(2023)Fukuda, de~Vet, Skevington, Bastianon,
  Fernández, Wu, McCaffrey, Naruse, Parsons \&
  Dorrell]{ar_Skevington_F006_TCFlowPower}
{\sc \au{Fukuda, S.}, \au{de~Vet, M.}, \au{Skevington, E.}, \au{Bastianon, E.},
  \au{Fernández, R.}, \au{Wu, X.}, \au{McCaffrey, W.}, \au{Naruse, H.},
  \au{Parsons, D.} \& \au{Dorrell, R.}} \yr{2023}  \at{Inadequacy of fluvial
  energetics for describing gravity current autosuspension}.  \jt{Nature
  Communications}  \bvol{14},  \pg{2288}.

\bibitem[Garcia(2008)]{ibk_Garcia_2008}
{\sc \au{Garcia, M.~H.}} \yr{2008} {\em Sedimentation Engineering: Processes,
  Measurements, Modeling and Practice\/}, chap. Seiment transport and
  morphodynamics,  \pg{pp. 21--146}.  \publ{ASCE}.

\bibitem[García(1993)]{ar_Garcia_1993}
{\sc \au{García, M.~H.}} \yr{1993}  \at{Hydraulic jumps in sediment‐driven
  bottom currents}.  \jt{Journal of Hydraulic Engineering}  \bvol{119}~(10),
  \pg{1094--1117}.

\bibitem[Guo(2020)]{ar_Guo_2020}
{\sc \au{Guo, J.}} \yr{2020}  \at{Empirical model for shields diagram and its
  applications}.  \jt{Journal of Hydraulic Engineering}  \bvol{146}~(6),
  \pg{04020038}.

\bibitem[Hogg \& Skevington(2021)]{ar_Skevington_F002_SemiInf_Wall}
{\sc \au{Hogg, A.~J.} \& \au{Skevington, E. W.~G.}} \yr{2021}  \at{Dam-break
  reflection}.  \jt{Quarterly Journal of Mechanics and Applied Mathematics}
  \bvol{74},  \pg{441--465}.

\bibitem[Hoult(1972)]{ar_Hoult_1972}
{\sc \au{Hoult, D.~P.}} \yr{1972}  \at{Oil spreading on the sea}.  \jt{Annual
  Review of Fluid Mechanics}  \bvol{4},  \pg{341--368}.

\bibitem[Hsu {\em et~al.\/}(2008)Hsu, Kuo, Lo, Tsai, Doo, Ku \&
  Sibuet]{ar_Hsu_2008}
{\sc \au{Hsu, S.~K.}, \au{Kuo, J.}, \au{Lo, C.~L.}, \au{Tsai, C.~H.}, \au{Doo,
  W.~B.}, \au{Ku, C.~Y.} \& \au{Sibuet, J.~C.}} \yr{2008}  \at{Turbidity
  currents, submarine landslides and the 2006 pingtung earthquake off sw
  taiwan}.  \jt{Terrestrial, Atmospheric and Ocenanic sciences journal}
  \bvol{19},  \pg{767--772}.

\bibitem[Huppert(2006)]{ar_Huppert_2006}
{\sc \au{Huppert, H.~E.}} \yr{2006}  \at{Gravity currents: a personal
  perspective}.  \jt{Journal of Fluid Mechanics}  \bvol{554},  \pg{299--322}.

\bibitem[Islam \& Imran(2010)]{ar_Islam_2010}
{\sc \au{Islam, M.~A.} \& \au{Imran, J.}} \yr{2010}  \at{Vertical structure of
  continuous release saline and turbidity currents}.  \jt{Journal of
  Geophysical Research: Oceans}  \bvol{115}~(C8).

\bibitem[Jansen {\em et~al.\/}(2021)Jansen, MacIntyre, Barrett, Chin, Cortés,
  Forrest, Hrycik, Martin, McMeans, Rautio \& Schwefel]{ar_Jansen_2021}
{\sc \au{Jansen, J.}, \au{MacIntyre, S.}, \au{Barrett, D.~C.}, \au{Chin,
  Y.-P.}, \au{Cortés, A.}, \au{Forrest, A.~L.}, \au{Hrycik, A.~R.},
  \au{Martin, R.}, \au{McMeans, B.~C.}, \au{Rautio, M.} \& \au{Schwefel, R.}}
  \yr{2021}  \at{Winter limnology: How do hydrodynamics and biogeochemistry
  shape ecosystems under ice?}  \jt{JGR Biogeosciences}  \bvol{126}~(6),
  \pg{e2020JG006237}.

\bibitem[Kadivar {\em et~al.\/}(2021)Kadivar, Tormey \&
  McGranaghan]{ar_Kadivar_2021}
{\sc \au{Kadivar, M.}, \au{Tormey, D.} \& \au{McGranaghan, G.}} \yr{2021}
  \at{A review on turbulent flow over rough surfaces: Fundamentals and
  theories}.  \jt{International Journal of Thermofluids}  \bvol{10},
  \pg{100077}.

\bibitem[Kaneda \& Yamamoto(2021)]{ar_Kaneda_2021}
{\sc \au{Kaneda, Y.} \& \au{Yamamoto, Y.}} \yr{2021}  \at{Velocity gradient
  statistics in turbulent shear flow: an extension of kolmogorov's local
  equilibrium theory}.  \jt{Journal of Fluid Dynamics}  \bvol{A13},  \pg{929}.

\bibitem[Knapp(1938)]{ar_Knapp_1938}
{\sc \au{Knapp, R.~T.}} \yr{1938}  \at{Energy-balance in stream-flows carrying
  suspended load}.  \jt{EoS Transactions of the American Geophysical Union}
  \bvol{19}~(1),  \pg{501--505}.

\bibitem[Koller {\em et~al.\/}(2022)Koller, Manica \& Fedele]{ar_Koller_2022}
{\sc \au{Koller, D.~K.}, \au{Manica, R.} \& \au{Fedele, J.~J.}} \yr{2022}
  \at{Comparative hydraulic and sedimentologic study of ripple formation using
  experimental turbidity currents and saline currents}.  \jt{Journal of
  Sedimentary Research}  \bvol{92}~(7),  \pg{601–618}.

\bibitem[Lee \& Moser(2015)]{ar_Lee_2015}
{\sc \au{Lee, M.} \& \au{Moser, R.~D.}} \yr{2015}  \at{Direct numerical
  simulation of turbulent channel flow up to {$\Rey_\tau \approx 5200$}}.
  \jt{Journal of Fluid Dynamics}  \bvol{774},  \pg{395--415}.

\bibitem[de~Leeuw {\em et~al.\/}(2018)de~Leeuw, Eggenhuisen \&
  Cartigny]{ar_Leeuw_2017}
{\sc \au{de~Leeuw, J}, \au{Eggenhuisen, J.~T.} \& \au{Cartigny, M. J.~B.}}
  \yr{2018}  \at{Linking submarine channel–levee facies and architecture to
  flow structure of turbidity currents: insights from flume tank experiments}.
  \jt{Sedimentology}  \bvol{65}~(3),  \pg{931--935}.

\bibitem[Lewis(1994)]{ar_Lewis_1994}
{\sc \au{Lewis, K.~B.}} \yr{1994}  \at{The 1500-km-long hikurangi channel:
  trench-axis channel that escapes its trench, crosses a plateau, and feeds a
  fan drift}.  \jt{Geo-Marine Letter}  \bvol{14},  \pg{19--28}.

\bibitem[Liu {\em et~al.\/}(2019)Liu, Godbole, Lu, Michal \&
  Linton]{ar_Liu_2019}
{\sc \au{Liu, X.}, \au{Godbole, A.}, \au{Lu, C.}, \au{Michal, G.} \&
  \au{Linton, V.}} \yr{2019}  \at{Investigation of the consequence of
  high-pressure co2 pipeline failure through experimental and numerical
  studies}.  \jt{Applied Energy}  \bvol{250},  \pg{32--47}.

\bibitem[Ma {\em et~al.\/}(2024)Ma, Parker, Cartigny, Viparelli, Balachander,
  Fu \& Luchi]{ar_Ma_2024}
{\sc \au{Ma, H.}, \au{Parker, G.}, \au{Cartigny, M.}, \au{Viparelli, E.},
  \au{Balachander, S.}, \au{Fu, X.} \& \au{Luchi, R.}} \yr{2024}  \at{Two-layer
  formulation for long-runout turbidity currents: theory and bypass flow case}.
   \jt{Earth ArXiv} .

\bibitem[Maffioli {\em et~al.\/}(2016)Maffioli, Brethouwer \&
  Lindborg]{ar_Maffioli_2016}
{\sc \au{Maffioli, A.}, \au{Brethouwer, G.} \& \au{Lindborg, E.}} \yr{2016}
  \at{Mixing efficiency in stratified turbulence}.  \jt{Journal of Fluid
  Mechanics}  \bvol{798},  \pg{R3}.

\bibitem[Maggi {\em et~al.\/}(2023)Maggi, Negretti, Hopfinger \&
  Adduce]{ar_Maggi_2023}
{\sc \au{Maggi, M.~R.}, \au{Negretti, M.~E.}, \au{Hopfinger, E.~J.} \&
  \au{Adduce, C.}} \yr{2023}  \at{Turbulence characteristics and mixing
  properties of gravity currents over complex topography}.  \jt{Physics of
  Fluids}  \bvol{35},  \pg{016607}.

\bibitem[van Maren {\em et~al.\/}(2009)van Maren, Winterwerp, Wang \&
  Pu]{ar_Maren_2009}
{\sc \au{van Maren, D.~S.}, \au{Winterwerp, J.~C.}, \au{Wang, Z.~Y.} \& \au{Pu,
  Q.}} \yr{2009}  \at{Suspended sediment dynamics and morphodynamics in
  theyellow river, china}.  \jt{Sedimentology}  \bvol{56},  \pg{785–806}.

\bibitem[Martin {\em et~al.\/}(2019)Martin, Negretti \&
  Hopfinger]{ar_Martin_2019}
{\sc \au{Martin, A.}, \au{Negretti, M.~E.} \& \au{Hopfinger, E.~J.}} \yr{2019}
  \at{Development of gravity currents on slopes under different interfacial
  instability conditions}.  \jt{Journal of Fluid Mechanics}  \bvol{880},
  \pg{180 -- 208}.

\bibitem[Mashayek {\em et~al.\/}(2021)Mashayek, Caulfield \&
  Alford]{ar_Mashayek_2021}
{\sc \au{Mashayek, A.}, \au{Caulfield, C.P.} \& \au{Alford, M.H.}} \yr{2021}
  \at{Goldilocks mixing in oceanic shear-induced turbulent overturns}.
  \jt{Journal of Fluid Mechanics}  \bvol{928},  \pg{A1}.

\bibitem[Mazzuoli \& Uhlmann(2017)]{ar_Mazzuoli_2017}
{\sc \au{Mazzuoli, M.} \& \au{Uhlmann, M.}} \yr{2017}  \at{Direct numerical
  simulation of open-channel flow over a fully rough wall at moderate relative
  submergence}.  \jt{Journal of Fluid Dynamics}  \bvol{824},  \pg{722--765}.

\bibitem[Meiburg {\em et~al.\/}(2015)Meiburg, Radhakrishnan \&
  Nasr-Azadani]{ar_Meiburg_2015}
{\sc \au{Meiburg, E.}, \au{Radhakrishnan, S.} \& \au{Nasr-Azadani, M.}}
  \yr{2015}  \at{Modeling gravity and turbidity currents: Computational
  approaches and challenges}.  \jt{Applied Mechanics Reviews}  \bvol{67}~(4),
  \pg{040802}.

\bibitem[Michon {\em et~al.\/}(1955)Michon, Goddet \&
  Bonnefille]{bk_Michon_1955}
{\sc \au{Michon, X.}, \au{Goddet, J.} \& \au{Bonnefille, R.}} \yr{1955} {\em
  Etude Th{\'e}orique et Exp{\'e}rimentale des Courants de Densit{\'e}\/}.
  \publ{Laboratoire national d'hydraulique}.

\bibitem[Momen {\em et~al.\/}(2017)Momen, Aheng, Bou-Zeid \&
  Stone]{ar_Momen_2017}
{\sc \au{Momen, M.}, \au{Aheng, A.}, \au{Bou-Zeid, E.} \& \au{Stone, H.~A.}}
  \yr{2017}  \at{Inertial gravity currents produced by fluid drainage from an
  edge}.  \jt{Journal of Fluid Mechanics}  \bvol{827},  \pg{640--663}.

\bibitem[Negretti {\em et~al.\/}(2017)Negretti, Flòr \&
  Hopfinger]{ar_Negretti_2017}
{\sc \au{Negretti, M.~E.}, \au{Flòr, J.-B.} \& \au{Hopfinger, E.~J.}}
  \yr{2017}  \at{Development of gravity currents on rapidly changing slopes}.
  \jt{Journal of Fluid Mechanics}  \bvol{833},  \pg{70 -- 97}.

\bibitem[Nikora {\em et~al.\/}(2004)Nikora, Koll, McEwan, McLean \&
  Dittrich]{ar_Nikora_2004}
{\sc \au{Nikora, V.}, \au{Koll, K.}, \au{McEwan, I.}, \au{McLean, S.} \&
  \au{Dittrich, A.}} \yr{2004}  \at{Velocity distribution in the roughness
  layer of rough-bed flows}.  \jt{Journal of Hydraulic Engineering}
  \bvol{130}~(10),  \pg{1036--1042}.

\bibitem[Odier {\em et~al.\/}(2014)Odier, Chen \& Ecke]{ar_Odier_2014}
{\sc \au{Odier, P.}, \au{Chen, J.} \& \au{Ecke, R.~E.}} \yr{2014}
  \at{Entrainment and mixing in a laboratory model of oceanic overflow}.
  \jt{Journal of Fluid Mechanics}  \bvol{746},  \pg{498 -- 535}.

\bibitem[Olu {\em et~al.\/}(2017)Olu, Decker, Pastor, Caprais, Khripounoff,
  Morineaux, Baziz, Menot \& Rabouille]{ar_Olu_2017}
{\sc \au{Olu, K.}, \au{Decker, C.}, \au{Pastor, L.}, \au{Caprais, J.-C.},
  \au{Khripounoff, A.}, \au{Morineaux, M.}, \au{Baziz, M.~Ain}, \au{Menot, L.}
  \& \au{Rabouille, C.}} \yr{2017}  \at{Cold-seep-like macrofaunal communities
  in organic- and sulfide-rich sediments of the congo deep-sea fan}.  \jt{Deep
  Sea Research Part II: Topical Studies in Oceanography}  \bvol{142},
  \pg{180--196}.

\bibitem[Orlandi(2019)]{ar_Orlandi_2019}
{\sc \au{Orlandi, P.}} \yr{2019}  \at{Turbulent kinetic energy production and
  flow structures in flows past smooth and rough walls}.  \jt{Journal of Fluid
  Dynamics}  \bvol{866},  \pg{897--928}.

\bibitem[Osborn(1980)]{ar_Osborn_1980}
{\sc \au{Osborn, T.~R.}} \yr{1980}  \at{Estimates of the local rate of vertical
  diffusion from dissipation measurements}.  \jt{Journal of Physical
  Oceanography}  \bvol{10}~(1),  \pg{83–89}.

\bibitem[Packman \& Jerolmack(2004)]{ar_Packman_2004}
{\sc \au{Packman, A.~I.} \& \au{Jerolmack, D.}} \yr{2004}  \at{The role of
  physicochemical processes in controlling sediment transport and deposition in
  turbidity currents}.  \jt{Marine Geology}  \bvol{204}~(1),  \pg{1--9}.

\bibitem[Parker {\em et~al.\/}(1986)Parker, Fukushima \&
  Pantin]{ar_Parker_1986}
{\sc \au{Parker, G.}, \au{Fukushima, Y.} \& \au{Pantin, H.~M.}} \yr{1986}
  \at{Self-accelerating turbidity currents}.  \jt{Journal of Fluid Mechanics}
  \bvol{171},  \pg{145--181}.

\bibitem[Parker {\em et~al.\/}(1987)Parker, Garcia, Fukushima \&
  Yu]{ar_Parker_1987}
{\sc \au{Parker, G.}, \au{Garcia, M.}, \au{Fukushima, Y.} \& \au{Yu, W.}}
  \yr{1987}  \at{Experiments on turbidity currents over an erodible bed}.
  \jt{Journal of Hydraulic Research}  \bvol{25}~(1),  \pg{123--147}.

\bibitem[Pohl {\em et~al.\/}(2020)Pohl, Eggenhuisen, Kane \&
  Clare]{ar_Pohl_2020}
{\sc \au{Pohl, F.}, \au{Eggenhuisen, J.~T.}, \au{Kane, I.~A.} \& \au{Clare,
  M.~A.}} \yr{2020}  \at{Transport and burial of microplastics in deep-marine
  sediments by turbidity currents}.  \jt{Environmental Science and Technology}
  \bvol{54}~(7),  \pg{4180–4189}.

\bibitem[Pope(2000)]{bk_Pope_TF}
{\sc \au{Pope, S.~B.}} \yr{2000} {\em Turbulent Flows\/}.  \publ{Cambridge
  University Press}.

\bibitem[Reece {\em et~al.\/}(2024)Reece, Dorrell \& Straub]{ar_Reece_2024}
{\sc \au{Reece, J.~K.}, \au{Dorrell, R.~M.} \& \au{Straub, K.~M.}} \yr{2024}
  \at{Circulation of hydraulically ponded turbidity currents and the filling of
  continental slope minibasins}.  \jt{Nature Communications}  \bvol{15},
  \pg{2075}.

\bibitem[Reichardt(1951)]{ar_Reichardt_1951}
{\sc \au{Reichardt, H.}} \yr{1951}  \at{Vollständige darstellung der
  turbulenten geschwindigkeitsverteilung in glatten leitungen}.
  \jt{Zeitschrift für angewandte Mathematik und Physik}  \bvol{31}~(7),
  \pg{208--219}.

\bibitem[Rottman {\em et~al.\/}(1985)Rottman, Simpson \& Hunt]{ar_Rottman_1985}
{\sc \au{Rottman, J.~W.}, \au{Simpson, J.~E.} \& \au{Hunt, J. C.~R.}} \yr{1985}
   \at{Unsteady gravity current flows over obstacles: Some observations and
  analysis related to the phase {II} trials}.  \jt{Journal of Hazardous
  Materials}  \bvol{11}~(1-4),  \pg{325--340}.

\bibitem[Saint-Venant(1871)]{ar_SaintVenant_1871}
{\sc \au{Saint-Venant, A. J. C.~de}} \yr{1871}  \at{Th{\'{e}}orie du mouvement
  non permanent des eaux avec applications aux crues des rivi{\`{e}}res et
  {\'{a}} l'introduction des mar{\'{e}}es dans leur lit}.  \jt{Comptes rendus
  de l'Acad{\'{e}}mie des Sciences de Paris}  \bvol{73},  \pg{148--154}.

\bibitem[Salinas {\em et~al.\/}(2020)Salinas, Balachandar, Shringarpure,
  Fedele, Hoyal \& Cantero]{ar_Salinas_2020}
{\sc \au{Salinas, J.}, \au{Balachandar, S.}, \au{Shringarpure, M.}, \au{Fedele,
  J.}, \au{Hoyal, D.} \& \au{Cantero, M.}} \yr{2020}  \at{Soft transition
  between subcritical and supercritical currents through intermittent cascading
  interfacial instabilities}.  \jt{Proceedings of the National Academy of
  Sciences}  \bvol{117}~(31),  \pg{18278--18284}.

\bibitem[Salinas {\em et~al.\/}(2021)Salinas, Balachandar \&
  Cantero]{ar_Salinas_2021}
{\sc \au{Salinas, J.~S.}, \au{Balachandar, S.} \& \au{Cantero, M.~I.}}
  \yr{2021}  \at{Control of turbulent transport in supercritical currents by
  three families of hairpin vortices}.  \jt{Physical Review Fluids}  \bvol{6},
  \pg{063801}.

\bibitem[Salinas {\em et~al.\/}(2023)Salinas, Balachandar, Z{\'u}{\~n}iga,
  Shringarpure, Fedele \& Cantero]{ar_Salinas_2023}
{\sc \au{Salinas, J.~S.}, \au{Balachandar, S.}, \au{Z{\'u}{\~n}iga, S.~L.},
  \au{Shringarpure, M.}, \au{Fedele, J.} \& \au{Cantero, D. Hoyaland M.~I.}}
  \yr{2023}  \at{On the definition, evolution, and properties of the outer edge
  of gravity currents: A direct-numerical and large-eddy simulation study}.
  \jt{Physics of Fluids}  \bvol{35},  \pg{016610}.

\bibitem[Salinas {\em et~al.\/}(2019{\natexlab{{\em a\/}}})Salinas, Cantero,
  Shringarpure \& Balachandar]{ar_Salinas_2019a}
{\sc \au{Salinas, J.~S.}, \au{Cantero, M.~I.}, \au{Shringarpure, M.} \&
  \au{Balachandar, S.}} \yr{2019{\natexlab{{\em a\/}}}}  \at{Propoerties of the
  body of a turbidity current at near-normal conditions: 1. effect of bed
  slope}.  \jt{JGR Oceans}  \bvol{124},  \pg{7989--8016}.

\bibitem[Salinas {\em et~al.\/}(2019{\natexlab{{\em b\/}}})Salinas, Cantero,
  Shringarpure \& Balachandar]{ar_Salinas_2019b}
{\sc \au{Salinas, J.~S.}, \au{Cantero, M.~I.}, \au{Shringarpure, M.} \&
  \au{Balachandar, S.}} \yr{2019{\natexlab{{\em b\/}}}}  \at{Propoerties of the
  body of a turbidity current at near-normal conditions: 2. effect of
  settling}.  \jt{JGR Oceans}  \bvol{124},  \pg{7989--8016}.

\bibitem[Salinas {\em et~al.\/}(2022)Salinas, Z{\'u}{\~n}iga, Cantero,
  Shringarpure, Fedele, Hoyal \& Balachandar]{ar_Salinas_2022}
{\sc \au{Salinas, J.~S.}, \au{Z{\'u}{\~n}iga, S.}, \au{Cantero, M.~I.},
  \au{Shringarpure, M.}, \au{Fedele, J.}, \au{Hoyal, D.} \& \au{Balachandar,
  S.}} \yr{2022}  \at{Slope dependence of self-similar structure and
  entrainment in gravity currents}.  \jt{JGR Oceans}  \bvol{934},  \pg{R4}.

\bibitem[Savoye {\em et~al.\/}(2009)Savoye, Babonneau, Dennielou \&
  Bez]{ar_Savoye_2009}
{\sc \au{Savoye, B.}, \au{Babonneau, N.}, \au{Dennielou, B.} \& \au{Bez, M.}}
  \yr{2009}  \at{Geological overview of the angola–congo margin, the congo
  deep-sea fan and its submarine valleys}.  \jt{Deep-Sea Research II: Topical
  Studies in Oceanography}  \bvol{56},  \pg{2169–2182}.

\bibitem[Schlichting \& Gersten(2016)]{bk_Schlichting_BLT}
{\sc \au{Schlichting, H.} \& \au{Gersten, K.}} \yr{2016} {\em Boundary-Layer
  Theory\/}, 9th edn.  \publ{Springer}.

\bibitem[Sen {\em et~al.\/}(2017)Sen, Dennielou, Tourolle, Arnaubec, Rabouille
  \& Olu]{ar_Sen_2017}
{\sc \au{Sen, A.}, \au{Dennielou, B.}, \au{Tourolle, J.}, \au{Arnaubec, A.},
  \au{Rabouille, C.} \& \au{Olu, K.}} \yr{2017}  \at{Fauna and habitat types
  driven by turbidity currents in the lobe complex of the congo deep-sea fan}.
  \jt{Deep Sea Research Part II: Topical Studies in Oceanography}  \bvol{142},
  \pg{167--179}.

\bibitem[Sequeiros {\em et~al.\/}(2018)Sequeiros, Mosquera \&
  Pedocchi]{ar_Sequerios_2018}
{\sc \au{Sequeiros, O.~E.}, \au{Mosquera, R.} \& \au{Pedocchi, F.}} \yr{2018}
  \at{Internal structure of a self-accelerating turbidity current}.  \jt{JGR
  Oceans}  \bvol{123}~(9),  \pg{6260--6276}.

\bibitem[Sher \& Woods(2015)]{ar_Sher_2015}
{\sc \au{Sher, D.} \& \au{Woods, A.W.}} \yr{2015}  \at{Gravity currents:
  entrainment, stratification and self-similarity}.  \jt{Journal of Fluid
  Mechanics}  \bvol{784},  \pg{130--162}.

\bibitem[Shringarpure {\em et~al.\/}(2012)Shringarpure, Cantero \&
  Balachandar]{ar_Shringarpure_2012}
{\sc \au{Shringarpure, M.}, \au{Cantero, M.~I.} \& \au{Balachandar, S.}}
  \yr{2012}  \at{Dynamics of complete turbulence suppression in turbidity
  currents driven by monodisperse suspensions of sediment}.  \jt{Journal of
  Fluid Mechnaics}  \bvol{712},  \pg{384--417}.

\bibitem[Simmons {\em et~al.\/}(2020)Simmons, Azpiroz-Zabala, Cartigny, Clare,
  Cooper, Parsons, Pope, Sumner \& Talling]{ar_Simmons_2020}
{\sc \au{Simmons, S.~M.}, \au{Azpiroz-Zabala, M.}, \au{Cartigny, M. J~.B.},
  \au{Clare, M.~A.}, \au{Cooper, C.}, \au{Parsons, D.~R.}, \au{Pope, E.~L.},
  \au{Sumner, E.~J.} \& \au{Talling, P.~J.}} \yr{2020}  \at{Novel acoustic
  method provides first detailed measurements of sediment concentration
  structure within submarine turbidity currents}.  \jt{Journal of Geophysical
  Research: Oceans}  \bvol{125}~(5),  \pg{e2019JC015904}.

\bibitem[Simpson(1982)]{ar_Simpson_1982}
{\sc \au{Simpson, J.~E.}} \yr{1982}  \at{Gravity currents in the laboratory,
  atmosphere, and ocean}.  \jt{Annual Review of Fluid Mechanics}  \bvol{14},
  \pg{213--234}.

\bibitem[Simpson(1997)]{bk_Simpson_GCEL}
{\sc \au{Simpson, J.~E.}} \yr{1997} {\em Gravity Currents in the Environment
  and the Laboratory\/}, 2nd edn.  \publ{Cambridge University Press}.

\bibitem[Skevington \& Hogg(2020)]{ar_Skevington_F001_Draining}
{\sc \au{Skevington, E. W.~G.} \& \au{Hogg, A.~J.}} \yr{2020}  \at{Unsteady
  draining of reservoirs over weirs and through constrictions}.  \jt{Journal of
  Fluid Mechanics}  \bvol{882}~(A9).

\bibitem[Skevington \& Hogg(2023)]{ar_Skevington_F003_CritWall_Overtopping}
{\sc \au{Skevington, E. W.~G.} \& \au{Hogg, A.~J.}} \yr{2023}  \at{The unsteady
  overtopping of barriers by gravity currents and dam-break flows}.
  \jt{Journal of Fluid Mechanics}  \bvol{960},  \pg{A27}.

\bibitem[Skevington \& Hogg(2024)]{ar_Skevington_F004_Backflow}
{\sc \au{Skevington, E. W.~G.} \& \au{Hogg, A.~J.}} \yr{2024}  \at{Gravity
  current escape from a topographic depression}.  \jt{Physical Review Fluids}
  \bvol{9},  \pg{014802}.

\bibitem[Skevington {\em et~al.\/}(2021)Skevington, Hogg \&
  Ungarish]{ar_Skevington_F005_RadSuper}
{\sc \au{Skevington, E. W.~G.}, \au{Hogg, A.~J.} \& \au{Ungarish, M.}}
  \yr{2021}  \at{Development of supercritical motion and internal jumps within
  lock-release radial currents and draining flows}.  \jt{Physical Review
  Fluids}  \bvol{6}~(6),  \pg{063803}.

\bibitem[Stoker(1957)]{bk_Stoker_WW}
{\sc \au{Stoker, J.~J.}} \yr{1957} {\em Water Waves, the Mathematical Theory
  with Applications\/}. {\em Pure and Applied Mathematics, a Series of Texts
  and Monographs\/} 4.  \publ{Interscience Publishers}.

\bibitem[Strang \& Fernando(2001)]{ar_Strang_2001}
{\sc \au{Strang, E.~J.} \& \au{Fernando, H. J.~S.}} \yr{2001}  \at{Entrainment
  and mixing in stratified shear flows}.  \jt{Journal of Fluid Mechanics}
  \bvol{428},  \pg{349--386}.

\bibitem[Talling {\em et~al.\/}(2023)Talling, Cartigny, Pope, Baker, Clare,
  Heijnen, Hage, Parsons, Simmons, Paull, Gwiazda, Lintern, Hughes~Clarke, Xu,
  Silva~Jacinto \& Maier]{ar_Talling_2023}
{\sc \au{Talling, P.~J.}, \au{Cartigny, M. J.~B.}, \au{Pope, E.}, \au{Baker,
  M.}, \au{Clare, M.~A.}, \au{Heijnen, M.}, \au{Hage, S.}, \au{Parsons, D.~R.},
  \au{Simmons, S.~M.}, \au{Paull, C.~K.}, \au{Gwiazda, R.}, \au{Lintern, G.},
  \au{Hughes~Clarke, J.~E.}, \au{Xu, J.}, \au{Silva~Jacinto, R.} \& \au{Maier,
  K.~L.}} \yr{2023}  \at{Detailed monitoring reveals the nature of submarine
  turbidity currents}.  \jt{Nature Reviews Earth \& Environment}  \bvol{4},
  \pg{642–658}.

\bibitem[Talling {\em et~al.\/}(2007)Talling, Wynn, Masson, Frenz, Cronin,
  Schiebel, Akhmetzhanov, Dallmeier-Tiessen, Benetti, Weaver, Georgiopoulou,
  Z{\"u}hlsdorff4 \& Amy]{ar_Talling_2007}
{\sc \au{Talling, P.~J.}, \au{Wynn, R.~B.}, \au{Masson, D.~G.}, \au{Frenz, M.},
  \au{Cronin, B.~T.}, \au{Schiebel, R.}, \au{Akhmetzhanov, A.~M.},
  \au{Dallmeier-Tiessen, S.}, \au{Benetti, S.}, \au{Weaver, P. P.~E.},
  \au{Georgiopoulou, A.}, \au{Z{\"u}hlsdorff4, C.} \& \au{Amy, L.~A.}}
  \yr{2007}  \at{Onset of submarine debris flow deposition far from original
  giant landslide}.  \jt{Nature}  \bvol{450},  \pg{541--544}.

\bibitem[Tesaker(1969)]{phd_Tesaker_1969}
{\sc \au{Tesaker, E.}} \yr{1969}  \at{Uniform turbidity current experiments}.
  PhD thesis, The Technical University of Norway.

\bibitem[Thakkar {\em et~al.\/}(2017)Thakkar, Busseb \&
  Sandhama]{ar_Thakkar_2017}
{\sc \au{Thakkar, M.}, \au{Busseb, A.} \& \au{Sandhama, N.}} \yr{2017}
  \at{Surface correlations of hydrodynamic drag for transitionally rough
  engineering surfaces}.  \jt{Journal of Turbulence}  \bvol{18}~(2),
  \pg{138--169}.

\bibitem[Toniolo {\em et~al.\/}(2006)Toniolo, Parker, Voller \&
  Beaubouef]{ar_Toniolo_2006}
{\sc \au{Toniolo, H.}, \au{Parker, G.}, \au{Voller, V.} \& \au{Beaubouef,
  R.T.}} \yr{2006}  \at{Depositional turbidity currents in diapiric minibasins
  on the continental slope: Experiments—numerical simulation and upscaling}.
  \jt{Journal of Sedimentary Research}  \bvol{76}~(5),  \pg{798–818}.

\bibitem[Ungarish(2020)]{bk_Ungarish_GCIv2}
{\sc \au{Ungarish, M.}} \yr{2020} {\em Gravity Currents and Intrusions,
  Analysis and Prediction\/}. {\em Environmental Fluid Mechanics\/} 1.
  \publ{World Scientific}.

\bibitem[Ungarish \& Hogg(2018)]{ar_Ungarish_2018}
{\sc \au{Ungarish, M.} \& \au{Hogg, A.~J.}} \yr{2018}  \at{Models of internal
  jumps and the fronts of gravity currents: unifying two-layer theories and
  deriving new results}.  \jt{Journal of Fluid Mechanics}  \bvol{846},
  \pg{654--685}.

\bibitem[Ungarish {\em et~al.\/}(2019)Ungarish, Zhu \& Stone]{ar_Ungarish_2019}
{\sc \au{Ungarish, M.}, \au{Zhu, L.} \& \au{Stone, H.~A.}} \yr{2019}
  \at{Inertial gravity current produced by the drainage of a cylindrical
  reservoir from an outer or inner edge}.  \jt{Journal of Fluid Mechanics}
  \bvol{874},  \pg{185--209}.

\bibitem[Varjavand {\em et~al.\/}(2015)Varjavand, Ghomeshi, Dalir, Farsadizadeh
  \& Gorgij]{ar_Varjavand_2015}
{\sc \au{Varjavand, P.}, \au{Ghomeshi, M.}, \au{Dalir, A.~H.},
  \au{Farsadizadeh, D.} \& \au{Gorgij, A.~D.}} \yr{2015}  \at{Experimental
  observation of saline underflows and turbidity currents, flowing over rough
  beds}.  \jt{Canadian Journal of Civil Engineering}  \bvol{42}~(11),
  \pg{834–844}.

\bibitem[Wahab {\em et~al.\/}(2022)Wahab, Hoyal, Shringarpure \&
  Straub]{ar_Wahab_2022}
{\sc \au{Wahab, A.}, \au{Hoyal, D.~C.}, \au{Shringarpure, M.} \& \au{Straub,
  K.~M.}} \yr{2022}  \at{A dimensionless framework for predicting submarine fan
  morphology}.  \jt{Nature Communications}  \bvol{13},  \pg{7563}.

\bibitem[Wells {\em et~al.\/}(2010)Wells, Cenedese \& Caulfield]{ar_Wells_2010}
{\sc \au{Wells, M.}, \au{Cenedese, C.} \& \au{Caulfield, C.~P.}} \yr{2010}
  \at{The relationship between flux coefficient and entrainment ratio in
  density currents}.  \jt{Journal of Physical Oceanography}  \bvol{40}~(12),
  \pg{2713–2727}.

\bibitem[Wells \& Dorrell(2021)]{ar_Wells_2021}
{\sc \au{Wells, M.~G.} \& \au{Dorrell, R.~M.}} \yr{2021}  \at{Turbulence
  processes within turbidity currents}.  \jt{Annual Review of Fluid Mechanics}
  \bvol{53},  \pg{59--83}.

\bibitem[Zúñiga {\em et~al.\/}(2024)Zúñiga, Balachandar, Yang, Zhang,
  Smith, Loppi, Cantero \& Kerkemeier]{ar_Zuniga_2024}
{\sc \au{Zúñiga, S.~L.}, \au{Balachandar, S.}, \au{Yang, Y.}, \au{Zhang, Y.},
  \au{Smith, K.}, \au{Loppi, N.}, \au{Cantero, M.~I.} \& \au{Kerkemeier, S.}}
  \yr{2024}  \at{Planar wall plumes bounded by vertical and inclined surfaces}.
   \jt{Physics of Fluids}  \bvol{36},  \pg{035173}.

\end{thebibliography}

\end{document}